\documentclass[11pt,a4paper]{article}
\usepackage{jheppub}

\DeclareMathOperator{\tr}{tr}

\DeclareMathOperator{\str}{str}

\newcommand{\del}{\partial}
\newcommand{\delbar}{\bar{\partial}}

\title{\boldmath Extended higher spin holography and Grassmannian models}

\author[a]{Thomas Creutzig,}
\author[b]{Yasuaki Hikida}
\author[c]{and Peter B. R\o nne}

\affiliation[a]{Fachbereich Mathematik,
Technische Universit\"{a}t Darmstadt,
 Schlo\ss gartenstr. 7,
64289 Darmstadt, Germany}
\affiliation[b]{Department of Physics, Rikkyo University,
Toshima, Tokyo 223-8521, Japan}
\affiliation[c]{University of Luxembourg, Mathematics Research Unit, FSTC,
Campus Kirchberg, 6, rue Coudenhove-Kalergi, L-1359 Luxembourg-Kirchberg, Luxembourg}

\emailAdd{tcreutzig@mathematik.tu-darmstadt.de}
\emailAdd{hikida@rikkyo.ac.jp}
\emailAdd{peter.roenne@uni.lu}

\abstract{We propose holographic dualities between higher spin gravity theories extended with
Chan-Paton factor on AdS$_3$ and a large $N$ limit of two dimensional Grassmannian models
with and without supersymmetry.
These proposals are natural extensions of the duality without Chan-Paton factor,
and the extensions are motivated by a higher dimensional version of the duality,
which implies a possible relation to superstring theory via ABJ theory.
As evidence for the proposals,
we show that the free limit of the Grassmannian models have the
higher spin symmetry expected from the dual gravity theory.
Furthermore, we construct currents in the 't Hooft limit of the supersymmetric Grassmannian
model and compare them with the currents from the bulk theory.
One-loop partition function of the supergravity theory is reproduced by the
't Hooft limit of the Grassmannian model
after decoupling a gauge sector.}


\arxivnumber{1306.0466}

\begin{document}

\maketitle
\flushbottom

\section{Introduction}

Superstring theory includes a lot of massive higher spin states and in the massless limit they
are believed to be described by higher spin gauge theory.
Since the higher spin states are characteristic to superstring theory,
it should be useful to study higher spin gauge theory to improve our understanding of
superstring theory.
Non-trivial examples of higher spin gauge theories are
Vasiliev theories on AdS spacetimes \cite{Vasiliev:2003ev}, and they have been useful in the construction of simplified versions of the AdS/CFT correspondence.
In \cite{Sezgin:2002rt,Klebanov:2002ja} it was proposed that the Vasiliev theory on AdS$_4$
is dual to a U$(N)$ or O$(N)$ vector model, and the proposal has been investigated quite actively.
A few years ago, the authors of \cite{Gaberdiel:2010pz} (see \cite{Gaberdiel:2012uj} for a review)
conjectured a lower dimensional version of the duality
between the higher spin gauge theory on AdS$_3$ found in \cite{Prokushkin:1998bq}
and a large $N$ minimal model. Several generalizations of the duality have been proposed in
\cite{Ahn:2011pv,Gaberdiel:2011nt,Creutzig:2011fe,Creutzig:2012ar,Beccaria:2013wqa,Gaberdiel:2013vva}.

In \cite{Aharony:2011jz,Giombi:2011kc} one-parameter families of conformal models have been constructed
by coupling Chern-Simons gauge fields to three dimensional  U$(N)$ or O$(N)$ vector matter.
If we take the limit of large Chern-Simons level $k$, then the theory reduces to that of free vector fields with
U$(N)$ or O$(N)$ invariant condition in \cite{Klebanov:2002ja}.
It was conjectured in \cite{Giombi:2011kc} that the dual gravity theory is a parity violating Vasiliev theory,
and the parameter characterizing the parity violation is related to the ratio $k/N$ for the large $N,k$
limit of the Chern-Simons vector model. The Vasiliev theory can be extended with supersymmetry and
$\text{U}(M)$ Chan-Paton factor, and the dual field theory is argued to be the supersymmetric
$\text{U}(N)_k \times \text{U}(M)_{-k}$ Chern-Simons theory with bi-fundamental matter for large $N,k$ but finite $M$ \cite{Chang:2012kt}.
The most important example of the theory is the ${\cal N}=6$ ABJ theory,
which should be dual to  type IIA superstring theory on AdS$_4 \times \mathbb{C}$P$^3$ with discrete torsion
for large $N,M,k$ \cite{Aharony:2008ug,Aharony:2008gk}.
In this way, the ABJ theory connects the vector-like model dual to higher spin gauge theory and
the matrix-like model dual to superstring theory.
Based on these two dualities, the authors of \cite{Chang:2012kt} discussed an interesting new
relation between higher spin gauge theory and superstring theory.

Motivated by the higher dimensional duality (or triality), we would like to construct
a holographic duality involving higher spin theory with Chan-Paton factor on AdS$_3$.%
\footnote{Recently, a holography involving a higher spin theory with U(2) Chan-Paton factor is proposed in
an intriguing paper \cite{Gaberdiel:2013vva}. The dual CFT is based on a Wolf space and has a large
${\cal N}=4$ superconformal symmetry. The ${\cal N}=4$ CFTs on Wolf spaces were suggested to be dual to supersymmetric higher spin theories in \cite{Henneaux:2012ny}.
A possible relation to superstring theory on
$\text{AdS}_3 \times $S$^3 \times $S$^3 \times $S$^1$ is also discussed in  \cite{Gaberdiel:2013vva}.}
Even in lower dimensions, the higher spin gauge theory can be easily extended with
supersymmetry and/or $\text{U}(M)$ Chan-Paton factor as discussed in \cite{Prokushkin:1998bq}.
In the bosonic case, we propose the dual theory is the Grassmannian coset model defined as
\begin{align}
 \frac{\text{su}(N+M)_k }{\text{su}(N)_{k} \oplus \text{su}(M)_{k}
 \oplus \text{u}(1)_{kNM(N+M)}}
 \label{GC}
\end{align}
 with central charge
\begin{align}
 c = \frac{(k^2 - 1) MN (2k + M + N)}{(N + M + k ) (N+k ) (M+k)} \, .
\end{align}
In order to compare with the classical gravity theory, we consider the 't Hooft limit with
large $N,k$, but keeping $M$ finite and the ratio
\begin{align}
 \lambda_{M} = \frac{k}{N+M+k}
 \label{thooft}
\end{align}
finite.
In the limit, the central charge behaves as $c \sim MN \lambda_M (1 - \lambda_M)$, thus
the theory is vector-like. On the other hand, if we take the limit with large $N,M,k$ but finite $N-M$ and $k/N$,
then the model is matrix-like as $c \sim {\cal O} (N^2)$.
Thus the Grassmannian model connects a vector-like model dual to higher spin gauge theory and
a matrix-like model dual to string theory like the ABJ theory, even though we do not know what
is the string dual of the Grassmannian model currently.

There is a level-rank duality for the coset \eqref{GC} and the
dual coset is given by \cite{Bowcock:1988vs,Altschuler:1988mg}%
\footnote{The possibility of holographic duality involving this form of coset was already pointed out in \cite{Gopakumar:2012gd}.
See also \cite{Ahn:2013ota}.}
\begin{align}
 \frac{\text{su}(k)_N \oplus \text{su}(k)_M }{\text{su}(k)_{N+M}} \, .
 \label{GGcoset}
\end{align}
This form of coset with $M=1$ is the model dual to the Vasiliev theory without Chan-Paton factor
\cite{Gaberdiel:2010pz}. Therefore, we can say that the Grassmannian model \eqref{GC} is a
natural extension for  $M \neq 1$ as the dual theory. Moreover, at the large level limit with $k \to \infty$
in \eqref{GC}, the Grassmannian coset reduces to a free system with $NM$ complex bosons (see e.g. \cite{Bakas:1990xu}).
We will show that the free system reproduces the asymptotic symmetry of the dual gravity theory with
the corresponding parameter. The one-loop partition functions are also examined.

We can consider the ${\cal N}=2$ supersymmetric higher spin theory with $\text{U}(M)$ Chan-Paton factor
from \cite{Prokushkin:1998bq} in a similar way.
We would like to propose that the dual CFT is given by the
${\cal N}=(2,2)$ Grassmannian Kazama-Suzuki model
\cite{Kazama:1988qp,Kazama:1988uz}
\begin{align}
 \frac{\text{su}(N+M)_k \oplus \text{so}(2NM)_1}{\text{su}(N)_{k+M} \oplus \text{su}(M)_{k+N}
 \oplus \text{u}(1)_{\kappa}}
 \label{scoset}
\end{align}
with $\kappa = NM(N+M)(N+M+k)$, whose central charge is
\begin{align}\label{eq:centralcharge}
 c = \frac{3NMk}{N+M+k} \, .
\end{align}
We are interested in the 't Hooft limit where $N,k \to \infty$, but%
\footnote{We use different notations for the 't Hooft parameter such that \eqref{thooft}
and \eqref{thoofts} with $M=1$ reduce to
those of \cite{Gaberdiel:2010pz} and \cite{Creutzig:2011fe}, respectively.}
\begin{align}
 \lambda = \frac{N}{N+M+k}
 \label{thoofts}
\end{align}
and $M$ finite.
{}From the central charge, we can see that the model again connects a vector-like model in the
't Hooft limit and a matrix-like model with large $N,M,k$.
The coset model with $M=1$ is dual to the full ${\cal N}=2$ higher spin supergravity without Chan-Paton factor
as argued in \cite{Creutzig:2011fe}.

In order to support our duality conjecture,
we discuss the free limit and partition functions as in the bosonic case.
Moreover, we investigate currents in the Grassmannian model in the 't Hooft limit, but not in the free limit.
From the bulk side we calculate several OPEs of these currents and successfully compare these to the CFT results.
In the 't Hooft limit with large $N,k$ but finite $M,\lambda$ and outside the free limit,
we see that extra bosonic su$(M)$ current should be coupled. Thus the cosets we are dealing with are
effectively of the form as
\begin{align}
 \frac{\text{su}(N+M)_k }{\text{su}(N)_{k} \oplus \text{su}(M)_{k}
 \oplus \text{u}(1)_{kNM(N+M)}} \oplus \ \text{su}(M)_{k}
\simeq  \frac{\text{su}(N+M)_k }{\text{su}(N)_{k}
 \oplus \text{u}(1)_{kNM(N+M)}}
\label{GC2}
\end{align}
for the bosonic case and
\begin{align}
 \frac{\text{su}(N+M)_k \oplus \text{so}(2NM)_1}{\text{su}(N)_{k+M} \oplus \text{su}(M)_{k+N}
 \oplus \text{u}(1)_{\kappa}}
\oplus \ \text{su}(M)_{k+N} \simeq \ \frac{\text{su}(N+M)_k \oplus \text{so}(2NM)_1}{\text{su}(N)_{k+M}
 \oplus \text{u}(1)_{\kappa}}
\label{scoset2}
\end{align}
for the supersymmetric case.
In \eqref{scoset2}  there is a subtlety since we add only bosonic $\text{su}(M)$.
In other words, if we write the coset in terms of ${\cal N}=1$ currents, then fermionic components of ${\cal N}=1$ $\text{su}(M)$ currents should be decoupled.%
\footnote{After the paper \cite{Gaberdiel:2013vva} appeared in the arXiv,
we have noticed that the model \eqref{scoset2} with $M=2$ coincides with their coset.}
With this form of coset \eqref{scoset2} we show that the gravity partition function
can be reproduced from the CFT partition function in the 't Hooft limit.

This paper is organized as follows:
In section \ref{Mvhst} we study the higher spin gravity theories with $\text{U}(M)$ Chan-Paton factor and
derive the partition functions of the gravity theories.
In section \ref{free} we propose that $MN$ complex free bosons (and $MN$ complex free fermions) are
dual to the higher spin theories with a particular mass parameter based on symmetry arguments.
Here we assign the $\text{U}(N)$ invariant condition, but the $\text{U}(M)$ symmetry is treated as a global symmetry.
Furthermore, we relate the free systems to the Grassmannian cosets \eqref{GC} and \eqref{scoset}.
We study the Grassmannian models in some details for the bosonic case in section \ref{BGm}
and for the supersymmetric case in section \ref{SGm}.
In particular we discuss the roles of the $\text{su}(M)$ symmetry.
We examine the characters of the coset models and reproduce the gravity partition functions in the free limits.
In section \ref{sec:producttheory} we construct currents in the Grassmannian model in the 't Hooft limit and compare them with those from the bulk theory.
Conclusion and discussions are given in section \ref{conclusion}.
In appendix \ref{glM} some properties of the $\text{gl}(M)$ extended higher spin algebra are given, and we construct $w_\infty [\lambda]$ algebras extended with a matrix algebra
from ghost systems in appendix \ref{shs}.
In appendix \ref{PF} we find agreement between the supergravity partition function
and that of \eqref{scoset2} in the 't Hooft limit but outside the free limit.

\section{Matrix valued higher spin theory}
\label{Mvhst}

In this section, we introduce the ${\cal N}=2$ higher spin gauge theory with U$(M)$ Chan-Paton factor
on AdS$_3$ found in \cite{Prokushkin:1998bq}.
The bosonic truncation of the higher spin theory with $M=1$ was used to construct a duality with a large $N$
minimal model \eqref{GGcoset}  in \cite{Gaberdiel:2010pz}.
Moreover, the total ${\cal N}=2$ higher spin supergravity with $M=1$ was argued to be dual to the $\mathbb{C}$P$^N$
Kazama-Suzuki model \eqref{scoset}  in \cite{Creutzig:2011fe}.
Since the equations of motion are written in terms of a $*$-product, we can easily replace the
fields by $M \times M$ matrix valued ones.
Using the reality condition of the fields discussed in \cite{Prokushkin:1998bq}, we can construct a higher spin gauge theory with
U$(M)$ Chan-Paton factor.
In the next section we introduce a bosonic truncation of the higher spin supergravity
with U$(M)$ Chan-Paton factor in \cite{Prokushkin:1998bq}, and in section \ref{N2sugra}
we examine the full ${\cal N}=2$ supergravity theory.
In particular, we obtain the one-loop partition functions of these higher spin theories
with U$(M)$ Chan-Paton factor, even though they easily follow from those without
Chan-Paton factor.

\subsection{A bosonic truncation}

Before including the Chan-Paton factor, let us start from a bosonic truncation of the higher spin
supergravity in \cite{Prokushkin:1998bq}.
The gravity theory includes gauge fields with spin $s=2,3,4,\ldots$ and massive complex scalar fields.
The gauge fields can be described by Chern-Simons gauge theory with the action \cite{Blencowe:1988gj}
\begin{align}
 S = S_\text{CS} [ A] - S_\text{CS} [\tilde A] \, ,
\end{align}
where
\begin{align}
 S_\text{CS} [A] = \frac{k_\text{CS}}{4 \pi} \int \text{tr} \left( A \wedge dA + \frac{2}{3} A \wedge A \wedge A \right) \, .
\end{align}
The gauge fields $A, \bar A$ take values in an infinite dimensional higher spin algebra
$\text{hs}[\lambda]$, where $\text{hs}[\lambda]$ can be truncated to sl$(N)$ at $\lambda = \pm N$.\footnote{To be more precise, the generators of spin higher than $N$ form an ideal, $\mathcal{I}_N$, at these integer values $\lambda =\pm N$. Factoring $\text{hs}[\lambda]$ by this ideal gives sl$(N)$, $\text{hs}[\lambda]/\mathcal{I}_N\cong sl(N)$. }
The generators of hs[$\lambda$] are given by
\begin{align}
V^s_m \qquad ( s=2,3,4,\ldots, m = - s+1 , -s + 2 , \ldots , s-1 ) \, ,
\end{align}
and the commutation relations are (see, e.g., \cite{Gaberdiel:2011wb})
\begin{align}
  [ V^s_m , V^t_n ] = \sum_{u=2,4,\ldots}^{s+t-|s-t|-1} g^{st}_u (m,n;\lambda) V_{m+n}^{s+t-u} \, .
\end{align}
We do not need the explicit form of the structure constants, but they can be found in \cite{Pope:1989sr}
 (see also \cite{Gaberdiel:2011wb}).
The products of gauge fields can be described by
the ``lone star product'' \cite{Pope:1989sr} as
\begin{align}
  V^s_m * V^t_n  = \frac12 \sum_{u=1,2,\ldots}^{s+t-|s-t|-1} g^{st}_u (m,n;\lambda) V_{m+n}^{s+t-u} \, .
  \label{lsp}
\end{align}
The one-loop partition function for spin $s$ gauge field was computed in \cite{Gaberdiel:2010ar} as
\begin{align}
 Z_B^{(s)} = \prod^{\infty}_{n=s} \frac{1}{|1-q^n|^2} \, .
 \label{zbs}
\end{align}
In order to compute thermal partition function,  the space-time is set to be Euclidean AdS$_3$.
The modulus of the boundary torus is $q = \exp \tau$.
The total partition function for the gauge sector is
\begin{align}
 Z_0 = \prod_{s=2}^\infty Z^{(s)}_B
 \label{bZ0}
\end{align}
by taking products of those for spin $s$ gauge fields.

Along with the higher spin gauge fields, the gravity theory includes two complex scalars
with mass
\begin{align}
M_B^2 = - 1 + \lambda ^2\, .
\end{align}
We may decouple one complex scalar depending on the
situation \cite{Chang:2011mz}.%
\footnote{
In the refined version of the conjecture, we should include only one complex scalar
\cite{Gaberdiel:2012ku,Gaberdiel:2012uj}. However, at the strict 't Hooft limit of the minimal
model, the original proposal with two complex scalars still works. It is argued that
non-perturbative objects of gravity theory play roles of one complex scalar outside the
't Hooft limit.
}
Now we set $0 \leq \lambda \leq 1$ and then  two types of boundary condition
for the scalar fields are allowed.
We choose to assign the standard boundary condition for one scalar and the alternative
boundary condition for the other scalar, then the conformal weights for operators dual
to the scalar fields are $(h_\pm,h_\pm)$ with
\begin{align}
h_\pm = \frac{1 \pm \lambda}{2} \, .
\label{bdcw}
\end{align}
The partition function for a complex scalar with dual conformal weight $(h,h)$ is
\cite{Giombi:2008vd,David:2009xg}
\begin{align}
 Z^h_\text{scalar} = \prod_{n,m = 0}^\infty \frac{1}{(1- q^{h+n} \bar q^{h + m})^2 } \, .
\end{align}
In order to compare the result with the CFT computation, it is useful to rewrite the
expression in terms of characters of Young diagrams as \cite{Gaberdiel:2011zw,Candu:2012jq}
\begin{align}
 Z^h_\text{scalar} = \sum_{\Lambda_l , \Lambda_r} | \text{ch}_{\Lambda_l} (U(h)) \text{ch}_{\Lambda_r} (U(h)) |^2 \, .
\label{bZh}
\end{align}
The character of $\Lambda$ is defined as
\begin{align}
 \text{ch}_\Lambda (U(h)) = \sum_{T \in \text{Tab}_\Lambda} \prod_{j \in T} q^{h+j} \, , \quad
 U(h)_{jj} = q^{h+j} \, , \label{bschur}
\end{align}
where Tab$_\Lambda$ denotes a Young tableau of a shape $\Lambda$.
In a Young tableau we assign a non-negative number $c_{i,j}$ to the box
 at the $i$-th row and the $j$-th column of $\Lambda$ with the rule that
$c_{i,j} \leq c_{i,j+1}$ and $c_{i,j} < c_{i+1,j}$.
Combined with the gauge sector, the total one-loop partition function of the gravity theory
with two complex scalars is
\begin{align}
\label{bulkbpf}
 Z^\text{Bulk} &= Z_0  Z^{h_+}_\text{scalar} Z^{h_-}_\text{scalar}   \\
                   &= Z_0 \sum_{\Lambda_l , \Lambda_r , \Xi_l , \Xi_r}
                   |\text{ch}_{\Lambda_l} (U(h_+)) \text{ch}_{\Lambda_r} (U(h_+)) \text{ch}_{\Xi_l} (U(h_-)) \text{ch}_{\Xi_r} (U(h_-)) | ^2 \, . \nonumber
\end{align}

It is easy to extend the above analysis to the case with  $\text{U}(M)$ Chan-Paton factor.
The fields are now of $M \times M$ matrix form instead of the functional form.
Here the reality condition is assigned as in \cite{Prokushkin:1998bq}.
The field equations for the master fields with $M=1$ are
written in terms of the $*$-product in  \cite{Prokushkin:1998bq}.
Since $*$-product has already non-abelian nature, the field equations for the fields with $M \neq 1$
can be obtained by using the product of the  $*$-product and the usual
matrix multiplication.
For the physical gauge fields in the master fields, the $*$-product
reduces to the lone star product in \eqref{lsp}.
Thus now the gauge fields take values in algebra generated by
\begin{align}
V_m^s \otimes t_a
\end{align}
with $t_a$ being a generator of $\text{gl}(M)$.
Notice that we should include $V_0^1 \otimes t_a$ since they cannot be decoupled
for $M \neq 1$ in general, but we can still decouple $V_0^1 \otimes 1_M$ where $1_M$ is the identity in $\text{gl}(M)$.
For $\lambda =1$ we can show that generators $V_0^1 \otimes t_a$ can be decoupled as well.
The scalar fields are also replaced by $M \times M$ matrix valued fields.
In the free limit, the only effect of introducing the $\text{U}(M)$ Chan-Paton factor is that the
number of fields simply is multiplied by $M^2$.
Therefore, the one-loop partition function is simply given by the $M^2$-th power of the one of
$M=1$; that is
\begin{align}
 Z^\text{Bulk}_M =  (Z^\text{Bulk} )^{M^2} (Z_B^{(1)})^{M^2-1}\, .
 \label{zmbulk}
\end{align}
The last factor comes from the fact that elements generated by $V_0^1 \otimes t_a$ with $t_a$ as a generator of $\text{sl}(M)$ do not decouple for $\lambda \neq 1$.

\subsection{The ${\cal N}=2$ supergravity}
\label{N2sugra}

Since the inclusion of the Chan-Paton factor can be made in the same way as in the bosonic case,
we first review the ${\cal N}=2$ higher spin supergravity in \cite{Prokushkin:1998bq}
without Chan-Paton factor.
The massless sector can be described by $\text{shs}[\lambda] \oplus \text{shs}[\lambda]$
Chern-Simons theory, where $\text{shs}[\lambda]$ can be truncated to the  Lie superalgebra
$\text{sl}(N+1|N)$ at $\lambda = N+1$.
See, e.g., \cite{Bergshoeff:1991dz,Creutzig:2011fe,Hanaki:2012yf,Moradi:2012xd}
for the details of $\text{shs}[\lambda]$ Lie superalgebra.
The gauge sector includes bosonic gauge fields with spin $s=2,3,\ldots$ generated by $V^{(s)+}_m$
$(m < |s|, m \in \mathbb{Z})$
and $s=1,2,\ldots$ generated by  $V^{(s)-}_m$ $(m < |s|, m \in \mathbb{Z})$. Moreover there are
two sets of fermionic gauge fields with spin $s= 3/2,5/2,\ldots$ generated by $V^{(s)\pm}_m$ $(m < |s|, m \in \mathbb{Z}+1/2)$.
The one-loop partition functions of  spin $s -1/2$ gauge fields
are given by \cite{Creutzig:2011fe}
\begin{align}
 Z_F^{(s -1/2)} = \prod_{n=s}^{\infty} |1 + q^{n -1/2}|^2 \, ,
\end{align}
thus the one-loop partition function of the massless sector is
\begin{align}
 {\cal Z}_0 = \prod_{s=2}^\infty Z^{(s)}_B (Z^{(s-1/2)}_F )^2 Z^{(s-1)}_B \, .
 \label{sZ0}
\end{align}
We can include the $\text{U}(M)$ Chan-Paton factor as in the bosonic case by replacing the
scalar coefficients in front of generators of $\text{shs}[\lambda] \oplus \mathbb{C}$ in gauge fields
by $M \times M$ matrix valued ones. The gauge fields thus take values in
algebra generated by $V_m^{(s)\pm} \otimes t_a$
with $\text{gl}(M)$ generator $t_a$. Here $V^{(1)+}_0 \otimes t_a$ are included, but
$V^{(1)+}_0 \otimes 1_M$ with the identity $1_M$ in gl$(M)$ is decoupled.
We can show that $V^{(1)+}_0 \otimes t_a$ can be excluded for $\lambda = 0$.

The massive sector includes four massive complex scalars and four massive Dirac fermions.
The masses are parameterized by a parameter $\lambda$ as  \cite{Prokushkin:1998bq}
\begin{align}
 ( M^B_+ )^2 = - 1 + (\lambda - 1 )^2 \, , \qquad
 ( M^B_- )^2 = - 1 + \lambda^2 \, , \qquad
 ( M_\pm^F )^2 =  ( \lambda - \tfrac12 )^2 \, ,
\end{align}
where two of the four bosons/fermions have the $+$ index and the others have the $-$ index.
We choose the boundary conditions such that the dual conformal dimensions
are \cite{Creutzig:2011fe,Candu:2012jq}
\begin{align}
 (\Delta^B_+ , \Delta^F_{\pm} , \Delta^B_- ) = (2 - \lambda , \tfrac32 - \lambda , 1- \lambda ) \, ,
 ~ (\lambda , \tfrac12 + \lambda , 1 + \lambda ) \, .
\end{align}
With this choice, the supersymmetry is enhanced to be ${\cal N}=(2,2)$ at the boundary.
The one-loop partition function of
a Dirac spin $1/2$ fermion with $(h,h-1/2)$ and $(h-1/2,h)$ is \cite{Creutzig:2011fe}
\begin{align}
 Z^h_\text{spinor} = \prod_{n,m = 0}^\infty (1+ q^{h+n} \bar q^{h -1/2 + m})^2 (1+ q^{h - 1/2 +n} \bar q^{h + m})^2 \, .
\end{align}
For an ${\cal N}=2$ multiplet we have
\begin{align}
 {\cal Z}^h_\text{matter} = Z^h_\text{scalar} (Z^{h+1/2}_\text{spinor})^2 Z^{h+1/2}_\text{scalar}\, .  \label{sZh0}
\end{align}
As in the bosonic case, we rewrite the partition function in terms of supercharacter of the Young diagram $\Lambda$
as \cite{Candu:2012jq}
 \begin{align}
 {\cal Z}^h_\text{matter} =
 \sum_{\Lambda_l , \Lambda_r} | \text{sch}_{\Lambda_l} ({\cal U}(h)) \text{sch}_{\Lambda_r} ({\cal U}(h)) |^2 \, ,
 \label{sZh}
\end{align}
where the supercharacter of $\Lambda$ is
\begin{align}
 \text{sch}_\Lambda ({\cal U} (h) ) = \sum_{T \in \text{STab}_\Lambda} \prod_{i \in T} q^{h+ i/2} \, ,
 \quad {\cal U} (h)_{ii} = (-1)^i q^{h+i/2} \, .
 \label{sschur}
\end{align}
The Young supertableau STab$_\Lambda$ is defined with a non-negative integer $c_{i,j}$
in each box at the $i$-th row and the $j$-th column of a Young diagram $\Lambda$.
The numbers should satisfy $c_{i,j} \leq c_{i,j+1}$ and $c_{i,j} \leq c_{i+1,j}$,
moreover $c_{i,j} < c_{i,j+1}$ if both $c_{i,j}$ and $c_{i,j+1}$ are odd and
$c_{i,j} < c_{i+1,j}$ if both $c_{i,j}$ and $c_{i+1,j}$ are even
(see, for instance, \cite{Candu:2012jq} for more details).

In total, the one-loop partition function of the gravity theory is
\begin{align}
{\cal Z}^\text{Bulk} & = {\cal  Z}_0 {\cal Z}^{\frac{\lambda}{2}}_\text{matter} {\cal Z}^{\frac{1- \lambda}{2}}_\text{matter}    \\
                 &= {\cal Z}_0 \sum_{\Lambda_l , \Lambda_r , \Xi_l , \Xi_r}
                   |\text{sch}_{\Lambda_l} ({\cal U}(\tfrac{\lambda}{2})) \text{sch}_{\Lambda_r} ({\cal U}(\tfrac{\lambda}{2})) \text{sch}_{\Xi_l} ({\cal U}(\tfrac{1-\lambda}{2})) \text{sch}_{\Xi_r} ({\cal U}(\tfrac{1-\lambda}{2})) |^2 \, . \nonumber
\end{align}
With the $U(M)$ Chan-Paton factor, all fields are now of $M \times M$ matrix form including the
fermionic fields. Therefore, the one-loop partition function is given by roughly
 the $M^2$-th power as
\begin{align}
 {\cal Z}^\text{Bulk}_M =  ({\cal Z}^\text{Bulk} )^{M^2} (Z_B^{(1)})^{M^2-1}\, .
 \label{zmbulks}
\end{align}
The last factor is the contribution from the spin 1 fields generated by $V_0^{(1)+} \otimes t_a$
with $\text{sl}(M)$ generator $t_a$, which do not decouple in general.

\section{Dual free CFTs}
\label{free}

In order to find CFTs dual to the gravity theories, it is useful to utilize their asymptotic symmetry.
For the gravity theory used in \cite{Gaberdiel:2010pz}, the gauge sector can be described by
$\text{hs}[\lambda] \oplus \text{hs}[\lambda]$ Chern-Simons theory.
Assigning the asymptotically AdS boundary condition, the classical symmetry around the AdS boundary
was found to be
$W_\infty[\lambda]$ symmetry \cite{Henneaux:2010xg,Campoleoni:2010zq,Gaberdiel:2011wb,Campoleoni:2011hg}. The $W_\infty[\lambda]$ symmetry algebra is non-linear and generated by
\begin{align}
W^s_m \, \qquad (s=2,3,\ldots, m \in \mathbb{Z}) \, .
\end{align}
In the large central charge limit, the wedge subalgebra generated by $W^s_m$ with
$m = - s+1 , - s+2 , \ldots , s-1$ reduces to the $\text{hs}[\lambda]$ Lie algebra
(see, e.g., \cite{Gaberdiel:2011wb}).
It was shown in \cite{Gaberdiel:2011wb,Gaberdiel:2012ku} that the $W_\infty[\lambda]$ symmetry algebra is realized by the coset
\eqref{GGcoset} with $M=1$ as
\begin{align}
 \frac{\text{su}(N)_k \oplus \text{su}(N)_1}{\text{su}(N)_{k+1}} \, .
 \label{GGcoset0}
\end{align}
We need to take the 't Hooft limit where $N,k \to \infty$ with
\begin{align}
 \lambda_\text{GG} = \frac{N}{N+k}
\end{align}
finite and we identify $\lambda_\text{GG} =\lambda$.

It is not easy to extend the analysis for the case with $M \neq 1$ since the symmetry
algebra of the coset \eqref{GGcoset} would be quite complicated.
However, the $W_\infty[\lambda]$ algebra becomes simple for $\lambda =0,1$,
and the algebra is realized by a system of $N$ complex free fermions for $\lambda=0$
or $k$ complex free bosons for $\lambda =1$.
In this section we mainly focus on the limit where coset models considered reduce to free theories.
The coset \eqref{GGcoset} is mapped to the Grassmannian coset \eqref{GC} with the level-rank duality
\cite{Bowcock:1988vs,Altschuler:1988mg} where $k$ and $N$ are exchanged.
After reviewing the known results with $M=1$ in the next subsection,
we argue in section \ref{bfreeCP} that the system reduces to $MN$ complex free bosons in the $\lambda_M = 1$ limit of
the 't Hooft parameter \eqref{thooft} in
the Grassmannian coset \eqref{GC}.
Moreover, we can see that the free system generates an algebra with correct wedge subalgebra for
the $M \neq 1$ cases. In section \ref{sfree} we examine the supersymmetric case.
For that case, we just need to consider  $MN$ complex free fermions
in addition to the $MN$ complex free bosons.
The free system arises in the $\lambda =0$ limit of the 't Hooft parameter \eqref{thoofts} in the
supersymmetric Grassmannian coset \eqref{scoset}.
In section \ref{umdec}, we analyze the symmetry of the coset \eqref{scoset} in the
't Hooft limit, but with $\lambda \neq 0$.

\subsection{Free limits of the $W_N$ minimal model}

Let us start from the large level $k$ limit of the coset \eqref{GGcoset0}, which means $\lambda_\text{GG}=0$.
Introducing $N$ free complex fermions $\psi_i$ $(i=1,2,\ldots,N)$ satisfying
\begin{align}
 \psi_i (z) \psi^\dagger_j (0) \sim \frac{\delta_{ij}}{z} \, ,
\end{align}
we can define currents
\begin{align}
 J^{(s)} (z) = \frac{2^{-s+1}(s-1)!}{(2s-3)!!} \sum_{i=1}^N \sum_{l=0}^{s-1} (-1)^l
  \begin{pmatrix} s- 1 \\ l \end{pmatrix} ^2 \partial^{s-l-1} \psi^\dagger_i \partial^{l} \psi_i \, ,
  \label{bw0}
\end{align}
where
\begin{align}
 J^{(s)} (z) = \sum_{m= - \infty}^\infty \frac{W^{s}_m}{z^{m+s}} \, .
 \end{align}
Here the spin of current is denoted by $s \, (=1,2,3,\cdots)$.
They generate a linear algebra $w_{1+\infty}$ with $c=2N$ \cite{Bergshoeff:1990yd}.
The wedge subalgebra is $\text{hs}[0] \oplus \text{u}(1)$, and decoupling the $\text{u}(1)$
current we obtain the non-linear algebra $W_\infty [0]$ from the linear algebra  $w_{1+\infty}$
\cite{FigueroaO'Farrill:1992cv}.
The dual operator with $h=1/2$ is given by
\begin{align}
 {\cal O} (z, \bar z ) = \sum_{i=1}^N  \psi_i (z) \otimes \bar \psi_i (\bar z)
\end{align}
and its complex conjugate. Here $\bar \psi_i$ are free complex fermions in the anti-holomorphic sector.
At $\lambda =0$, the two operators dual to scalar fields with different boundary conditions have the
same conformal weight $h=1/2$ as in \eqref{bdcw}.
Thus they are indistinguishable at $\lambda = 0$ and we can pick up only one of them  as discussed in  \cite{Gaberdiel:2011aa}.
By bosonizing the fermions we have $N$ free real bosons. The summation over the $i$-index corresponds
to the $\text{U}(N)$ invariant condition discussed in \cite{Gaberdiel:2011aa}.
In order to construct a modular invariant theory, we need to add sectors twisted by the $\text{U}(N)$
action. In this paper we neglect the sectors since they decouple from the $\text{U}(N)$ invariant sector in the 't Hooft limit.

Next we consider the symmetry algebra for the $\lambda_\text{GG}=1$ limit of the coset \eqref{GGcoset0}.
The $W_\infty[1]$ algebra reduces to the linear algebra $W_\infty^\text{PRS}$ of
\cite{Pope:1989sr}, and its wedge subalgebra is hs[1].
The central charge of the coset \eqref{GGcoset0} is $c = 2 k + {\cal O}(1/N)$,
and the algebra with central charge $c=2k$ can be realized by $k$ free complex bosons
as shown in
\cite{Bakas:1990ry}. We choose the normalization of the free bosons $\phi_i$ $(i=1,2,\ldots,k)$ and
their complex conjugates $\phi^\dagger_i$ such as to have OPEs
\begin{align}
 \phi_i (z) \phi^\dagger_j (0) \sim - \delta_{ij} \ln z ~.
\end{align}
Then, the generators of $W_\infty[1]$ with $c=2k$ are written as
\begin{align}
 J^{(s)} (z) = - \frac{2^{-s+1}s!}{(2s-3)!!}  \sum_{i=1}^k \sum_{l=0}^{s-2} (-1)^l  \frac{1}{s-1}
 \begin{pmatrix} s- 1 \\ l \end{pmatrix}
 \begin{pmatrix} s- 1 \\ s-l \end{pmatrix} \partial^{s-l-1} \phi^\dagger_i \partial^{l+1} \phi_i \,
 \label{bw1}
\end{align}
with $s=2,3,\ldots$.
In terms of free bosons, the operator dual to the scalar field can be given by
\begin{align}
 {\cal O} (z, \bar z ) = \sum_{i=1}^k \partial \phi_i (z) \otimes \bar \partial \phi_i (\bar z)
\end{align}
and its complex conjugate. The conformal dimension of the operator is $h=1$.
The other operator has dimension $h=0$ according to \eqref{bdcw},
and this value is the same as that of the vacuum. Thus it is natural to pick up one complex
scalar field whose dual operator has conformal dimension $h=1$.

The free boson system in the dual coset description \eqref{GC} with $M=1$ can be understood as
\begin{align}
 \frac{\text{su}(N+1)_k}{\text{su}(N)_k \oplus \text{u}(1)_{k N(N+1)}} \, .
\end{align}
According to \cite{Bakas:1990xu}, the coset in the classical limit with
$k \to \infty$ has the same algebra $W_\infty [1]$ generated by $N$ free complex bosons.
It was claimed that the su$(N)_k$ current algebra is flattened out to be the $\text{u}(1)^{N^2 -1}$
current algebra in large $k$ limit. This implies that the limit of the Grassmannian model has
$\text{u}(1)^N \otimes \text{u}(1)^N$ currents generated by $N$ complex bosons.
{}From the complex bosons we can construct operators generating the $W_\infty[1]$ algebra
as in \eqref{bw1} with $k$ replaced by $N$.

\subsection{Adding the Chan-Paton factor}
\label{bfreeCP}

{}From the above consideration, we can guess that the CFT dual to the higher spin theory with
$\text{U}(M)$ Chan-Paton factor at $\lambda= 1$ is given by the free system of $NM$ complex bosons $\phi_{i A}$
with $i = 1,2,\ldots , N$ and $A=1,2,\ldots ,M$. We assign the following OPEs to the bosons
\begin{align}
 \phi_{iA} (z) \phi_{jB}^\dagger (0) \sim - \delta_{ij} \delta_{AB} \ln z \, .
 \label{fbops}
\end{align}
We extend the currents \eqref{bw1} for $M=1$ in a natural way as
\begin{align}
 [J^{(s)} (z)]_{AB} = - \frac{2^{-s+1}s!}{(2s-3)!!}  \sum_{i=1}^N \sum_{l=0}^{s-2} (-1)^l  \frac{1}{s-1}
 \begin{pmatrix} s- 1 \\ l \end{pmatrix}
 \begin{pmatrix} s- 1 \\ s-l \end{pmatrix} \partial^{s-l-1} \phi^\dagger_{iA} \partial^{l+1} \phi_{iB}
 \label{bw1m}
\end{align}
with $s=2,3,\ldots$.
Notice that spin $s=1$ currents do not appear and the wedge subalgebra is
given by $\text{hs}[1] \otimes {\cal M} $ with the matrix algebra ${\cal M}$ as desired.
The operators dual to $M \times M$ matrix valued scalar fields are
\begin{align}
 [ {\cal O} (z, \bar z ) ]_{AB} = \sum_{i=1}^N \partial \phi_{iA} (z) \otimes \bar \partial \phi_{iB} (\bar z)
 \label{freeps}
\end{align}
and their complex conjugates.

A point here is that we have contracted the $i$ index, but not the $A$ index,
and because of this we have $M \times M$ matrix valued currents invariant under the $\text{U}(N)$ symmetry.
In the case of ABJ theory, the theory includes $\text{U}(N)_k \times \text{U}(M)_{-k}$ Chern-Simons gauge
fields coupled with bi-fundamental fields $A^{(a)}_{\alpha i}$ and $B^{(b)}_{j \beta}$
with $a,b =1,2$, $i,j =1,2,\ldots,N$ and $\alpha , \beta = 1,2,\ldots, M$ \cite{Aharony:2008ug,Aharony:2008gk}.
When we consider the duality with a higher spin gauge theory, we should take the large $N,k$ limit, but
keep $M$  finite \cite{Chang:2012kt}.
The operators dual to the bulk higher spin fields are proposed to be of the form
$\sum_{i=1}^N A^{(a)}_{\alpha i}B^{(b)}_{i \beta}$, since gluing  for $i$-index is much stronger than that for $\alpha$-index when $N \gg M$.
Therefore, the $\text{U}(N)$ and $\text{U}(M)$ gauge groups play very different roles,
and this was interpreted in \cite{Chang:2012kt} as a confinement/deconfinement phase transition
of the ABJ theory in the bulk 't Hooft parameter $\lambda^\text{Bulk}= M/N$.

We would like to claim that the free boson system arises in the $\lambda_M =1$ limit
of the Grassmannian model \eqref{GC}
as for the $M=1$ case.
An element of $SU(M+N)$ in the numerator of the coset \eqref{GC} may be expressed by an
$(M+N) \times (M+N)$ matrix. Here and in the following argument we neglect the trace part just
for simplicity.
We embed $SU(N)$ and $SU(M)$ gauge groups in the denominator as
\begin{align}
 \begin{pmatrix}
  A & 0 \\
  0 & D
 \end{pmatrix} \in \text{SU}(M+N)
\end{align}
for $A \in \text{SU}(M)$ and $D \in \text{SU}(N)$.
Since $A,D$ are gauged out, there are off-diagonal components left.
It is natural to expect that these $NM$ complex elements reduce to $NM$ complex free boson
in large $k$ limit.
Indeed, we can also obtain the same conclusion by making use of the argument in
 \cite{Bakas:1990xu} as in the $M = 1$ case.

\subsection{Supersymmetric extension}
\label{sfree}

We can discuss the supersymmetric case in the same way.
Without Chan-Paton factor, the gauge fields of the ${\cal N}=2$ higher spin theory in \cite{Prokushkin:1998bq} take values in the
shs$[\lambda]$ algebra, and the asymptotic symmetry of the gravity theory is found to be the
${\cal N}=2$ $W_\infty [\lambda]$ algebra \cite{Creutzig:2011fe,Henneaux:2012ny,Hanaki:2012yf,Candu:2012tr}.
If we consider the large central charge limit, then the wedge subalgebra of  ${\cal N}=2$ $W_\infty [\lambda]$
should reduce to shs$[\lambda]$. Moreover, at $\lambda = 0$, it is known that ${\cal N}=2$ $W_\infty [0]$
can be generated by complex bosons and fermions \cite{Bergshoeff:1990yd}.
In other words, the ${\cal N}=2$ higher spin theory with $\lambda = 0$ should be dual to $N$ complex
free bosons and $N$ complex free fermions.

As in the bosonic case, we can easily extend the above argument to the case with
U$(M)$ Chan-Paton factor, since we just need to introduce $NM$ free fermions
 $\psi_{i A}$ with $i = 1,2,\ldots , N$ and $A=1,2,\ldots ,M$ satisfying
\begin{align}
 \psi_{iA} (z) \psi^\dagger_{jB} (0) \sim \frac{\delta_{ij} \delta_{AB}}{z}
\end{align}
in addition to the $NM$ free bosons satisfying \eqref{fbops}.
The bosonic currents are
\begin{align}
[ J^{(s)-} (z) ]_{AB} = \frac{2^{-s+1}(s-1)!}{(2s-3)!!} \sum_{i=1}^N \sum_{l=0}^{s-1} (-1)^l
  \begin{pmatrix} s- 1 \\ l \end{pmatrix} ^2 \partial^{s-l-1} \psi^\dagger_{iA} \partial^{l} \psi_{iB}
 \label{bw0m}
\end{align}
with $s=1,2,\ldots$ as a natural extension of \eqref{bw0} and $[ J^{(s)+} (z) ]_{AB} = [ J^{(s)} (z) ]_{AB} $
in \eqref{bw1m} with $s=2,3,\ldots$. The fermionic ones are
\begin{align}
 [ F^{(s)} (z) ]_{AB} = \frac{2^{-s+2} (s-1)!}{(2s-1)!!} \sum_{l=0}^{s-2} (-1)^l
  \begin{pmatrix} s- 2 \\ l \end{pmatrix}
 \begin{pmatrix} s- 1 \\ l \end{pmatrix} \partial^{s-l-1} \phi^\dagger_{iA} \partial^{l} \psi_{iB}
\end{align}
and $[ F^{(s)\dagger} (z) ]_{AB}$ is simply obtained by replacing $\phi^\dagger_{iA}$ and $\psi_{iB}$
with $\phi_{iA}$ and $\psi^\dagger_{iB}$, respectively.
Here the label $s$ for $[ F^{(s)} (z) ]_{AB}$ and $[ F^{(s)\dagger} (z) ]_{AB}$ take values $s=2,3,\ldots$.
At $\lambda = 0$ we see that $[J^{(1)+}(z)]_{AB}$ do not appear.

{}From the free fields, we can construct operators as
\begin{align}
& [{\cal O} (z, \bar z) ]_{AB}^{(\frac12 , \frac12)} = \sum_{i=1}^N \psi_{iA} (z) \otimes \bar \psi_{iB} (\bar z) \, ,
 \qquad
  [{\cal O} (z, \bar z) ]_{AB}^{(1 , 1)} = \sum_{i=1}^N \partial \phi_{iA} (z) \otimes \bar \partial \phi_{iB} (\bar z) \, , \\
& [{\cal O} (z, \bar z) ]_{AB}^{(1 , \frac12)} = \sum_{i=1}^N \partial \phi_{iA} (z) \otimes \bar \psi_{iB} (\bar z) \, ,
 \qquad
  [{\cal O} (z, \bar z) ]_{AB}^{(\frac12, 1)} = \sum_{i=1}^N \psi_{iA} (z) \otimes \bar \partial \phi_{iB} (\bar z) \, ,
\end{align}
and their complex conjugates. They form one of two ${\cal N}=2$ multiplets with
\begin{align}
 (\Delta^B_+ , \Delta^F_{\pm} , \Delta^B_-) = (2 ,\tfrac32 , 1) \, ,
\end{align}
and this implies that the gravity theory dual to the free theory
includes only half of the sets of scalars and fermions.

We can see that the free system arises in the $\lambda = 0$ limit of the 't Hooft parameter \eqref{thoofts}
in the supersymmetric Grassmannian model \eqref{scoset}.
There is an affine $\text{so}(2NM)_1$ algebra in the numerator of the coset \eqref{scoset},
which yields $NM$ complex fermions. The other parts are the same as the bosonic
coset in \eqref{GC} in the large $k$ limit, and thus lead to $NM$ complex bosons.
Totally we have the free system with $NM$ complex bosons and $NM$ complex fermions.

\subsection{Beyond the free 't Hooft limit}
\label{umdec}

In this subsection, we will see how a $\text{U}(M)$-valued $\mathcal N=2$ super $W$-algebra
appears in the 't Hooft limit of the coset for arbitrary $\lambda$. The construction depends crucially on normalization of fields.
Recall that we want to have fields that have norm scaling with $N$. For example the Virasoro field has
to have norm $c/2$. But the coset algebra is usually constructed using fields having a well-defined norm, that is a
finite norm. Our strategy will thus be to find the coset fields having finite norm, and then
rescale them to have norm scaling with $N$. It will turn out that the coset fields already
provide a $\text{U}(M)$-valued algebra, and the rescaled fields will then generate the desired
$\mathcal N=2$ super $W$-algebra with independent strong generators of dimension
$1,1,3/2,3/2,2,2,...$.

Let us first consider the coset \eqref{scoset} at finite $N$.
Normally, an efficient way to determine the spectrum of the coset algebra is to take the
large $k$ limit and invoke invariance under deformation to finite $k$. In this limit, the coset can be described
by $2NM$ free bosons $\partial\phi_{iA},\partial\phi^\dagger_{Ai}$ and $2NM$ free fermions $\psi_{iA},\psi^\dagger_{Ai}$ in the tensor product of the fundamental representation of SU($N$) and SU($M$)
and its conjugate, with OPEs
\begin{equation}
 \partial\phi_{iA}(z,\bar z)\partial\phi^\dagger_{Bj}(w,\bar w)= \frac{\delta_{AB}\delta_{ij}}{(z-w)^2}\, ,\qquad
\psi_{iA}(z) \psi^\dagger_{Bj}(w)= \frac{\delta_{AB}\delta_{ij}}{(z-w)}\, ,
\end{equation}
as mentioned above.
The coset algebra is then the invariant subalgebra under the action of SU($N$)$\times$SU($M$)$\times$ U($1$).
Using Weyl's first theorem of invariant theory for the unitary group \cite{We} one obtains that the coset algebra is generated by normal ordered products of type
\begin{equation}
\begin{split}
&\sum_{i=1}^N\sum_{A=1}^M :\partial^a \phi_{iA}(z)\partial^b\phi^\dagger_{Ai}(z):\, ,\qquad
\sum_{i=1}^N\sum_{A=1}^M :\partial^a \psi_{iA}(z)\partial^b\phi^\dagger_{Ai}(z):\, ,\\
&\sum_{i=1}^N\sum_{A=1}^M :\partial^a \phi_{iA}(z)\partial^b\psi^\dagger_{Ai}(z):\, ,\qquad
\sum_{i=1}^N\sum_{A=1}^M :\partial^a \psi_{iA}(z)\partial^b\psi^\dagger_{Ai}(z):\, .
\end{split}
\end{equation}
This gives a superalgebra with generator of spin $1,3/2,3/2,2,2,5/2,5/2,3,...$, i.e. an $\mathcal N=2$ super $W$-algebra.

If we consider the case of large $N$, we have to be careful.
First define the currents of the $\text{su}(N)_k$ subalgebra by $K^a(z)$, those of the $\text{su}(M)_k$
subalgebra by $J^\alpha(z)$
and the remaining currents carry the representation $\text{f}_N\otimes \bar {\text{f}}_M\oplus \bar{\text{f}}_N\otimes \text{f}_M$,
where $\text{f}_L$ denotes the fundamental representation of $\text{su}(L)$ and the bared expression its conjugate. We can thus label them
by $I^{Ai},I^{iA},i=1,...,N$ and $A=1,...,M$.
Further, the fermions $:\psi_{iA}\psi^\dagger_{Aj}:$ define $\text{u}(N)$ currents of level $M$,
whose $\text{su}(N)_M$ subalgebra currents (the traceless ones) we denote by $k^a$.
Analogously, the fermions $j^{AB}=:\psi_{iA}\psi^\dagger_{Bi}:$ define $\text{u}(M)$ currents of level $N$,
\begin{equation}
j^{AB}(z)j^{CD}(w)\sim \frac{N\delta_{AD}\delta_{BC}}{(z-w)^2}+\frac{\delta_{BC}j^{AD}(w)-\delta_{AD}j^{CB}(w)}{(z-w)}
\end{equation}
whose $\text{su}(M)_N$ subalgebra currents we denote by $j^\alpha$.
In addition, the $\text{u}(1)$ defined by the fermions is denoted by $h$ and the one of the affine algebra by $H$.
The coset algebra is then defined to be the subalgebra that commutes with the currents
\begin{equation}\label{eq:curr}
 K^a(z)+k^a(z), \qquad J^\alpha(z)+j^\alpha(z)
\end{equation}
and with the $\text{u}(1)$-current $H+h$.

Consider now the large $N,k$ limit with $\lambda=N/(N+M+k)$ fixed and $M$ finite.
For finite $N$, the choice of basis of currents is irrelevant, but we require to take the limit in such a way, that currents have finite norm.
The invariants under $K^a(z)+k^a(z)$ as the level is $k+M$ become in the large $k$ limit
the $\text{SU}(N)$-group invariants as before.
But for the $J^\alpha+j^\alpha$ we have to be more careful as the level of both $J^\alpha$
and $j^\alpha$ is taken to infinity; we have to rescale the currents
\begin{equation}
X^\alpha(z)=\frac{J^\alpha(z)+j^\alpha(z)}{\sqrt{k+N}}
\end{equation}
and also $H+h$ has to be rescaled by $1/\sqrt{k+N}$.
Now, the fields
\begin{equation}\label{eq:cosfields1}
\begin{split}
&\frac{1}{N}\sum_{i=1}^N:\partial^a \psi_{iA}(z)\partial^bI^{Bi}(z):\, ,\qquad
\frac{1}{N}\sum_{i=1}^N :\partial^a I^{iA}(z)\partial^b\psi^\dagger_{Bi}(z):
\end{split}
\end{equation}
commute with $X^\alpha(z)$ and the rescaled $\text{u}(1)$-current,
and they have finite norm.
These give us two $\text{U}(M)$-valued fermionic fields of dimension $3/2,5/2,...$.
The $J^\alpha$ together with $H$ form $\text{u}(M)$ of level $k$; let us denote those currents by
matrix indices $J^{AB}$. Then
\begin{equation}
\sqrt{\frac{1-\lambda}{k}}J^{AB}(z)-\sqrt{\frac{\lambda}{N}}j^{AB}(z)
\end{equation}
commute with the $X^{AB}$.
In summary, in the limit $k,N\rightarrow\infty$, we found $\text{U}(M)$-valued
coset fields of dimension $1,3/2,3/2,2,5/2,5/2,3,7/2,...$ of finite norm. We remark that these are certainly
not all coset fields.

We are interested in taking the limit, where the norm of fields scales as $N$.
We thus rescale our coset fields \eqref{eq:cosfields1} by $\sqrt{N}$,
\begin{equation}\label{eq:cosfields2}
\begin{split}
G^{AB}_{ab}&=\frac{1}{\sqrt{N}}\sum_{i=1}^N:\partial^a \psi_{iA}(z)\partial^bI^{Bi}(z):\, ,\qquad
\bar G^{AB}_{ab}=\frac{1}{\sqrt{N}}\sum_{i=1}^N :\partial^a I^{iA}(z)\partial^b\psi^\dagger_{Bi}(z):\, .
\end{split}
\end{equation}
But now, some OPE computations reveal that we get closed operator product algebra, if we include the
fields
\begin{equation}
J^{AB} \, ,  \quad V^{AB}_{ab}= \sum_{i=1}^N :\partial^a \psi_{iA}(z)\partial^b\psi^\dagger_{Bi}(z):\, ,
\quad W^{AB}_{ab}=\frac{1}{N} \sum_{i=1}^N :\partial^a I^{Ai}(z)\partial^bI^{iB}(z): \, .
\end{equation}
That is we get $\text{U}(M)$-valued fields of dimension $1,1,3/2,3/2,2,2,...$ as desired.
Let us provide one example of an OPE illustrating this:
\begin{equation}
\begin{split}
G^{AB}_{00}(z)\bar G^{CD}_{00}(w) &\sim
\frac{k\delta_{AD}\delta_{BC}}{(z-w)^{3}}+\frac{\delta_{AD}J^{BC}(w)+\frac{k}{N}\delta_{BC}V^{AD}_{00}(w)}{(z-w)^2}+\\
&\qquad +\frac{\delta_{AD}W^{BC}_{00}(w)+\frac{k}{N}\delta_{BC}V^{AD}_{10}(w)}{(z-w)}\, .
\end{split}
\end{equation}
Finally, we mention that one can decouple one $\text{u}(1)$-current.

\section{Bosonic Grassmannian model}
\label{BGm}

In this section, we study the bosonic Grassmannian model \eqref{GC}
\begin{align}
 \frac{\text{su}(N+M)_k}{\text{su}(N)_k \oplus \text{su}(M)_k \oplus \text{u}(1)_{\kappa }}
 \label{bgm}
\end{align}
with $\kappa = k N M (N+M)$.
We consider the 't Hooft limit where $N,k \to \infty$, but keeping $M$ and the 't Hooft parameter
$\lambda_M$ defined in \eqref{thooft} finite.
In the previous section, we have observed that the coset reduces to the system of $MN$ complex free bosons
at $\lambda_M =1$, and the $\text{U}(M)$ invariant condition is not assigned to construct
conserved currents.
In the next subsection, we first
introduce the bosonic Grassmannian coset \eqref{bgm} and
then in section \ref{umdecon} we see how
the ``$\text{U}(M)$ deconfinement'' is realized in the coset language.
In section \ref{charcp} and \ref{charfree}
we reproduce the gravity partition function from the coset model \eqref{bgm} in the $\lambda_M = 1$ limit.

\subsection{Primary states}

In the Grassmannian coset \eqref{bgm}, the denominator $\text{su}(M)  \oplus \text{su}(N)  \oplus \text{u}(1)$ is embedded into
$\text{su}(N+M)$. The way of embedding is determined by those of
$\text{SU}(M) \times \text{SU}(N)   \times \text{U}(1) $ into $\text{SU}(N+M)$.
We use
\begin{align}
 \imath_1 (v,u,w) =
 \begin{pmatrix}
   w^N u & 0\\
  0 & \bar w^M v
 \end{pmatrix} \in \text{SU}(N+M) \, , \label{emb1}
\end{align}
where $u \in \text{SU}(M)$, $v \in \text{SU}(N)$ and $w \in \text{U}(1)$.
In terms of $\text{su}(N+M)$, the $\text{u}(1)$ current $K$ is expressed as $\text{diag}(N,\ldots,N,-M,\ldots,-M)$,
which has the OPE $K(z)K(0) \sim k M N (N+M) z^{-2}$.

The states of the Grassmannian model \eqref{bgm} are labeled by the representations of the algebra
appearing in the coset expression. Let $\Lambda_L$ be the highest-weights of
$\widehat{\text{su}}(L)$ of level $k$. We then use the convention that we also
denote the corresponding irreducible highest-weight module by $\Lambda_L$.
For $u(1)_k$ we use $m \in \mathbb Z_k$. Then the states of the coset are
obtained by the module decomposition as
\begin{align}
 \Lambda_{N+M}  = \bigoplus_{\Lambda_N , \Lambda_M , m}
  (\Lambda_{N+M}; \Lambda_N , \Lambda_M , m ) \otimes
  \Lambda_N \otimes \Lambda_M \otimes m \, .
  \label{decomposeb}
\end{align}
Using the orthogonal basis $\epsilon_i$ $(i=1,2,\ldots,N+M)$ with $\epsilon_i \cdot \epsilon_j = \delta_{i,j}$,
the highest weights are expressed as
\begin{align}
& \Lambda_{N+M} = \sum_{i=1}^{N+M} \lambda^{N+M}_i \epsilon_i
 - \frac{|\Lambda_{N+M}|}{N+M} \sum_{j=1}^{N+M} \epsilon_j \, , \\
& \Lambda_{M} = \sum_{i=1}^{M} \lambda^{M}_i \epsilon_i
 - \frac{|\Lambda_{M}|}{M} \sum_{j=1}^{M} \epsilon_j \, , \quad
 \Lambda_{N} = \sum_{i=1}^{N} \lambda^{N}_i \epsilon_{i+M}
 - \frac{|\Lambda_{N}|}{N} \sum_{j=1}^{N} \epsilon_{j+M} \, . \nonumber
\end{align}
Here $\lambda^L_j$ is the number of boxes in the $j$-th row of the Young diagram corresponding to
$\Lambda_L$, and $|\Lambda_L|$ is the sum of the boxes of the Young diagram.
The $\text{u}(1)$ charge is embedded in $\text{su}(N+M)$ as
\begin{align}
 \omega_m = \frac{m}{NM(N+M)} \left(  N \sum_{i=1}^M \epsilon_i - M \sum_{j=1}^N \epsilon_{j+M}\right)
\end{align}
which satisfy $\omega_m (K) = m$. The embedding is possible only when
\begin{align}
 \Lambda_{N+M} - \Lambda_M - \Lambda_N - \omega_m  \in \Delta_{N+M}
\end{align}
with $\Delta_{N+M}$ as the root system of $\text{su}(N+M)$.
This condition leads to selection rules as
\begin{align}
 \frac{|\Lambda_{N+M}|}{N+M} - \frac{|\Lambda_M|}{M} + \frac{m}{M(N+M)} \in \mathbb{Z} \, , \quad
 \frac{|\Lambda_{N+M}|}{N+M} - \frac{|\Lambda_N|}{N} - \frac{m}{N(N+M)} \in \mathbb{Z} \, ,
 \label{selectionb}
\end{align}
where $m$ is defined modulo $\kappa$. In principle,
we have to take into account field identifications \cite{Gepner:1989jq}, but for large $k$ limit
we can neglect it.

The conformal dimension of the state $(\Lambda_{N+M} ; \Lambda_N , \Lambda_M , m )$
is
\begin{align}
 h = n + h^{M+N,k}_{\Lambda_{M+N}}  - h^{N,k+M}_{\Lambda_N} - h^{M,k+N}_{\Lambda_M}
 - h_m \label{cwg}
\end{align}
with an integer $n$. Here
\begin{align}
 h^{L,K}_{\Lambda} = \frac{C_2 (\Lambda_L)}{K+L} \, , \qquad h_m = \frac{m^2}{2 \kappa}  \label{cw}
\end{align}
with $C_2 (\Lambda_L)$ as the second Casimir of $\text{su}(L)$.
Notice that $h^{M,k+N}_{\Lambda_M} = 0$ in the large $k$ limit, thus
the states of all possible $\Lambda_M$ degenerate.

\subsection{States dual to bulk matter fields}
\label{umdecon}

In order to construct the primary states of the full CFT, we need to take a product of chiral
and anti-chiral primary states.
In the free limit with $\lambda_M = 1$, we constructed operators  as in \eqref{freeps}, where the $\text{U}(N)$
invariant condition is assigned, but  $\text{U}(M)$ indices are still free.
We would like to see how this condition is realized in the coset description.

We consider the coset of the form  $\mathfrak{g} / \mathfrak{h}$, then the coset primary fields may be given as
$\Psi_{RL}$, where $R$ and $L$ are representations of algebras $\mathfrak{g}$ and $\mathfrak{h}$.
The fields can be constructed by those of affine algebras $\hat {\mathfrak{g}}$ and $\hat {\mathfrak{h}}$
\cite{Gawedzki:1988hq,Gawedzki:1988nj,Bratchikov:2000mh,Papadodimas:2011pf}.
For $\hat {\mathfrak{g}}$, we define $\chi^i_{\hat {\mathfrak{g}}, RL}$ which are in the affine algebra built from $R$
and transforms in $L$. Here $i=1,2,\ldots, \text{dim} L$. The grade in the affine algebra is related to
the integer $n$ in \eqref{cwg}.
For $\hat {\mathfrak{h}}$ we define $\chi^{\hat {\mathfrak{h}}^*, \bar L}_i$, which transform in $\bar L$. Then the coset
fields are
\begin{align}
 \Psi_{RL} = \sum_{i=1}^{\text{dim} L} \chi^i_{\hat {\mathfrak{g}} , RL} \chi^{\hat {\mathfrak{h}}^*, \bar L}_i \, .
\end{align}
For our case, we should set
\begin{align}
 \mathfrak{g} = \text{su} (N+M)_k  \, , \qquad
 \mathfrak{h} =  \text{su} (N)_{k} \oplus \text{su} (M) _{k} \oplus \text{u} (1)_\kappa \, .
\end{align}

Let us consider the states dual to the bulk matter fields.
Generic primary states of the coset are given by the products of the two sectors as
\begin{align}
  (\Lambda_{N+M} ; \Lambda_N , \Lambda_M , m ) \otimes
   (\Lambda_{N+M} ' ; \Lambda_N ' , \Lambda_M ' , m ') \, .
\end{align}
The two simplest examples are
\begin{align}
 (\text{f};0,\text{f},N) \otimes (\bar{\text{f}},0,\bar{\text{f}},N) \, , \qquad
 (0;\bar{\text{f}},\text{f},N+M) \otimes (0;\text{f},\bar{\text{f}},N+M) \, ,
 \label{states}
\end{align}
where $\text{f}$ $(\bar{\text{f}})$ denotes (anti-)fundamental representation.
There are also complex conjugated ones with exchanging $\text{f}$ and $\bar{\text{f}}$.
We choose the simplest sets for $\Lambda_{N+M}, \Lambda_N$ and  $\Lambda_{N+M} ', \Lambda_N '$,
but $\Lambda_M , \Lambda_M '$ and $m, m'$ are chosen such as to satisfy the selection rules \eqref{selectionb}
in the 't Hooft limit.  See \eqref{largeNu1} and \eqref{lm} below.

The field corresponding to the first state in \eqref{states} is
\begin{align}
 \Psi_{(\text{f};0,\text{f},N) } = \sum_{i=1}^{M} \chi^i_{\hat{\text{su}}(N+M) , (\text{f};0,\text{f},N) }
  \chi^{\hat{\text{su}}(M), \bar{\text{f}}}_i \chi^{\hat{\text{u}}(1), N}  \, .
\end{align}
With large $k \to \infty$, but finite $M$, we expect that the $\text{su}(M)$ sector decouples. Thus the chiral  sector may be given by
\begin{align}
 \chi^i_{\hat g , (\text{f};0,\text{f},N) }  \chi^{\hat{\text{u}}(1), N}
\end{align}
with label $i= 1, 2, \ldots , M$ for the fundamental representation of $\text{su}(M)$.
Multiplying the anti-chiral sector we have $(i, \bar \jmath)$ labels
with $\bar \jmath = 1,2,\ldots , M$ for the anti-fundamental representation of $\text{su}(M)$.
Thus the state can be identified with a scalar field with $\text{U}(M) $ Chan-Paton factor.
The field corresponding to the second state in \eqref{states} is
\begin{align}
 \Psi_{(0;\bar{\text{f}},\text{f},N+M) } = \sum_{l=1}^N \sum_{i=1}^{M} \chi^{l,i}_{\hat{\text{su}}(N+M)  , (0;\bar{\text{f}},\text{f},N+M) }
  \chi^{\hat{\text{su}}(N), \text{f}}_l \chi^{\hat{\text{su}}(M), \bar{\text{f}}}_i \chi^{\hat{\text{u}}(1), N+M}  \, .
\end{align}
Decoupling the $\text{su}(M)$ sector, the field becomes
\begin{align}
  \sum_{l=1}^N  \chi^{l,i}_{\hat{\text{su}}(N+M)  , (0;\bar{\text{f}},\text{f},N+M) }
  \chi^{\hat{\text{su}}(N), \text{f}}_l \chi^{\hat{\text{u}}(1), N+M}  \, .
  \label{state2}
\end{align}
Multiplying the anti-chiral sector, the field is dual to another bulk scalar field with $\text{U}(M) $ Chan-Paton factor.

We may see the relation to the free limit in section \ref{free} as follows.
Above we assumed that the $\text{su}(M)$ sector decouples in the large $k$ limit.
However, for the $\text{su}(N)$ sector, we first take the 't Hooft limit with
$N,k \to \infty$ and $\lambda_M = k/(N+k+M)$ finite, then take $\lambda_M \to 1$ limit.
In this limit, the $\text{su}(N)$ sector becomes free, but the $\text{SU}(N)$
invariant constraint should be left, see \cite{Gaberdiel:2011aa}.
The state in \eqref{state2} may reduce to the expression in section \ref{free}
by integrating over  $\chi^{\hat{\text{su}}(N), \text{f}}_l $.

\subsection{The character of the chiral part}
\label{charcp}

The CFT is given by the product of chiral and anti-chiral sectors,
and the combination is fixed such that the torus amplitude is modular invariant.
One of the conditions is $h - \bar h \in \mathbb{Z}$ for bosonic states, where $h$ and $\bar h$ are
the conformal dimension of the states for the chiral and anti-chiral sectors.
For finite $k$, this is a quite strong condition and a typical one is given by a diagonal
modular invariant, where charge conjugated representations are paired.
However, for infinite $k$, we have seen that states with all possible $\Lambda_M$ have
the same conformal weight. Therefore, the above condition is satisfied even without
using the diagonal modular invariant with respect to the $\Lambda_M$ label.
In this subsection we examine the chiral part of the partition function.
In the next subsection we consider how we should pair the chiral and the anti-chiral parts,
where we first take the $\lambda_M = 1$ limit of the 't Hooft parameter in \eqref{thooft},
and then discuss the generalization with $\lambda_M \neq 1$.

The characters of $\text{su}(L)_k$ are denoted as
\begin{align}
 \text{ch}^{L,k}_\Lambda (q, e^H) = \text{tr}_\Lambda q^{L_0} e^H \, .
 \label{chsu}
\end{align}
Here $L_0$ is the zero mode of Virasoro generator and $H$ is an element of
Cartan subalgebra of $\text{su}(L)$.
The character of $\text{u}(1)_\kappa$ is
\begin{align}
 \Theta^{\kappa}_m (q,w) = \text{tr}_m q^{L_0} w^{J_0}
  = \sum_{l \in \mathbb{Z}} w^{m + \kappa l} \frac{q^{\frac{1}{2 \kappa} (m + \kappa l )}}{\prod_{n=1}^\infty (1 - q^n)}
 \, . \label{chu1}
\end{align}
Then the character of the coset \eqref{bgm}
\begin{align}
 b^{N,M,k}_{\tilde \Xi } (q) = \text{tr}_{\tilde \Xi} q^{L_0} \, , \qquad
 \tilde \Xi =  (\Lambda_{N+M} ; \Lambda_N , \Lambda_M , m )
 \label{chcoset}
\end{align}
is given by the decomposition \eqref{decomposeb} as
\begin{align}
 \text{ch}_{\Lambda_{N+M}}^{N+M,k} (q , \imath_1 (u, v,w))
  = \sum_{\Lambda_N , \Lambda_M , m} b^{N,M,k}_{\tilde \Xi} (q)
 \text{ch}_{\Lambda_{N}}^{N,k} (q , v )  \text{ch}_{\Lambda_{M}}^{M,k} (q , u ) \Theta^\kappa_m (q,w) \, .
 \end{align}
The embedding is defined in \eqref{emb1}.

Let us study the large $k$ behavior of the coset character \eqref{chcoset}.
At large $k$, the $\text{su}(L)_k$ character behaves as
 (see, e.g., \cite{Bouwknegt:1992wg,Gaberdiel:2011zw,Candu:2012jq})
\begin{align}
 \text{ch}^{L,k}_\Lambda (q,e^H) \sim
 \frac{q^{h^{L,k}_\Lambda}\text{ch}^L_\Lambda (e^H)}{\prod_{n=1}^\infty [(1-q^n)^{L-1} \prod_{\alpha \in \Delta_L}(1 - q^n e^{\alpha (H)}) ] } \, ,
 \label{asymsu}
\end{align}
where
$\Delta_L$ represent the roots of $\text{su}(L)$ and $\text{ch}^L (e^H)$ is the character of the finite Lie algebra $\text{su}(L)$.
Similarly, the $\text{u}(1)_\kappa$ character behaves as
\begin{align}
 \Theta^\kappa_m (q,w) \sim \frac{q^{h^\kappa_m} w^m}{\prod_{n=1}^\infty (1 - q^n)}  \, .
 \label{asymu1}
\end{align}
Combining all the factors in the coset \eqref{bgm}, we have
\begin{align}
 \text{ch}^{N+M}_{\Lambda_{N+M}} (\imath_1 (u,v,w)) \vartheta (q , \imath_2 (u,v,w))
  = \sum_{\Lambda_N, \Lambda_M , m} a^{N,M}_{\tilde \Xi}  (q) \text{ch}^{N}_{\Lambda_N} (v) \text{ch}^{M}_{\Lambda_M} (u)
   w^m \, ,
   \label{eqm}
\end{align}
where $a^{N,M}_{\tilde \Xi} (q)$ is related to the leading term at large $k$ as
\begin{align}
 b^{N,M,k}_{\tilde \Xi} (q) \sim q^{h^{N,M,k}_{\tilde \Xi}} a^{N,M}_{\tilde \Xi} (q) \, , \qquad
 h^{N,M,k}_{\tilde \Xi} = h^{N+M,k}_{\Lambda_{N+M}} - h^{N,k}_{\Lambda_{N}}- h^{M,k}_{\Lambda_{M}}
 - h ^\kappa_m \, .
\end{align}
The factor
\begin{align}
 \vartheta (q ,  \imath_2 ( u,v,w) ) = \prod_{n=0}^\infty \prod_{i=1}^N \prod_{A=1}^M
 \frac{1}{(1 - \bar w^{N+M} \bar u_A v_i q^{n+1}) (1 - w^{N+M} u_A \bar v_i q^{n+1})}
\end{align}
can be interpreted as the partition function of  $NM$ complex free bosons $\phi_{iA}$,
$\phi_{jB}^\dagger$  with $i =1,2,\ldots,N$ and $A=1,2,\ldots,M$.
Here $\phi_{iA}$ belongs to $(\text{f} , \bar {\text{f}})$ for $\text{su}(N) \oplus\text{su}(M)$
and has $-(M+N)$ charge for $\text{u}(1)$. Moreover, $\phi_{jB}^\dagger$ belongs to
$(\bar{\text{f}},  \text{f})$ for $\text{su}(N) \oplus \text{su}(M)$
and has $(M+N)$ charge for $\text{u}(1)$.

At large $N$ it is convenient to express the $\Lambda_L$ of $\text{su}(L)$
in terms of a pair of two Young diagrams as  $\mathbf{\Lambda}_L = (\Lambda_L^l , \Lambda_L^r )$.
Denoting $|\mathbf{\Lambda}_L|_- = |\Lambda_L^r| - |\Lambda_L^l|$, the selection rule \eqref{selectionb} uniquely fixes
\cite{Candu:2012jq}
\begin{align}
 m =  N |\mathbf{\Lambda}_{N+M}|_- - (N+M)  |\mathbf{\Lambda}_{N}|_-
\label{largeNu1}
\end{align}
in the large $N$ limit. This also leads to
\begin{align}
 |\Lambda_M| = |\mathbf{\Lambda}_{N+M}|_- - |\mathbf{\Lambda}_N|_-  \label{lm}
\end{align}
modulo $M$.
The $w$-dependent factor in \eqref{eqm} may be removed
as
\begin{align}
  \widetilde{\text{ch}}^{N+M}_{\mathbf{\Lambda}_{M+N}} (\imath_1 (u,v,w)) \vartheta (q , \imath_2 (u,v,w))
  = \sum_{\mathbf{\Lambda}_N, \Lambda_M } a^{N,M}_{\mathbf{\Lambda}_{M+N};\mathbf{\Lambda}_{N}, \Lambda_M}  (q) \widetilde{\text{ch}}^{N}_{\mathbf{\Lambda}_N} (v \bar w^{N+M}) \text{ch}^{M}_{\Lambda_M} (u)
   \label{eqm1}
\end{align}
by  rewriting the $\text{su}(L)$ character by the $\text{u}(L)$ one as
\begin{align}
 \text{ch}^L_{\mathbf{\Lambda}_L } (v) w^{|\mathbf{\Lambda}_L|_-} = \widetilde{\text{ch} }^L_{\mathbf{\Lambda}_L} (v w) \, .
\label{chu}
\end{align}
In the right hand side, the pair of two Young diagrams labels the irreducible representation of u$(L)$.

\subsection{The character at the free limit}
\label{charfree}

Let us set $\mathbf{\Lambda}_{M+N}=0$, then we have
\begin{align}
 \vartheta (q , \imath_2 (u,v,w))
  = \sum_{\mathbf{\Lambda}_N, \Lambda_M }a^{N,M}_{0;\mathbf{\Lambda}_{N}, \Lambda_M}  (q) \widetilde{\text{ch}}^{N}_{\mathbf{\Lambda}_N} (v \bar w^{N+M}) \text{ch}^{M}_{\Lambda_M} (u) \, .
\end{align}
Thus this sector reduces to  $MN$ complex free bosons in the bi-fundamental
representation of $\text{su}(N) \oplus \text{su}(M)$, and the free bosons are
organized in terms of representations of $\text{u}(N) \oplus \text{su}(M)$ in the coset model
\eqref{bgm}.  This is consistent with the fact that the coset reduces to the free boson system
in the $\lambda_M = 1$ limit.%
\footnote{A similar relation between a free boson  theory and the $\lambda = N/(N+k) \to 0$ limit of
the coset \eqref{GGcoset} is obtained in \cite{Gaberdiel:2011aa}.
See also \cite{Fredenhagen:2012rb,Fredenhagen:2012bw}.}
For the other value of $\lambda_M$, we should also consider
the sector with $\mathbf{\Lambda}_{M+N} \neq 0$ as we will discuss below.
In this subsection, we mainly study the partition function of the free boson theory as a coset partition function in the $\lambda_M = 1$ limit. Here the $\text{U}(N)$ invariant condition is assigned, but the
$\text{SU}(M)$ symmetry is treated as a global symmetry like a flavor symmetry.
Later we discuss the case with $\lambda_M \neq 1$.

We denote the  bosonic modes by $j_{aA} = \partial \phi_{aA}$  with
$a =1,2, \ldots , N$ as the index of the $\text{u}(N)$ fundamental representation and
$A=1,2,\ldots , M$ as the index of $\text{su}(M)$ anti-fundamental representation.
There are also complex conjugates as $ j_{aA}^\dagger = \partial \phi_{aA}^\dagger$.
The full Fock space of the free system is spanned by
\begin{align}
 \prod_{l} j^{\dagger a_l C_l}_{- t_l - 1}
 \prod_m j^{b_m D_m} _{- u_m - 1} | 0 \rangle \, ,
\end{align}
where $ t_l , u_m$ are non-negative integers and $| 0 \rangle$ is the vacuum state.
The $\text{u}(N)$ invariants are then given by the linear combinations of the ``basic'' invariants as \cite{We}
\begin{align}
  \prod_{t,u = 0}^\infty \left( \sum_{a=1}^N j^{\dagger a A_{tu}}_{-t - 1 } j^{a B_{tu}}_{-u - 1}\right)^{L_{tu}}
 | 0 \rangle \, ,
 \nonumber
\end{align}
where $L_{tu}$ are non-negative integers.
Since there are $M^2$ times more bi-linears than the ones for $M=1$,
the contribution from the $\text{u}(N)$ invariants
 is computed as (see (3.68) of  \cite{Candu:2012jq})
\begin{align}
Z_0^\text{CFT} (q) = \left( \prod_{s=2}^\infty Z_B^{(s)} \right)^{M^2} \, ,
\end{align}
which is the one-loop partition function for the massless gauge sector of the gravity theory, see \eqref{bZ0}.
Recall that some of the spin 1 gauge fields decouple at $\lambda_M=1$.

For $\text{u}(N)$ non-invariants, we should pair chiral and anti-chiral parts.%
\footnote{For more proper treatment, see appendix \ref{PF}.}
{}From the experience of the free limit in section \ref{free}, we propose to
 combine
\begin{align}
  j^{\dagger a A}_{-t - 1 } \bar \jmath^{a B}_{-u - 1} \, , \qquad
 \bar  \jmath^{\dagger a A}_{-t - 1 }  j^{a B}_{-u - 1} \, ,
\end{align}
where the $\text{u}(N)$ index $a$ is the same (but not summed over here) for the chiral and anti-chiral sectors. In other words, for a fixed $\text{su}(M)$ label $A$, states of the form
\begin{align}
 \prod_{l=1}^{n_j} j^{a_l A} _{- u_l - 1}| 0 \rangle
\end{align}
produces a $\text{u}(N)$ tensor of the shape $\Lambda^l$. Then we take a pair with the
anti-chiral state from $\bar \jmath^{\dagger b_m B}$ corresponding to the $\text{u}(N)$ tensor of the shape $\bar {\Lambda^l}$.
If we sum over all possible $u_l$ while keeping $n_j = |\Lambda^l|$ fixed, then the partition function
is computed as (see (2.50) of \cite{Candu:2012jq} for a free fermion system)
\begin{align}
 | \text{ch}_{\Lambda^l} (U(1)) |^2 \, ,
\end{align}
where the character is defined in \eqref{bschur}.
Since we have $M^2$ sets with $A,B=1,2,\ldots,M$, we have
\begin{align}
Z^\text{CFT} (q) &=
\left ( \sum_{\mathbf{\Lambda}_N}  |  \text{ch}_{\Lambda^{l}_N}  ({U} (1)) \text{ch}_{\Lambda^{r}_N} ({U} (1)) |^2 \right)^{M^2}
 Z_0^\text{CFT} (q)  \\
  &=  \left( Z_0 Z^1_\text{scalar} \right)^{M^2} \, , \nonumber
\end{align}
where we have taken into account the $\text{u}(N)$ invariants discussed above.
This CFT partition function is the same as the bulk one, but including only one
complex scalar with $h_+ = 1$. This is consistent with the argument in section \ref{free}.

From the experience of the $M=1$ case in \cite{Gaberdiel:2011zw,Candu:2012jq},
we should include the sectors with $\mathbf{\Lambda}_{N+M} \neq 0$ for $\lambda_M \neq 1$,
and the dual gravity theory should have two complex scalars with two different boundary
conditions. The detailed analysis can be found in appendix \ref{PF}
(but only for the supersymmetric case).
For $\lambda_M \neq 1$, the decoupling of spin 1 gauge fields does not occur on the gravity side,
and this effect can be easily included in the CFT side.
These decoupled spin 1 gauge fields correspond to  extra su$(M)$ currents in the CFT as discussed in
section \ref{umdec}.
The negative-mode contribution to the character  is simply given by
\begin{align}
 \text{ch} '_{M}(q, e^{H}) = \frac{1}{\prod_{n=1}^\infty \prod_{\alpha \in \Delta_M} (1 - q^n e^{\alpha (H)})} \, .
\end{align}
Taking $H \to 0$ limit, we have
\begin{align}
| {\text{ch} '}^{M}(q, 1) |^2  =  \left(\prod_{n=1}^\infty\frac{1}{|1- q^n|^2}\right)^{M^2 - 1} = (Z_B^{(1)})^{M^2 - 1} \, ,
\label{nmsum}
\end{align}
which reproduces the extra factor in \eqref{zmbulk}.

\section{Supersymmetric Grassmannian model}
\label{SGm}

We now move on to study the supersymmetric Grassmannian model \eqref{scoset}
\begin{align}
 \frac{\text{su}(N+M)_k \oplus \text{so}(2NM)_1}{\text{su}(N)_{k+M} \oplus \text{su}(M)_{k+N}
 \oplus \text{u}(1)_{\kappa}}
 \label{scosetp}
\end{align}
with $\kappa = MN (N+M)(N+M+k)$.
This coset is proposed to be dual
to the ${\cal N}=2$ higher spin supergravity in \cite{Prokushkin:1998bq} with $\text{U}(M)$ Chan-Paton factor.
We study the 't Hooft limit of the coset where $N,k \to \infty$ while keeping $M$ and the 't Hooft parameter
\eqref{thoofts}
finite.
In the next section we review the supersymmetric Grassmanian model \eqref{scosetp},
see e.g. \cite{Naculich:1997ic}.
In section \ref{scharcp} we see that the coset is related to a system with $MN$ complex free bosons and
$MN$ complex free fermions in the language of the partition function.
In section \ref{scfl} we reproduce the gravity partition function
from the $\lambda = 0$ limit of the coset \eqref{scosetp}.

\subsection{Primary states}

In the coset \eqref{scosetp}
the denominator $\text{su}(M)  \oplus \text{su}(N)  \oplus \text{u}(1) $ is embedded into
$\text{su}(N+M) \oplus \text{so}(2MN)$. The way of embedding is determined by those of
$\text{SU}(M) \times \text{SU}(N)   \times \text{U}(1) $ into $\text{SU}(N+M)$ and $\text{SO}(2NM)$.
For $\text{SU}(N+M)$ we use \eqref{emb1}.
For $\text{so}(2NM)_1$ we may use the system of $2NM$ free Majorana fermions,
which can be given by $NM$ free complex fermions. We denote the fermions as
$\psi_{i  A}$ and the complex conjugates as $\psi^\dagger_{i A}$.
Here $i$ is the index for the $\text{su}(N)$ fundamental representation
and $A$ is the index for the  $\text{su}(M)$ anti-fundamental representation.
The $\text{u}(1)$ charges for the fermions are set as $-(N+M)$ for $\psi_{i A}$
and $(N+M)$ for  $\psi^\dagger_{i A}$.
We denote the corresponding embedding as $\imath_2 (u,v,w)$.

The states of the Grassmannian model \eqref{scosetp} are labeled by the representations of the algebras
appearing in the coset expression. We denote $\Lambda_{L}$ as the highest weight of the representation for
$\text{su}(L)$ and $m \in \mathbb{Z}_\kappa$ for $\text{u}(1)_\kappa$  as in the bosonic case.
For affine $\text{so}(2NM)_1$ we use $\omega = -1,0,1,2$, where $\omega = 0,2$ denotes the identity and
the vector representations and $\omega = 1, - 1$ denotes the spinor and the co-spinor representations.
Then the states of the coset are obtained by the decomposition as
\begin{align}
 \Lambda_{N+M} \otimes \omega = \bigoplus_{\Lambda_N , \Lambda_M , m}
  (\Lambda_{N+M}, \omega ; \Lambda_N , \Lambda_M , m ) \otimes
  \Lambda_N \otimes \Lambda_M \otimes m \, .
  \label{decompose}
\end{align}
There are selection rules as (see, e.g., \cite{Naculich:1997ic})
\begin{align} \label{selection}
&m =  N |\Lambda_{N+M}| - (N+M) |\Lambda_N| + N (N+M) (a + \tfrac12 M \epsilon ) \,  , \\
&m = - M |\Lambda_{N+M}| + (N+M) |\Lambda_M| + M (N+M) (b + \tfrac12 N \epsilon ) \nonumber
\end{align}
modulo $\kappa$. Here $a$ and $b$ are integers defined modulo $M(N+M+k)$ and $N(N+M+k)$,
respectively, and $\epsilon = 0$ for $\omega = 0,2$ and $\epsilon = 1$ for $\omega = \pm 1$.
We have to take field identifications in account \cite{Gepner:1989jq}, but for large $k$ limit
we can neglect it.
The conformal dimension of the state $(\Lambda_{N+M}, \omega ; \Lambda_N , \Lambda_M , m )$
is
\begin{align}
 h = n + h^{M+N,k}_{\Lambda_{M+N}} + h_\omega^{2MN} - h^{N,k+M}_{\Lambda_N} - h^{M,k+N}_{\Lambda_M}
 - h_m
\end{align}
with an integer $n$ and \eqref{cw}.
Moreover, $h^{2L}_\omega = \omega / 4$ for $\omega = 0 , 2$ and $h^{2L}_\omega = L/8$ for
$\omega = \pm 1$.
In the following we consider the sector denoted by NS, which is the direct sum of the
identity and the vector representations as in \cite{Creutzig:2011fe,Candu:2012jq}.

\subsection{The character of chiral part}
\label{scharcp}

As in the bosonic case, we examine first the chiral part of the partition function and later consider how we should
pair the chiral and the anti-chiral parts.
For the $2NM$ free fermions we use
\begin{align}
 \theta (q , \imath_2 (u,v,w)) &= \text{tr}_\text{NS} q^{L_0} \imath_2 (u,v,w) \label{chtheta}\\
  &= \prod_{n=0}^\infty \prod_{i=1}^N \prod_{A=1}^M (1 + \bar w^{N+M} \bar u_A v_i q^{n+\frac12})
  (1 + w^{N+M}  u_A \bar v_i q^{n+\frac12}) \, , \nonumber
\end{align}
where $u_A , v_i$ are eigenvalues of $u,v$.
Then the character of the coset \eqref{scoset}
\begin{align}
 sb^{N,M,k}_{\tilde \Xi } (q) = \text{tr}_{\tilde \Xi} q^{L_0} \, , \qquad
 \tilde \Xi =  (\Lambda_{N+M} ; \Lambda_N , \Lambda_M , m )
 \label{chcosets}
\end{align}
is given by the decomposition \eqref{decompose} as
\begin{align}
 &\text{ch}_{\Lambda_{N+M}}^{N+M,k} (q , \imath_1 (u, v,w)) \theta (q , \imath_2 (u,v,w)) \\
 & \qquad = \sum_{\Lambda_N , \Lambda_M , m} sb^{N,M,k}_{\tilde \Xi} (q)
 \text{ch}_{\Lambda_{N}}^{N,k+M} (q , v )  \text{ch}_{\Lambda_{M}}^{M,k+N} (q , u ) \Theta^\kappa_m (q,w)
 \nonumber
 \end{align}
 with \eqref{chsu} and \eqref{chu1}.
At large $k$ we have
\begin{align}
 \text{ch}^{N+M}_{\Lambda_{N+M}} (\imath_1 (u,v,w)) s\vartheta (q , \imath_2 (u,v,w))
  = \sum_{\Lambda_N, \Lambda_M , m} sa^{N,M}_{\tilde \Xi}  (q) \text{ch}^{N}_{\Lambda_N} (v) \text{ch}^{M}_{\Lambda_M} (u)
   w^m \, ,
\end{align}
where we have used \eqref{asymsu} and \eqref{asymu1}.
Here $sa^{N,M}_{\tilde \Xi} (q)$ is related to the leading term at large $k$ as
\begin{align}
 sb^{N,M,k}_{\tilde \Xi} (q) \sim q^{h^{N,M,k}_{\tilde \Xi}} sa^{N,M}_{\tilde \Xi} (q) \, , \qquad
 h^{N,M,k}_{\tilde \Xi} = h^{N+M,k}_{\Lambda_{N+M}} - h^{N,k+M}_{\Lambda_{N}}- h^{M,k+N}_{\Lambda_{M}}
 - h ^\kappa_m \, .
\end{align}
Moreover, we have used
\begin{align}
 s \vartheta (q , \imath_2 (u,v,w)) = \prod_{n=0}^\infty \prod_{i=1}^N \prod_{A=1}^M
 \frac{(1 + \bar w^{N+M} \bar u_A v_i q^{n+\frac12}) (1 + w^{N+M} u_A \bar v_i q^{n+ \frac12})}{(1 - \bar w^{N+M} \bar u_A v_i q^{n+1}) (1 - w^{N+M} u_A \bar v_i q^{n+1})} \, .
\label{chstheta}
\end{align}
This is the partition function of $MN$ complex free bosons and $MN$ complex free fermions in the
bi-fundamental representations of $\text{su}(N) \oplus \text{su}(M)$.
At large $N$, we may rewrite this as
\begin{align}
 \widetilde{\text{ch}}^{N+M}_{\mathbf{\Lambda}_{N+M}} (\imath_1 (u,v,w)) s \vartheta (q , \imath_2 (u,v,w))
  = \sum_{\mathbf{\Lambda}_N, \Lambda_M } sa^{N,M}_{\mathbf{\Lambda}_{M+N};\mathbf{\Lambda}_N , \Lambda_M}  (q) \widetilde{\text{ch}}^{N}_{\mathbf{\Lambda}_N} (v \bar w^{N+M}) \text{ch}^{M}_{\Lambda_M} (u) \, .
\end{align}
Here we set $m =  N |\mathbf{\Lambda}_{N+M}|_- - (N+M)  |\mathbf{\Lambda}_{N}|_-$ as in \eqref{largeNu1}
and used \eqref{chu}.

\subsection{The character at the free limit}
\label{scfl}

Setting $\mathbf{\Lambda}_{N+M} = 0$, we have
\begin{align}
 s \vartheta (q , \imath_2 (u,v,w))
  = \sum_{\mathbf{\Lambda}_N , \Lambda_M} sa^{N,M}_{0 ; \mathbf{\Lambda}_N,\Lambda_M} (q)
   \widetilde{\text{ch}}^{N}_{\mathbf{\Lambda}_{N}} (v \bar w^{N+M}) \text{ch}^{M}_{\Lambda_M} (u)  \, .
\end{align}
Therefore $sa^{N,M}_{0 ; \mathbf{\Lambda}_N,\Lambda_M} (q) $ can be obtained by decomposing the
free system in terms of the $\text{u}(N)$ representation $\mathbf{\Lambda}_{N}$ and
the $\text{su}(M)$ representation $\Lambda_M$.
We need to assign the $\text{U}(N)$ invariant condition, but we treat $\text{SU}(M)$ as a global symmetry.
The free system corresponds to the case with $\lambda =0$ of the 't Hooft parameter \eqref{thoofts}.
For  $\lambda \neq 0$, we should consider the sectors with $\mathbf{\Lambda}_{N+M} \neq 0$ as in the
bosonic case, see appendix \ref{PF}.

We denote the fermionic and bosonic modes by $\psi_{aA}$ and $j_{aA}$  with
$a =1,2, \ldots , N$ as the index of the $\text{u}(N)$ fundamental representation and
$A=1,2,\ldots , M$ as the index of the $\text{u}(M)$ anti-fundamental representation.
There are also complex conjugates $\psi^\dagger_{aA}$ and $j^\dagger_{aA}$.
The full Fock space of the free system is spanned by
\begin{align}
 \prod_j \psi^{\dagger a_j A_j}_{-r_j-\frac12}
  \prod_k \psi^{b_k B_k}_{-s_k-\frac12}
 \prod_{l} j^{\dagger c_l C_l}_{- t_l - 1}
 \prod_m j^{d_m D_m} _{- u_m - 1} | 0 \rangle \, ,
\end{align}
where $r_j , s_k , t_l , u_m$ are non-negative integers.
The $\text{u}(N)$ invariants are then given by the linear combinations of the ``basic'' invariants as
\begin{align}
 &\prod_{r,s = 0}^\infty \left( \sum_{a=1}^N \psi^{\dagger a A^1_{rs}}_{-r - \frac12} \psi^{a B^1_{rs}}_{-s - \frac12}\right)^{K_{rs}}
 \prod_{t,u = 0}^\infty \left( \sum_{a=1}^N j^{\dagger a A^2_{tu}}_{-t - 1 } j^{a B^2_{tu}}_{-u - 1}\right)^{L_{tu}}
  \\
 & \qquad  \times \prod_{t,s = 0}^\infty \left( \sum_{a=1}^N j^{\dagger a A^3_{ts}}_{-t - 1} \psi^{a B^3_{ts}}_{-s - \frac12}\right)^{P_{ts}}
 \prod_{r,u = 0}^\infty \left( \sum_{a=1}^N \psi^{\dagger a A^4_{ru}}_{-r - \frac12} j^{a B^4_{tu}}_{-u - 1}\right)^{Q_{ru}}
| 0 \rangle \, ,
 \nonumber
\end{align}
where $K_{rs},L_{tu}$ are non-negative integers while  $P_{ts},Q_{ru} = 0 ,1$.
Since there are $M^2$ times more bi-linears than the ones for $M=1$,
the contribution from $\text{u}(N)$ invariants are
(see (3.68) of  \cite{Candu:2012jq})
\begin{align}
 {\cal Z}^\text{CFT}_0 (q) =
\left ( \prod_{s=2}^\infty Z^{(s)}_B (Z^{(s-1/2)}_F )^2 Z^{(s-1)}_B  \right )^{M^2} \, ,
\end{align}
which is the one-loop partition function for the massless gauge sector of the gravity theory,
see \eqref{sZ0}.

As in the bosonic case,
we take pairs such that they are charge conjugates with
$\text{u}(N)$, but we do not have to take care for the $\text{su}(M)$ sector.
{}From the experience of the free limit in section \ref{free}, we propose to
combine
\begin{align}
  \psi^{\dagger  a A}_{-r - \frac12} \bar \psi^{a B}_{-s - \frac12} \, , \qquad
  \bar \psi^{\dagger a A}_{-r - \frac12} \psi^{a B}_{-s - \frac12} \, , \qquad
  j^{\dagger a A}_{-t - 1 } \bar \jmath^{a B}_{-u - 1} \, , \qquad
 \bar  \jmath^{\dagger a A}_{-t - 1 }  j^{a B}_{-u - 1} \, , \\
 j^{\dagger a A}_{-t - 1} \bar \psi^{a B}_{-s - \frac12}\, , \qquad
 \bar \jmath^{\dagger a A}_{-t - 1} \psi^{a B}_{-s - \frac12}\, , \qquad
 \psi^{\dagger a A}_{-r - \frac12} \bar \jmath^{a B}_{-u - 1} \, \qquad
 \bar \psi^{\dagger a A}_{-r - \frac12} j^{a B}_{-u - 1} \, ,\nonumber
\end{align}
where the $\text{u}(N)$ index $a$ is the same for chiral and anti-chiral sector.
{}Following the analysis for the bosonic case and section 3.4 of \cite{Candu:2012jq},
we have
\begin{align}
{\cal Z}^\text{CFT} (q) &= \left( \sum_{\mathbf{\Lambda}_N}|\text{sch}_{(\Lambda^{l}_N)^t}  ({\cal U} (\tfrac12)) \text{sch}_{(\Lambda^{r}_N)^t} ({\cal U} (\tfrac12 )) |^2    \right)^{M^2}  {\cal Z}_0^\text{CFT} (q) \\
 & = \left(  {\cal Z}_0 {\cal Z}_\text{matter}^{1/2} \right) ^{M^2} \, .\nonumber
\end{align}
Here the supercharacter $\text{sch}_{\Lambda}  ({\cal U}_1 ) $ is defined in \eqref{sschur}.
The partition function reproduces the gravity one, but with matter fields dual to an ${\cal N}=2$ multiplet,
and this is consistent with the argument for the free theory in section \ref{free}.

For $\lambda \neq 0$, we need to consider the sectors with $\mathbf{\Lambda}_{N+M} \neq 0$.
Moreover, we have to include bosonic su$(M)$ currents dual to the spin 1 gauge fields generated
by $V_0^{(1)+} \otimes t_a$ with $\text{sl}(M)$ generator $t_a$.
This amounts to multiplying with the negative mode contribution of  $\text{su}(M)$ currents \eqref{nmsum},
which is consistent with the gravity partition function in \eqref{zmbulks}.
In appendix \ref{PF} we show that the gravity partition function
with matter fields dual to two sets of ${\cal N}=2$ multiplet in \eqref{zmbulks}
can be reproduced by the supersymmetric Grassmanian model with  $\lambda \neq 0$
 once extra $\text{su}(M)$ factor is added as in \eqref{scoset2}.

\section{Supersymmetric Grassmannian model as a product theory}\label{sec:producttheory}

In this section we again consider the Grassmannian Kazama-Suzuki model, but now in a BRST approach. In the first section we construct the Grassmanian model as a product theory of WZNW models with a BRST constraint. In section \ref{subsec:firstorder} we introduce a first order formulation for the product theory in order to compare with the free theory which is done in section \ref{subsec:freehigherspin}. In section \ref{subsec:currents} all the currents of spin-one, 3/2 and two are constructed explicitly in the large $N$ limit, and the OPEs of the extended supercharges are calculated. Finally, in section \ref{subsec:opebulk} the OPEs of the spin-one and spin-3/2 currents are derived from the bulk side, including non-linear terms, and they are successfully compared to the results from the CFT side.

\subsection{The product theory}\label{subsec:theproductheory}
Following \cite{FigueroaO'Farrill:1997jj} we can write a supersymmetric coset model $\mathfrak{g}/\mathfrak{h}$ as a product theory. Here we assume that $\mathfrak{g}$ has invariant bi-linear form $\Omega$ which restricted to $\mathfrak{h}$ is non-degenerate. We choose a basis $\{t_\alpha\}$ of $\mathfrak{g}$ such that the basis of $\mathfrak{h}$ is given by the subset $\{t_a\}$ and the remaining generators outside the coset are denoted with barred Greek letters $\{t_{\bar\alpha}\}$. The product theory is, after removing Kugo-Ojima quartets, consisting of the WZNW model $(\mathfrak{g},\Omega-\tfrac12\kappa^{(\mathfrak{g})})$ where $\kappa^{(\mathfrak{g})}_{\alpha\beta}={{f_{\alpha\gamma}}^\delta f_{\beta\delta}}^\gamma $ is the Killing metric, the WZNW model $(\mathfrak{h},-\Omega|_{\mathfrak{h}}-\tfrac12\kappa^{(\mathfrak{h})})$, $\dim \mathfrak{h}$ ghosts $b_a, c^a$ and, finally, $\dim\mathfrak{g}-\dim\mathfrak{h}$ fermions $\psi_{\bar\alpha}$ with metric $\Omega_{\bar\alpha\bar\beta}$. The spectrum is the 
BRST cohomology generated by the current
\begin{align}\label{eq:BRST}
    j_{\textrm{BRST}}=(J_a+\tilde J_a)c^a+\frac12 \Omega^{\bar\alpha\bar\gamma}{f_{a\bar\alpha}}^{\bar\beta}\psi_{\bar\beta}\psi_{\bar\gamma} c^a-\frac12 {f_{ab}}^c b_c c^a c^b\ ,
\end{align}
where $J_a$ are currents of the $\mathfrak{g}$ WZNW model in the $\mathfrak{h}$-direction and $\tilde J_a$ of the $\mathfrak{h}$ WZNW model.

For the Grassmannian Kazama-Suzuki model we start with $\Omega=(k+M+N)\tr$, and the content is thus as follows
\begin{itemize}
  \item A $\widehat{\text{su}}(N+M)_k$ WZNW model with matrix element $g$ and central charge $c[M+N,k]=k((N+M)^2-1)/(N+M+k)$\ .
  \item A $\widehat{\text{su}}(N)_{-2N-M-k}$ WZNW model with matrix element $h_N$ and central charge\newline $c[N,-2N-M-k]=(2N+M+k)(N^2-1)/(N+M+k)$\ .
  \item A $\widehat{\text{su}}(M)_{-N-2M-k}$ WZNW model with matrix element $h_M$ and central charge\newline $c[M,-N-2M-k]=(N+2M+k)(M^2-1)/(N+M+k)$\ .
  \item A
  scalar $\phi_h$ with level $-NM(N+M)(M+N+k)$, that is with action \newline  $S=-\frac{N M(N+M)(N+M+k)}{4\pi}\int d^2 z\ \del\phi_h\bar\del\phi_h$ and central charge $c_{\textrm{scalar}}=1$\ .
  \item $2NM$ fermions ${\psi^{A}}_i,\bar\psi^i_{\phantom{i}A}$ with $i=1,\ldots,N$, $A=1,\ldots,M$, OPEs (after renormalization) ${\psi^{A}}_i(z)\bar\psi^j_{\phantom{j}B}(w)\sim\delta^{j}_i\delta^{B}_A$ and total central charge $c_{\textrm{ferm.}}=NM$\ .
  \item Three families of ghost systems $c^a_{(N)},b_{(N)a}$, $c^d_{(M)},b_{(M)d}, c_{(\phi)},b_{(\phi)}$ with $a=1,\ldots,N^2-1$ and $d=1,\ldots,M^2-1$. The total central charge is $c_{\textrm{ghosts}}=-2(N^2+M^2-1)$\ .
\end{itemize}
We can check that the total central charge is
\begin{align}\label{}
  c&=c[M+N,k]+c[N,-2N-M-k]+c[M,-N-2M-k]+c_{\textrm{scalar}}+c_{\textrm{ferm.}}+c_{\textrm{ghosts}}\nonumber\\
  &=\frac{3NMk}{N+M+k}\ ,
\end{align}
which agrees with \eqref{eq:centralcharge}.

\subsection{First order formulation}\label{subsec:firstorder}

To compare with the free theory (see appendix \ref{shs} and below) it is useful to introduce a first order formalism for the $\widehat{\text{su}}(N+M)_k$ WZNW model which has the action
\begin{align}
 S^{\textrm{WZNW}}_k (g) = -\frac{k}{4 \pi} \int_{\Sigma} d ^2z
 \tr (g^{-1} \partial g , g^{-1} \bar \partial g)
 - \frac{k}{2 4 \pi} \int_{B} \tr (g^{-1} d g
  [g^{-1} d g , g^{-1} d g ])
\end{align}
with $\partial B = \Sigma$. We make an sl$(N+M)$ like 3-decomposition of the group element $g = g_{1} g_0 g_{-1}
$ where
\begin{align}
  g_{1} &= e^{{\gamma^{i}}_A t_{i,N+A}} ~,
  &g_0 &= e^{i\phi t_\phi} \left(
                            \begin{array}{cc}
                              g_N & 0 \\
                              0 & g_M \\
                            \end{array}
                          \right)
   ~,
    &g_{-1} &=  e^{\bar \gamma^{A}_{\phantom{A}i} t_{N+A,i}}
    ~.
\end{align}
Here $(t_{IJ})_{KL}=\delta_{IK}\delta_{JL}$, $I,J,K,L=1,\ldots,N+M$, $i=1,\ldots,N$, $A=1,\ldots,M$, $t_\phi=M\sum_{i=1}^N t_{ii}-N\sum_{l=1}^M t_{N+l,N+l}$, and $g_N,g_M$ are respectively group elements of su$(N)$, su$(M)$.

We can now use the Polyakov-Wiegmann identity
\begin{align}
 S^{\textrm{WZNW}}_k (gh)  =  S^{\textrm{WZNW}}_k (g) +  S^{\textrm{WZNW}}_k (h)
 - \frac{k}{2 \pi} \int d^2 z
   \tr( g^{-1} \bar \partial g  \partial h h^{-1} )
\end{align}
to express the action as
\begin{align}
  S^{\textrm{WZNW}}_k (g_{1} g_0 g_{-1})  =&  S_k^{\textrm{WZNW}} (h_N)+ S_k^{\textrm{WZNW}} (h_M)+\frac{kNM(N+M)}{4 \pi} \int d^2 z \del\phi\delbar\phi
 \nonumber\\
 &- \frac{k}{2 \pi} \int d^2 z \tr( g_1^{-1} \bar \partial g_1
    g_0 \partial g_{-1} (g_{-1})^{-1} g_0^{-1} ) ~,
\end{align}
where
\begin{align}\label{}
  - \frac{k}{2 \pi} \int d^2 z \tr( g_1^{-1} \bar \partial g_1
    g_0 \partial g_{-1} g_{-1}^{-1} g_0^{-1} )=- \frac{k}{2 \pi}\int d^2 z e^{-(N+M)i\phi}\delbar{\gamma^i}_A{(g_M)^A}_B\del{\bar\gamma^B}_{\phantom{B}j}{(g^{-1}_N)^j}_i \ .
\end{align}

We can now integrate in dimension one fields $\beta^A_{\phantom{A}i}$ and $\bar\beta^i_{\phantom{i}A}$, and we propose that the action takes the following first order form
\begin{align}
  S^{\textrm{WZNW}} (\phi,\gamma,\beta,\bar\gamma,\bar\beta)=&S_0+S_{\text{int}} \ ,
\end{align}
where
\begin{multline}
   S_0=S_{k+M}^{\textrm{WZNW}} (h_N)+ S_{k+N}^{\textrm{WZNW}} (h_M)\\
+ \frac{1}{2\pi} \int d^2 z \,(
     \tfrac{(k+N+M)NM(N+M)}{2}  \partial \phi \bar \partial \phi
     + \tfrac{iNM(N+M)}{8} \sqrt g R \phi
     +\beta^A_{\phantom{A}i} \delbar{\gamma^i}_A
         + \bar \beta^i_{\phantom{i}A} \del{\bar\gamma^A}_{\phantom{A}j} )
\end{multline}
and
\begin{align}\label{}
    S_{\text{int}}=&\frac{1}{2\pi k} \int d^2 z\, e^{(N+M)i\phi}{\beta^A}_i{(g_N)^i}_j{\bar\beta^j}_{\phantom{j}B}{(g^{-1}_M)^B}_A\ .
\end{align}
Here the renormalization can be read of from the equations of motion
\begin{align}\label{}
  {\beta^A}_i&= -k e^{-(N+M)i\phi}{(g_M)^A}_B{\del\bar\gamma^B}_{\phantom{B}j}{(g^{{-1}}_N)^j}_i\ ,\nonumber \\
  {\bar\beta^i}_{\phantom{i}A}&= -k e^{-(N+M)i\phi}{(g^{-1}_N)^i}_j{\delbar\gamma^j}_{\phantom{j}B}{(g_M)^B}_A\ ,
\end{align}
which shows that each of the $NM$ ${\beta^A}_i$ comes with a path integral measure contribution of
\begin{align}
\delta S=  \frac{(N+M)^2}{4\pi} \int d^2 z \partial \phi \bar \partial \phi
  + \frac{(N+M)i}{16 \pi} \int d^2z  \sqrt{g} {\cal R } \phi ~,
\end{align}
and for the WZNW terms we e.g. have that for each $i=1,\ldots,N$, ${\beta^A}_i$ will increase the level of the su$(M)$ WZNW model with one. We choose not to renormalize $\phi$ since in this formulation it has radius $1$ for $N,M$ mutually prime. One can check that the central charge and conformal dimensions are correct for this action.

\subsection{The free higher spin CFT}\label{subsec:freehigherspin}

In \cite{Bergshoeff:1991dz} a free CFT consisting of a $\beta\gamma$ and a $bc$ system was constructed representing the linear supersymmetric $sw_\infty[\lambda]$ algebra. We propose that the extension to general $M$ is done by introducing ${\beta^A}_i,{\gamma^i}_A$-systems with conformal weights $(1-\lambda/2,\lambda/2)$ and ${b^A}_i,{c^i}_A$-systems with weights $(1/2-\lambda/2,1/2+\lambda/2)$ where $i=1,\ldots, N$ and $A=1,\ldots,M$, and the systems have standard OPEs
\begin{align}\label{}
    {\gamma^i}_A(z){\beta^B}_j(w)\sim \frac{\delta^B_A\delta^i_j}{z-w}\ ,\qquad {b^A}_i (z) {c^j}_B (w) \sim\frac{\delta^A_B\delta^j_i}{z-w}\ .
\end{align}
We can then define the currents ($s$ integer)
\begin{align}
  J_{\text{lin}}^{(s)+}[t_a]  =& -(-1)^{s-1}\sum_{i=0}^{s-1} a^i (s, \lambda) \partial^{s - 1 - i} \{{\beta^B}_l {(t_a)^A}_B (\partial^i {\gamma^l}_A )\} \nonumber\\
 &- (-1)^{s-1}\sum_{i=0}^{s-1} a^i (s , \lambda + 1) \partial^{s-1-i} \{ {b^B}_l {(t_a)^A}_B(\partial^i {c^l}_A) \} \, , \nonumber\\
  \tilde J_{\text{lin}}^{(s)-}[t_a]=& (-1)^{s-1} \frac{s-1+\lambda}{2s-1}\sum_{i=0}^{s-1} a^i (s, \lambda) \partial^{s - 1 - i} \{ {\beta^B}_l {(t_a)^A}_B (\partial^i {\gamma^l}_A )\} \nonumber \\
 &- (-1)^{s-1}\frac{s-\lambda}{2s-1}
 \sum_{i=0}^{s-1} a^i (s , \lambda + 1) \partial^{s-1-i} \{ {b^B}_l {(t_a)^A}_B(\partial^i {c^l}_A) \} \, , \nonumber \\
  J_{\text{lin}}^{(s-1/2)\pm}[t_a]=& \pm(-1)^{s-1/2}\sum_{i=0}^{s-1} \alpha^i (s, \lambda) \partial^{s - 1 - i} \{ (\partial^i {\gamma^l}_A ) {(t_a)^A}_B {b^B}_l\}\nonumber\\
 &-(-1)^{s-1/2}
 \sum_{i=0}^{s-2} \beta^i (s , \lambda) \partial^{s-2-i} \{ (\partial^i {c^l}_A)  {(t_a)^A}_B {\beta^B}_l\} \label{eq:suggestedfreecurrents}
\end{align}
where $t_a\in \mathrm{u}(M)$ and where
\begin{align}
 a^i (s , \lambda) = \begin{pmatrix} s-1 \\ i \end{pmatrix} \frac{(-  \lambda - s + 2)_{s-1-i}}{(s+i)_{s-1-i} } \, , \nonumber \\
 \alpha^i (s , \lambda) = \begin{pmatrix} s-1 \\ i \end{pmatrix} \frac{(-  \lambda - s + 2)_{s-1-i}}{(s+i-1)_{s-1-i} } \, , \\
 \beta^i (s , \lambda) = \begin{pmatrix} s-2 \\ i \end{pmatrix} \frac{(-  \lambda - s + 2)_{s-2-i}}{(s+i)_{s-2-i} }\ . \nonumber
\end{align}
We propose that these currents generate the extended linear $sw_\infty[\lambda]$ algebra with $J_{\text{lin}}^{(s)\pm}[t_a]$ corresponding to the generator\footnote{The tilde on the generator $\tilde V^{(s)-}$ is to denote that it is not directly $V^{(s)-}$ in our basis (see app. \ref{glM} for notation), but the generator orthogonal to $V^{(s)+}$ under the supertrace metric.} $V^{(s)\pm}\otimes t_a$. Indeed for $M=1$ it reduces to the system in \cite{Bergshoeff:1991dz} up to the traced $i$-index. Note here that we have a factor of $2$ on $\lambda$ compared to \cite{Bergshoeff:1991dz}, and for later use we have taken $\beta\mapsto i\gamma$, and $\gamma\mapsto i\beta$ and $b\leftrightarrow c$, and finally used the anti-automorphism $\sigma(V_m^{(s)\pm})=(\pm)^{2s}(-1)^{s-1}V_m^{(s)\pm}$. There is however one subtlety: In the $M=1$ case the $J_{\text{lin}}^{(1/2)+}=-J_{\text{lin}}^{(1/2)-}$ and $J_{\text{lin}}^{(1)+}$ form a decoupled short multiplet of the remaining algebra, however in this case we know we 
cannot decouple the extra spin-one generator (see app. \ref{glM}), but we have to decouple the spin-half generator.

Let us now go back to our product theory for the Grassmannian using the first order formalism for the su$(N+M)$ factor. The point is now that if we ignore the background charge for $\phi$, the model takes the form of the ${\beta^A}_i,{\gamma^i}_A$-systems, the corresponding fermions ${\psi^{A}}_i,\bar\psi^i_{\phantom{i}A}$ and three topological $\mathfrak{g}/\mathfrak{g}$ models, namely the ones based on $\hat{\text{u}}(1)_{(k+N+M)(N+M)NM}$ and (supersymmetric) $\widehat{\text{su}}(N)_{N+M+k}$ and $\widehat{\text{su}}(M)_{N+M+k}$. The models are then related by the interaction term. This splitting is also respected by the BRST current \eqref{eq:BRST}. The reason is that it only contains the su$(N+M)$ currents in the su$(N)$, su$(M)$ and u(1) directions. These currents are in the first order formalism
\begin{align}\label{eq:currentssumsun}
    J_a^{(N+M,k)}=&J_a^{(N,k+M)}+:{\beta^A}_i{(t_a)^i}_j {\gamma^j}_A:\ ,\qquad \textrm{for }t_a\in su(N)\subset su(N+M)\ , \nonumber \\
    J_d^{(N+M,k)}=&J_d^{(M,k+N)}-:{\beta^A}_i{(t_d)^B}_A {\gamma^i}_B:\ ,\qquad \textrm{for }t_d\in su(M)\subset su(N+M)\ , \nonumber\\
    J_{t_\phi}^{(N+M,k)}=&J_\phi^{(1,NM(N+M)(N+M+k))}+(M+N){\beta^A}_i {\gamma^i}_A
\end{align}
where $J_a^{(N+M,k)}$ is the current of $\widehat{\text{su}}(N+M)_k$ and $J_a^{(N,k+M)}$ is the current of $\widehat{\text{su}}(N)_{k+M}$ etc. Note that the $\beta\gamma$-currents here generate $\widehat{\text{su}}(N)_{-M}$ and $\widehat{\text{su}}(M)_{-N}$ respectively. For completeness the remaining currents $J_{t_{i,N+A}},J_{t_{N+A,i}}$ take the following form
\begin{align}\label{}
  J_{t_{i,N+A}}^{(N+M,k)}=&  {\beta^{A}}_{i}\     \ ,\nonumber\\
  J_{t_{N+A,i}}^{(N+M,k)}=&  J_{t_{N+A,N+B}}{\gamma^{i}}_{B}-J_{t_{j,i}}{\gamma^{j}}_{A}-:{\beta^B}_j {\gamma^j}_A{\gamma^{i}}_{B}:-k\partial {\gamma^{i}}_{A}\nonumber \\
  =&J_d^{(M,k+N)}g_{(M)}^{de}\tr(t_et_{N+A,N+B}){\gamma^{i}}_{B}-J_a^{(N,k+M)} g_{(N)}^{ab}\tr(t_bt_{j,i}){\gamma^{j}}_{A}\nonumber \\ &-\frac{1}{MN}J_\phi^{(1,NM(N+M)(N+M+k))}{\gamma^{i}}_{A}-:{\beta^B}_j {\gamma^j}_A{\gamma^{i}}_{B}:-k\partial {\gamma^{i}}_{A} \ ,
\end{align}
with right-nested normal ordering. Where $J_{t_{N+A,N+B}}$ and $J_{t_{j,i}}$ are combinations of su$(N)$, su$(M)$ and $\partial\phi$ currents.

We can now write the BRST currents \eqref{eq:BRST} as
\begin{align}
j_{\textrm{BRST}}=&\ K_a c^a\ , \label{eq:BRSTexplicit}
\end{align}
where $K_a$ are currents of level zero given explicitly by
\begin{align}\label{}
K_a =&\ J_a^{(N,k+M)}+{\beta^A}_i{(t_a)^i}_j {\gamma^j}_A+ \tilde J_a^{(N,-k-2N-M)} +{\psi^{A}}_i{(t_a)^i}_j\bar\psi^j_{\phantom{i}A}+\frac12 {f_{ab}}^c b_c c^b  \ ,\nonumber \\ &\qquad \textrm{for }t_a\in su(N)\ , \nonumber \\
K_d=&\  J_d^{(M,k+N)}-{\beta^A}_i{(t_d)^B}_A {\gamma^i}_B+ \tilde J_d^{(M,-k-N-2M)} -{\psi^{B}}_i{(t_d)^A}_B\bar\psi^i_{\phantom{i}A}+\frac12 {f_{de}}^f b_f c^e \ ,\nonumber \\ &\qquad \textrm{for }t_d\in su(M)\ ,  \nonumber
 \\
K_\phi=&\  J_\phi^{(1,NM(N+M)(N+M+k))}+(M+N){\beta^A}_i {\gamma^i}_A+ \tilde J_\phi^{(1,-NM(N+M)(N+M+k))} \nonumber \\&+(M+N){\psi^{A}}_i\bar\psi^i_{\phantom{i}A}\ ,\qquad \nonumber \\ &\qquad \textrm{for }t_\phi\in u(1)\ .
\end{align}
For finite $N,M,k$ this means that BRST invariant currents built out of ${\beta^A}_i,{\gamma^j}_B$ would need contraction in both the $i$ and $A$ index.

Let us first consider the case $\lambda=0$, i.e. $k$ infinite with $N/k\rightarrow0$ and $M$ kept finite. If it were not for the possible infinite number of fields we could localize the action to its classical value. The background charge is further effectively zero in this limit (if we take $N\lambda\rightarrow0$) and the u(1) systems are thus together with the su$(N)$ and su$(M)$ factors topological, at least when acting in certain states. The left over $\beta\gamma$ systems have exactly the right dimensions compared to the free field theory generating the linear $W$-algebra.

For general $\lambda$ we have to be more careful to get the correct dimensions for the $\beta\gamma$ systems, since we cannot obtain this by simple field redefinitions. However, we can do it by adding the following BRST exact term to the action
\begin{align}\label{}
    Q_{\textrm{BRST}}\cdot\frac{\lambda}{(M+N)4\pi} \int d^2 z \,\bar\del\ln |\rho|^2 b_\phi =\frac{\lambda}{(M+N)4\pi} \int d^2 z \,\bar\del\ln |\rho|^2 K_\phi \ ,
\end{align}
which introduces background charges for the u(1) scalars, the $\beta\gamma$ ghosts and the fermions. Here $\sqrt g R=-4\del\bar\del\ln |\rho|^2$. Indeed, if we bosonize as
\begin{align}\label{}
 {\beta^A}_i =  \partial {Y^A}_i e^{- {X^A}_i +  {Y^A}_i}~, \qquad
 {\gamma^i}_A = e^{{X^A}_i -  {Y^A}_i}  ~, \qquad :{\beta^A}_i {\gamma^i}_A:=\del {X^A}_i\, ,
\end{align}
\begin{align}\label{}
    {\psi^{A}}_i=e^{-{Z^{A}}_i}\ ,\qquad \bar\psi^i_{\phantom{i}A}=e^{{Z^{A}}_i}\ ,\qquad :{\psi^{A}}_i\bar\psi^i_{\phantom{i}A}:=-\del {Z^{A}}_i\ ,
\end{align}
and use
\begin{align}\label{}
    J_\phi^{(1,NM(N+M)(N+M+k))} &=-iNM(N+M)(N+M+k)\del\phi \ , \nonumber\\
    \tilde J_\phi^{(1,-NM(N+M)(N+M+k))}&=iNM(N+M)(N+M+k)\del\phi_h\ ,
\end{align}
we see that the total background charge for ${X^A}_i$ becomes $-1/2+\lambda/2$ and for ${Z^{A}}_i$ it becomes $-\lambda/2$, i.e. the conformal dimensions for the $\beta\gamma$ and $bc$ system take the wanted values. The field $\phi$ now has background charge $iNM^2/2$, and $\phi_h$ has background charge $iN^2M/2$. We can calculate the central charge and check that we get $c=3NM(1-\lambda)-3\lambda M^2=  3NMk / (N+M+k )$ again. Thus the free theory is naturally embedded in the product theory.

In taking the large $k$ limit, we can choose to take $\beta\mapsto k\beta$ (since it basically had this $k$
in its definition). This means that the action for the $\beta\gamma$ systems also localize to the classical solutions. To keep the currents finite, we then divide them with $k$. We thus also have to divide $j_{\textrm{BRST}}$ with $k$. In all, we have the same expression for the currents \eqref{eq:currentssumsun}, but where the OPEs of $\gamma$ with $\beta$ now is proportional to $1/k$ and hence vanishing. We can thus forget the constraints from the $\widehat{\text{su}}(M)_{-N}$ currents, but we cannot forget the constraints from the $\widehat{\text{su}}(N)_{-M}$ currents. The reason is that we have an infinite amount of currents. If we, as an example, want to know if ${\beta^{A_0}}_{i_0}{\gamma^{j_0}}_{B_0}$ is in the BRST cohomology, we see that it will be mapped to a state with an infinite number of fields, due to the infinite number of generators $t_a\in su(N)\subset su(N+M)$, divided by $k$ which is a state of undetermined norm. To ensure invariance we thus can only consider currents of the form  $\sum\
\del^n{\beta^{A}}_{i}{\gamma^{i}}_{B}$, i.e. precisely of the form suggested in \eqref{eq:suggestedfreecurrents} generating the higher spin algebra. However, due to the derivatives these will only be BRST-invariant up to derivatives of the ghosts, and will not keep the interaction term invariant. In the next section we will show how at low spins we can dress the free currents with su$(N)$, su$(M)$ and u$(1)$ currents such that they are approximately BRST invariant in the 't Hooft limit.

\subsection{Currents}\label{subsec:currents}

We now want to extend the free currents from last subsection such that they obey BRST invariance and keep the interaction term invariant, i.e. that they are true symmetries of the theory. We denote the currents dual to $V^{(s)\sigma}_m$ on the bulk side by $J^{(s)\sigma}$. In principle, we already know some BRST invariant currents, namely, the u(1) current, the supercharges and the stress-energy tensor. These can be expressed in their full versions including the ghosts or BRST-equivalent in the coset description where they are constructed only out of currents from the su$(N+M)$ factor and the fermions.

Our strategy will be to basically follow the latter coset description. That is, we write up the most general linear combination of terms of a given spin, not including the fields coming from the gauging, and also not including the ghosts, and demand BRST invariance. That is BRST invariance up to finite contributions in the su$(M)$ direction. It is natural to expect that such a coset description is possible, but it is not a priori clear and we will have one exception below.

To ensure invariance of the interaction term in the action we can only build higher spin currents from the fermions ${b^A}_i$, ${c^i}_A$, and from the full $\hat{\text{su}}(M+N)_k$ currents including the $\beta\gamma$ terms, which we now denote $J^{(N)}_a$, $J^{(M)}_d$, $J_\phi$, ${J^A}_i\equiv J_{t_{i,N+A}}$ and ${J^i}_A\equiv J_{t_{N+A,i}}$. Here upper indices transform in the fundamental representation, and lower indices in the anti-fundamental representation. We can create the currents using products and derivatives of these constituents, but we have to contract all $i$ indices to ensure the first order OPEs with the BRST currents are zero, and in the case of finite $N,M,k$ both $i$ and $A$ indices should be contracted. To organize the fermionic currents and easily calculate their OPEs, it is a good idea to introduce the u(1) level $MN$ current $I_\phi={b^A}_i{c^i}_A$, the su$(N)$ level $M$ current $I^{(N)}_a={b^A}_i{(t_a)^i}_j {c^j}_A$, and the su$(M)$ level $N$ current $I^{(M)}_d=-{b^A}_i{(t_d)^B}_A {
c^i}_B$ which are all primaries. The BRST currents are now simply of the form
\begin{align}\label{}
    j_{\textrm{BRST}}=\sum_{\text{su}(N)}(J^{(N)}_a+I^{(N)}_a)c^a+\sum_{\text{su}(M)}(J^{(M)}_d+I^{(M)}_d)c^d+(J_\phi+(M+N) I_\phi)c^\phi+\ldots
\end{align}
where the extra terms are from the denominator-factors and ghosts.

\subsection*{Spin-one}
 Let us start with the spin-one currents. For $t_d\in \mathrm{su}(M)$ we have the most general linear combination
\begin{align}\label{}
J^{(1)}=  \sum_d  N^d_1  J_d^{(M)}
 +  \sum_d  N^d_2
    I^{(M)}_d +  N^\phi_1  J_\phi
 +  N^\phi_2
    I_\phi\, .
\end{align}
The second order of the OPEs with the currents $K_a$ from \eqref{eq:BRSTexplicit} are generically divergent in the large $N$ limit, and can spoil BRST invariance even when dividing the BRST current with $k$. We thus demand that these second order terms are zero, and this fixes the coefficient $N^d_2=-\frac{k}{N} N^d_1$. We thus get one independent solution for each $t_d\in \mathrm{su}(M)$
\begin{align}\label{}
J^{(1)-}[t_d]=&-\lambda  J_d^{(M)}
 +  \frac{\lambda k}{N}
    I^{(M)}_d \nonumber \\
    &=  \lambda   {\beta^B}_i {(t_d)^A}_B  {\gamma^i}_A-\lambda J_d^{(M,k+N)}
 - (1-\lambda)
    {b^B}_i {(t_d)^A}_B {c^i}_A \, ,
\end{align}
where we have chosen an overall normalization $\lambda$. We see that the result is exactly as in \eqref{eq:suggestedfreecurrents} with the simple addition of the su$(M)$ current. In the u(1) case we of course have to add $J_\phi$ instead
\begin{align}\label{}
J^{(1)-}[1]=&\frac{\lambda}{M+N} J_\phi-\frac{\lambda k}{M+N}I_\phi\nonumber\\
=&  \lambda   {\beta^A}_l   {\gamma^l}_A+\frac{\lambda}{M+N} J_\phi^{(1,NM(N+M)(N+M+k))}
 - (1-\lambda)
    {b^A}_l {c^l}_A \, .
\end{align}
This current is BRST invariant also in the finite case, where the coefficient in front of the $bc$ current is $-\lambda k/(M+N)$. This is up to normalization exactly the u(1) charge of the $\mathcal{N}=2$ algebra from \cite{FigueroaO'Farrill:1995pv}.

We now introduce the notation $J_{1_M}^{(M)}=-J_\phi/(M+N)$ and $I_{1_M}^{(M)}=-I_\phi$ since these currents in the large $N$ limit have the same OPEs as we expect for the enlargement of $J_{d}^{(M)}$ to u$(M)$. We can then compactly write
\begin{align}\label{eq:spinonevminus}
  J^{(1)-}[t_d]=&-\lambda  J_d^{(M)}
 +  (1-\lambda)
    I^{(M)}_d
\end{align}
for all $t_d\in\text{u}(M)$. The action of the su$(M)$ currents $K_d$ on these currents is just first order OPEs giving the su$(M)$ rotation. The BRST current thus has the following OPE
\begin{align}\label{eq:spinonechangeunderBRST}
\frac{1}{k}j_{\textrm{BRST}}(z) J^{(1)-}[t_d](w)\sim \frac{\sum_{\textrm{u}(M)} {f_{ed}}^gJ^{(1)-}[t_g] c^e(w)}{k(z-w)}\ .
\end{align}

The OPEs of $J^{(1)}_1[t_d]$ with themselves obviously do not close, and we need a second set of spin one currents. If we now transcend the coset formalism and allow factors of $\tilde J_d^{(M,-k-N-2M)}$ in our linear combinations, we also get the following possibility in the large $N$ limit
\begin{align}\label{}
J^{(1)+}[t_d]=&J_d^{(M)}
 + I^{(M)}_d +\tilde J_d^{(M,-k-N-2M)}\nonumber \\
=&-   {\beta^B}_i {(t_d)^A}_B  {\gamma^i}_A+ J_d^{(M,k+N)}
 -  {b^B}_i {(t_d)^A}_B {c^i}_A+\tilde J_d^{(M,-k-N-2M)} \ .
\end{align}
This again fits exactly with \eqref{eq:suggestedfreecurrents} with the simple addition of the su$(M)$ currents, and its OPE with the BRST current is like in \eqref{eq:spinonechangeunderBRST}. Note that we could also make an operator with $t_d=1_d$, but since this is a central element, we do not need to include it on the bulk side, and from the point of view of the CFT it would be $K_\phi/(M+N)$ which goes like $K_\phi/k$ and is thus exact. Also note that the field $J^{(1)+}[t_d]$ would not be seen in a standard coset analysis since we used $\tilde J_d^{(M,-k-N-2M)}$. Obviously, this does not help for the closure of the $J^{(1)-}[t_d]$ currents, however, in the large $N$ limit the field $\tilde J_d^{(M,-k-N-2M)}$ becomes classical, and we will from now on remove it by hand. In this case the algebra will close.

\subsection*{Spin-3/2 currents, large $N$ case}

We now consider the spin-3/2 generators. Let us first note that we have no spin-1/2 BRST invariant generator, and thus the short multiplet of spin-1/2 and spin-one generators present in the $M=1$ free case \eqref{eq:suggestedfreecurrents} gets truncated to the top component. Spin-3/2 generators in the large $N$ limit can be built from ${b^B}_i {J^i}_A$ and ${c^i}_A{J^B}_i$. We suggest
\begin{align}\label{eq:spin32currents}
  J^{(3/2)\pm}[t_d]=\pm\frac{1}{k+N}{b^B}_i {(t_d)^A}_B{J^i}_A+ {c^i}_B {(t_d)^B}_A{J^A}_i\ ,
\end{align}
where $t_d\in\text{u}(M)$. The coefficients have been chosen to fit with \eqref{eq:suggestedfreecurrents}, when we only keep the $\beta\gamma$- and $bc$-systems. In the comparison we have to remove a term of the form $\gamma\del b$ in the free theory current which cannot appear in the CFT (like the spin-1/2 currents) and a term in the CFT of the form $\gamma\gamma\beta b$ which cannot appear in the free theory. In this case, and with the chosen coefficients we get a perfect match.  Finally, we can also recast this is ordinary supercharge notation as u$(M)$ extensions of the supercurrents
\begin{align}\label{eq:cftsupercurrents}
    G^{\pm}[t_d]=\frac{1}{\sqrt{2}}(J^{(3/2)+}[t_d]\pm J^{(3/2)-}[t_d])\ ,
\end{align}
where $G^{\pm}[1_M]$ are the actual supercurrents since these are BRST invariant - also in the finite case.

\subsection*{Spin-two currents finite case}
Let us now consider the spin-two currents. We will first consider the case of finite $N,M,k$. We need invariance in the $i$ and $A$ indices. The possible terms in the spin-two operator are thus
\begin{align}
  &J_\phi J_\phi\ ,\ g_{(N)}^{ab}J^{(N)}_a J^{(N)}_b\ ,\ g_{(M)}^{de}J^{(M)}_d J^{(M)}_e\ ,\ {J^A}_i{J^i}_A\ ,\ \del J_\phi\ ,\nonumber \\
   &J_\phi I_\phi\ ,\   g_{(N)}^{ab}J^{(N)}_a I^{(N)}_b\ ,\ g_{(M)}^{de}J^{(M)}_d I^{(M)}_e\ ,\nonumber\\ &I_\phi I_\phi\ ,\ g_{(N)}^{ab}I^{(N)}_aI^{(N)}_b \ ,\ T_{bc}\ ,\ \del I_\phi \ ,\label{eq:spin2constituents}
\end{align}
where $g_{ab}=\tr t_at_b$. For the fermions we have used that there are only two independent terms for terms with four fermions. The reason is the normal ordering for the four fermion terms can be chosen as $(({\beta^A}_i {\gamma^i}_B)({\beta^C}_i {\gamma^i}_D))$ up to terms containing two fermions with one derivative. There are now only two ways to contract the su$(M)$ indices. Explicitly we find that
\begin{align}\label{}
  g_{(M)}^{de}I^{(M)}_dI^{(M)}_e =2(M+N)T_{bc}-\frac{M+N}{MN} I_\phi I_\phi-g_{(N)}^{ab}I^{(N)}_aI^{(N)}_b \ ,
\end{align}
which is seen using the relation
\begin{align}\label{}
  g_{(M)}^{de}I^{(M)}_dI^{(M)}_e=(({b^A}_i {c^i}_B)({b^B}_i {c^i}_C))-\frac{1}{M}I_\phi I_\phi\ .
\end{align}
There is however one caveat: When $M=1$ we see that $g_{(N)}^{ab}I^{(N)}_aI^{(N)}_b$ can be rewritten in terms of $I_\phi I_\phi$.

We can now use the well-known OPE for a simple Lie algebra
\begin{align}\label{}
  J_a(z) g^{bc}J_b J_c(w)\sim 2(k+\hat g)  \frac{J_a(w)}{(z-w)^2}
\end{align}
where $\hat g$ is the dual Coxeter number and $k$ the level. We can now easily get the possible spin-two symmetries. Firstly we have
\begin{align}\label{}
  J^{(2)}_1=&T_{\textrm{su}(N+M)}-\frac{g_{(N)}^{ab}J^{(N)}_a J^{(N)}_b}{2(k+N)}-\frac{g_{(M)}^{de}J^{(M)}_d J^{(M)}_e}{2(k+M)}-\frac{J_\phi J_\phi}{2NM(N+M)k}\nonumber \\
  =&\frac{1}{2(k+N+M)}(2{J^A}_i{J^i}_A-\del J_\phi-\frac{1}{NMk}J_\phi J_\phi-\frac{M}{k+N} g_{(N)}^{ab}J^{(N)}_a J^{(N)}_b\nonumber\\
  &\qquad\qquad\qquad\qquad-\frac{N}{k+M} g_{(M)}^{de}J^{(M)}_d J^{(M)}_e) \ .
\end{align}
For the invariance the following OPEs are useful
\begin{align}\label{}
  J^{(N)}_a(z){J^A}_i{J^i}_A(w)\sim& \frac{M J^{(N)}_a}{(z-w)^2} \ ,\nonumber \\
  J^{(M)}_d(z){J^A}_i{J^i}_A(w)\sim& \frac{N J^{(M)}_d}{(z-w)^2} \ ,\nonumber \\
  J_\phi(z){J^A}_i{J^i}_A(w)\sim&\frac{(N+M)NMk}{(z-w)^3}+ \frac{(N+M) J_\phi}{(z-w)^2}  \ .
\end{align}

We can also get a spin-two current including the fermions:
\begin{align}\label{}
  J^{(2)}_2=&T_{bc}-\left(\frac{1}{NM(N+M)k}J_\phi I_\phi-\frac{1}{2NM(N+M)k^2}J_\phi J_\phi\right)\nonumber \\
  & \phantom{T_{bc}}-\left(\frac{1}{k}g_{(N)}^{ab}J^{(N)}_a I^{(N)}_b-\frac{M}{2(k+N)k}g_{(N)}^{ab}J^{(N)}_a J^{(N)}_b\right)\nonumber \\
  &\phantom{T_{bc}}-\left(\frac{1}{k}g_{(M)}^{de}J^{(M)}_d I^{(M)}_e-\frac{N}{2(k+M)k}g_{(M)}^{de}J^{(M)}_d J^{(M)}_e\right) \ .
\end{align}
Surprisingly, a third spin-two current is also possible when $M>1$ where the term $g_{(N)}^{ab}I^{(N)}_aI^{(N)}_b$ also can be used
\begin{align}\label{}
    J^{(2)}_3=& \frac{1}{2(M+N)}g_{(N)}^{ab}I^{(N)}_aI^{(N)}_b -\left(\frac{1}{k}g_{(N)}^{ab}J^{(N)}_a I^{(N)}_b-\frac{M}{2(k+N)k}g_{(N)}^{ab}J^{(N)}_a J^{(N)}_b\right)\ .
\end{align}
These are all the possible spin-two operators which we see as follows. Given a general linear combination of the operators in \eqref{eq:spin2constituents}, we can use $J^{(2)}_1,J^{(2)}_2,J^{(2)}_3$ to ensure that the coefficients of ${J^A}_i{J^i}_A$, $T_{bc}$ and $g_{(N)}^{ab}I^{(N)}_aI^{(N)}_b$ are zero. Further we can use $J^{(1)-}J^{(1)-}$ and $\del J^{(1)-}$ to set the coefficient of $I_\phi I_\phi$ and $\del I_\phi$ to zero. The coefficient of $\del J_\phi$ must then be zero since it is the only term with a third order OPE with $K_\phi$. It is easy to see that the remaining terms can not give invariants of the BRST operator.

We now want to consider the large $N$ limit. From the bulk side we expect to find two u$(M)$ extended spin-two currents. It might seem worrisome that we found three spin-two currents in the finite $N$ case, however we see that $J^{(2)}_3$ can be rewritten in terms of the invariant contraction of the extra spin one currents $J^{(1)-}[t_a]$, and thus should not be counted as a new current.

\subsection*{Spin-two currents, large $N$ case}

In the large $N$ limit, we again do not have to worry about the first order OPEs with $K_d$ for $t_d\in\textrm{su}(M)$. We thus have more possible spin-two operators
\begin{align}
  &J_\phi J_\phi\ ,\ J_\phi J^{(M)}_d\ ,\ g_{(N)}^{ab}J^{(N)}_a J^{(N)}_b\ ,\ J^{(M)}_d J^{(M)}_e+J^{(M)}_e J^{(M)}_d\ ,\ {J^A}_i{J^i}_B\ ,\ \del J_\phi\ ,\del J^{(M)}_d\ ,\nonumber \\
   &J_\phi I_\phi\ ,\ J_\phi I^{(M)}_d \ ,\ J^{(M)}_d I_\phi \ ,\ g_{(N)}^{ab}J^{(N)}_a{b^A}_i{(t_b)^i}_j{c^j}_B \ ,\ J^{(M)}_d I^{(M)}_e\ ,\nonumber\\
   &I_\phi I_\phi\ ,\ I_\phi I^{(M)}_d\ ,\ I^{(M)}_d I^{(M)}_e+I^{(M)}_e I^{(M)}_d \ ,\ \del I_\phi \ ,\ \del I^{(M)}_d  \ ,\ (\del{b^A}_i) {c^i}_B -{b^A}_i \del {c^i}_B\ .\label{eq:infinitespin2constituents}
\end{align}
Products of non-commuting currents have been kept in symmetric combinations for linear independence since we also include their derivatives. We can now find the following u$(M)$ extended spin-two operators for $t_d\in\textrm{su}(M)$
\begin{align}\label{}
  J^{(2)}_1[t_d]=&{J^B}_i{(t_d)^A}_B  {J^i}_A+\frac{N(k+M)}{2k+M}\del  J^{(M)}_d+\frac{1}{Mk}J_\phi J^{(M)}_d-\frac{N}{2k+M} J^{(M)}_f\tr (t^ft_dt^g)J^{(M)}_g \nonumber \\
  =& {J^B}_i{(t_d)^A}_B  {J^i}_A+\frac{N}{2}\del  J^{(M)}_d+\frac{1}{Mk}J_\phi J^{(M)}_d-\frac{N}{2(2k+M)} J^{(M)}_f\tr (t_d \{t^f,t^g\})J^{(M)}_g\ ,
\end{align}
and for $t_d=1$ we, of course, get the same result as above
\begin{align}\label{}
  J^{(2)}_1[1_M]=&{J^A}_i  {J^i}_A-\frac{1}{2}\del J_\phi-\frac{1}{2NMk}J_\phi J_\phi-\frac{M}{2(k+N)} g_{(N)}^{ab}J^{(N)}_a J^{(N)}_b\nonumber\\
  &-\frac{N}{2(k+M)} g_{(M)}^{de}J^{(M)}_d J^{(M)}_e \ .
\end{align}
We can also write this in a unified form for all $t_d\in \text{u}(M)$ with the notation from \eqref{eq:spinonevminus} in the large $N$ limit (we will also denote explicitly when the sum over generators is for u$(M)$ and not only su$(M)$)
\begin{align}\label{eq:J21uM}
  J^{(2)}_1[t_d]=&
   {J^B}_i{(t_d)^A}_B  {J^i}_A+\frac{N}{2}\del  J^{(M)}_d-\frac{\tr t_d}{2(k+N)} g_{(N)}^{ab}J^{(N)}_a J^{(N)}_b\nonumber\\
  &-\frac{N}{4k} \sum_{\text{u}(M)\text{ generators}}J^{(M)}_f\tr (t_d \{t^f,t^g\})J^{(M)}_g\ .
\end{align}

Here the following OPEs for $t_d\in\textrm{u}(M)$ are useful
\begin{align}\label{}
  J_\phi(z){J^B}_i{(t_d)^A}_B  {J^i}_A(w)\sim&\frac{(M+N)Nk\tr(t_d)}{(z-w)^3}+(M+N)\frac{\tr(t_d)\frac{1}{M}J_\phi-N\tr(t_d t^f)J_f}{(z-w)^2}\ ,\nonumber \\
  J^{(N)}_a(z){J^B}_i{(t_d)^A}_B  {J^i}_A(w)\sim& \frac{\tr(t_d)J^{(N)}_a}{(z-w)^2}\ ,\nonumber \\
  J^{(M)}_e(z){J^B}_i{(t_d)^A}_B  {J^i}_A(w)\sim&\frac{-kN\tr(t_et_d)}{(z-w)^3}+\frac{-\tr(t_et_d)\frac{1}{M}J_\phi+N\tr(t_d t_e t^f)J^{(M)}_{f}}{(z-w)^2}\ ,\nonumber \\
  &+ \frac{{J^B}_i{([t_e,t_d])^A}_B  {J^i}_A}{z-w}\ ,
\end{align}
and
\begin{align}\label{}
  J^{(M)}_e(z) J^{(M)}_f\tr (t^ft_dt^g)J^{(M)}_g(w)\sim&\frac{kM\tr(t_et_d)}{(z-w)^3}\nonumber \\
  &+\frac{k\tr(\{t_e,t_d\}t^f)J^{(M)}_f+M\tr(t_et_dt^f)J^{(M)}_f+\tr(t_d)J^{(M)}_e}{(z-w)^2}\nonumber \\
  &+\frac{J^{(M)}_f\tr (t^f[t_e,t_d]t^g)J^{(M)}_g}{z-w}
\end{align}
where we have used the sl$(N)$ identity $\sum_{d}t_d A t^d=\tr A \ 1_{M}-\frac{1}{M}A$. This gives vanishing OPEs with $K_a, K_\phi$ whereas for $K_e$ with $t_e\in\textrm{su}(M)$ we have first order OPEs, and we get
\begin{align}\label{eq:changeunderNRST}
\frac{1}{k}j_{\textrm{BRST}}(z) V^{(2)}_1[t_d](w)\sim \frac{\sum_e V^{(2)}_1[[t_e,t_d]] c^e(w)}{k(z-w)}\ ,
\end{align}
as in \eqref{eq:spinonechangeunderBRST}.

Considering again a general linear combination of the operators in \eqref{eq:infinitespin2constituents} we can use $J^{(2)}_1[t_d]$ to assume that we do not have any ${J^A}_i{J^i}_B$ terms. Further, we can use products and derivatives of $J^{(1)-}[t_a]$ to get that the only pure fermion terms are $(\del{b^A}_i) {c^i}_B -{b^A}_i \del {c^i}_B$. We can thus create the second u$(M)$ extended operator as
\begin{align}\label{}
  J^{(2)}_2[t_d]=&(\del{b^B}_i){(t_d)^A}_B  {c^i}_A -{b^B}_i {(t_d)^A}_B \del {c^i}_A+\frac{2}{NMk}J_\phi I^{(M)}_d +\frac{2}{Mk}J^{(M)}_d I_\phi -\frac{2}{Mk^2} J_\phi J^{(M)}_d\nonumber \\
    &-\frac{2}{k}g_{(N)}^{ab}J^{(N)}_a{b^B}_i{(t_b)^i}_j{(t_d)^A}_B {c^j}_A\nonumber \\
    &-\frac{1}{k}\big(J^{(M)}_f \tr(t_d\{t^f,t^g\})I^{(M)}_g-\frac{N}{2k+M}J^{(M)}_f \tr(t_d\{t^f,t^g\})J^{(M)}_g\big)\end{align}
for $t_d\in\textrm{su}(M)$ and as before
\begin{align}\label{}
  J^{(2)}_2[1_M]=&2T_{bc}-2\left(\frac{1}{NMk}J_\phi I_\phi-\frac{1}{2NMk^2}J_\phi J_\phi\right)\nonumber \\
  & \phantom{T_{bc}}-2\left(\frac{1}{k}g_{(N)}^{ab}J^{(N)}_a I^{(N)}_b-\frac{M}{2(k+N)k}g_{(N)}^{ab}J^{(N)}_a J^{(N)}_b\right)\nonumber \\
  &\phantom{T_{bc}}-2\left(\frac{1}{k}g_{(M)}^{de}J^{(M)}_d I^{(M)}_e-\frac{N}{2(k+M)k}g_{(M)}^{de}J^{(M)}_d J^{(M)}_e\right) \ .
\end{align}
Finally, we again write this in unified form for $t_d\in \text{u}(M)$ with the notation from \eqref{eq:spinonevminus} in the large $N$ limit
\begin{align}\label{eq:J22uM}
  J^{(2)}_2[t_d]=&(\del{b^B}_i){(t_d)^A}_B  {c^i}_A -{b^B}_i {(t_d)^A}_B \del {c^i}_A\nonumber \\
    &-\frac{2}{k}g_{(N)}^{ab}J^{(N)}_a{b^B}_i{(t_b)^i}_j{(t_d)^A}_B {c^j}_A+\frac{\tr t_d}{(k+N)k}g_{(N)}^{ab}J^{(N)}_a J^{(N)}_b\nonumber \\
    &-\frac{1}{k}\sum_{\text{u}(M)\text{ generators}}\big(J^{(M)}_f \tr(t_d\{t^f,t^g\})I^{(M)}_g-\frac{N}{2k}J^{(M)}_f \tr(t_d\{t^f,t^g\})J^{(M)}_g\big) \, .
\end{align}

Here we have used that for $t_d=1_M$, $T_{bc}[t_d]\equiv\tfrac12(\del{b^B}_i){(t_d)^A}_B  {c^i}_A -\tfrac12{b^B}_i {(t_d)^A}_B \del {c^i}_A$ is simply $T_{bc}$ and for $t_d\in\textrm{su}(M)$ we have the OPEs
\begin{align}\label{}
  I_\phi(z)T_{bc}[t_d](w)\sim&\frac{-I^{(M)}_d }{(z-w)^2}\ ,\nonumber\\
  I^{(N)}_a (z)T_{bc}[t_d](w)\sim&\frac{{b^B}_i{(t_a)^i}_j{(t_d)^A}_B{c^j}_A}{(z-w)^2}\ ,\nonumber \\
  I^{(M)}_e (z)T_{bc}[t_d](w)\sim&\frac{\tfrac12\tr(\{t_e,t_d\}t^f)I^{(M)}_f-\frac{1}{M}\tr(t_et_d)I_\phi}{(z-w)^2}+\frac{T_{bc}[[t_e,t_d]]}{z-w} \ .
\end{align}
Also note that we have a relation similar to \eqref{eq:changeunderNRST}.

It is easy to see that it is not possible to construct other BRST invariant operators.

For the comparison to the bulk in the next subsection, let us here calculate the OPE of the extended supercurrents from \eqref{eq:cftsupercurrents}
\begin{align}\label{eq:opesupercurrent1}
        G^{+}&[t_a](z)G^{-}[t_b](0)\sim\frac{\tfrac23 c \tr (t_at_b)}{M z^3}+\frac{2\lambda(1-\lambda){f_{ab}}^cJ^{(1)+}[t_c]+((1-2\lambda){f_{ab}}^c+{s_{ab}}^c)J^{(1)-}[t_c]}{z^2}
    \nonumber\\
    &+\frac{({s_{ab}}^c-{f_{ab}}^c)\tfrac{1}{k+N}J^{(2)}_1[t_c]+({s_{ab}}^c+{f_{ab}}^c)\tfrac{1-\lambda}{2}J^{(2)}_2[t_c]}{z}
    \nonumber\\
    &+\frac{\lambda(1-\lambda){f_{ab}}^c\del J^{(1)+}[t_c]+\tfrac12((1-2\lambda){f_{ab}}^c+{s_{ab}}^c)\del J^{(1)-}[t_c]}{z}\nonumber\\
    &-\frac{\lambda{f_{ab}}^c{s^{fg}}_c\big((1-\lambda)J^{(1)+}-J^{(1)-}\big)[t_f]\big((1-\lambda)J^{(1)+}-J^{(1)-}\big)[t_g]}{2k z}\nonumber\\
    &-\frac{\big({{f^g}_{b}}^c({s_{ac}}^f+{f_{ac}}^f)+\tfrac12({s_{ab}}^c+{f_{ab}}^c){f^{gf}}_c\big)\big(\lambda J^{(1)+}+J^{(1)-}\big)[t_f]\big((1-\lambda)J^{(1)+}-J^{(1)-}\big)[t_g]}{(k+N) z}\ ,
\end{align}
where the sums are over $\text{u}(M)$, ${f_{ab}}^c$ are the structure constants and ${s_{ab}}^c t_c=\{t_a,t_b\}$. The expression simplifies in the case of the actual supercurrents,
\begin{align}\label{eq:opesupercurrent2}
        G^{+}[1_M](z)G^{-}[1_M](0)\sim&\frac{\tfrac23 c }{ z^3}+\frac{J^{(1)-}[1_M]}{z^2} +\frac{\tfrac{2}{k+N}J^{(2)}_1[1_M]+(1-\lambda)J^{(2)}_2[1_M]+\del J^{(1)-}[1_M]}{z}
\ .
\end{align}
From this we see that the Virasoro tensor is
\begin{align}\label{}
    T=\frac{1}{k+N}J^{(2)}_1[1_M]+\frac{1-\lambda}{2}J^{(2)}_2[1_M]\ .
\end{align}
We note that the field $J^{(1)-}[1_M]$ is primary, whereas the fields $J^{(1)-}[t_a]$ in the su$(M)$ directions will be non-primary
\begin{align}\label{jminusnonprim}
    J^{(1)-}[t_a](z)T(0)\sim \frac{J^{(1)-}[t_a]}{z^2}-\frac{1-\lambda}{k}\frac{{f_a}^{bc}J^{(M)}_b I^{(M)}_c}{z}\ .
\end{align}
Of course, the fields $J^{(1)+}[t_a]$ are all non-primary having vanishing OPE with the stress-energy tensor.

The relation of $J^{(2)}_{1,2}$ to the basis $J^{(2)\pm}$ is not as straightforward in this case and will contain products of the spin-one operators, as we will see in next subsection. Also the relation to \eqref{eq:suggestedfreecurrents} is difficult due to normal ordering issues.

\subsection{OPEs of currents from the bulk/boundary correspondence}\label{subsec:opebulk}

In this section we will derive some of the boundary OPEs from the bulk side, and check the results with the currents derived in last subsection.

Let us first see which OPEs we get to linear order from the bulk side. These are most easily obtained in the following way using the notation and results of \cite{Chang:2011mz,Creutzig:2012xb}: We split the gauge field into the AdS$_3$ part $A_\text{AdS}$ and a small deformation $\Omega$
\begin{align}
 A=A_\text{AdS}+\Omega\ .
\end{align}
Here
\begin{align}
 A_\text{AdS} = e^\rho V_1^2 dz + V_0^2 d \rho  ~.
\label{background}
\end{align}
The coupling between the bulk and the boundary is then
\begin{align}\label{}
    \exp\big(-\frac{1}{2\pi}\int d^2z [(\Omega_{\bar z})^{(s)\sigma}_{s-1}]|_{\textrm{bdry}}J^{(s)\sigma}\big)\ .
\end{align}

The change under (gauge) transformations of the field is
\begin{align}
 &\delta A = d \Lambda + [A,\Lambda]_* \label{gaugetrans} ~.
\end{align}
As proposed in \cite{Creutzig:2012xb} the gauge transformation
\begin{align}
\begin{split}\label{eq:gaugetransrepeat}
 \Lambda^{(s)\pm}&=\epsilon^\pm_{s}\sum_{n=1}^{2s-1} \frac{1}{(n-1)!} (- \partial)^{n-1}
  \Lambda^{(s)} (z) e^{(s-n)\rho} V^{(s)\pm}_{s-n}
\end{split}
\end{align}
of some operator $\mathcal{O}$ corresponds on the bulk side to
\begin{align}\label{eq:changeonbdry}
    \frac{1}{2\pi i}\oint_0 dz \Lambda^{(s)} (z) J^{(s)\pm}(z)\mathcal{O}(0)\ .
\end{align}
Indeed for $\Lambda^{(s)}(z)=z^{s-1-m}$, $m=-s+1,\ldots,s-1$ these are the global transformations. To calculate the OPEs of currents with themselves, we can then proceed as follows. First we create a current insertion $J^{(s)\sigma}$ at $z=0$ by using \eqref{eq:gaugetransrepeat} with $\Lambda^{(s)}=1/z$ on the identity operator, respectively, on the AdS solution. This gives the following fields on the bulk side
\begin{align}\label{eq:omegasol}
    \Omega^{(s)\pm}_z&=\epsilon\frac{1}{(2s-2)!}\del^{2s-1}\Lambda^{(s)}(z)e^{-(s-1)\rho}V^{(s)\pm}_{-(s-1)}\ ,\\
    \Omega^{(s)\pm}_{\bar z}&=\epsilon\sum_{n=1}^{2s-1} \frac{1}{(n-1)!} (- \partial)^{n-1}
  \bar\del\Lambda^{(s)} (z) e^{(s-n)\rho} V^{(s)\pm}_{s-n}\sim \epsilon 2\pi\delta^{(2)}(z-w)e^{(s-1)\rho}V^{(s)\pm}_{s-1}+\ldots\ , \nonumber \\
    \Omega^{(s)\pm}_\rho&=0\ , \nonumber
\end{align}
where the important part is the leading term in $\Omega^{(s)\pm}_{\bar z}$ which is a delta function, indeed giving the wanted insertion. This is a solution to the linearized equation of motion
\begin{align}\label{eq:eqmomega}
 d\Omega+A_\text{AdS}\wedge_*\Omega+\Omega\wedge_*A_\text{AdS}=0\ .
\end{align}
To find the OPE with $J^{(s')\sigma'}$ we now perform the transformation \eqref{eq:gaugetransrepeat} with $\Lambda^{(s)}=1,z,\ldots,z^{s+s'-1}$. When we are not dealing with a global symmetry we have to remember that $A_\text{AdS}$ is not kept invariant, which gives the central extension on the CFT side. This gives an extra term (coming from varying both the bulk and the necessary extra boundary term, see \cite{Chang:2011mz})
\begin{align}
 \delta S=-\frac{k_{\text{CS}}}{2\pi}\int d^2 z e^{2\rho}\str(\Omega_{\bar z}\delta\Omega_z) \ .
\end{align}

Here the supertrace is the supertrace over higher spin algebra and the trace over the su$(M)$ part with a prefactor $1/M$. The supertrace is normalized to have $\str(V^{(2)+}_1 V^{(2)+}_{-1})=-1$, i.e. the total supertrace has $\str(V^{(2)+}_1\otimes 1_M , \,V^{(2)+}_{-1}\otimes 1_M)=-1$. An explicit formula was found in \cite{Creutzig:2012xb}
\begin{align}\label{eq:supertraceexpliciteven}
 &\str\left(P^\pm V^{s}_m,P^\pm V^{s}_{-m}\right)  \\
& =-\frac{2(-1)^{s-m-1}\Gamma(s+m)\Gamma(s-m)}{\lambda_+(1-\lambda_+)(2s-2)!}\frac{\Gamma(s)\sqrt{\pi}}{4^s\Gamma(s+1/2)}(1-\lambda_\pm)_{s-1}(1+\lambda_\pm)_{s-1}\lambda_\pm \nonumber
\end{align}
for $s \in \mathbb{Z}$ and
\begin{align}\label{eq:supertraceexplicitodd}
 &\str\left(P^\pm V^{s}_m,P^\mp V^{s}_{-m}\right) \\
&=-\frac{2(-1)^{s-m-1}\Gamma(s+m)\Gamma(s-m)}{\lambda_+(1-\lambda_+)(2s-2)!}\frac{\Gamma(s-\tfrac12)\sqrt{\pi}}{4^s\Gamma(s)}(1-\lambda_+)_{s-\frac12}(1+\lambda_+)_{s-\frac32}\lambda_+\nonumber
\end{align}
for $s \in \mathbb{Z}+1/2$. Here $(a)_n=\Gamma(a+n)/\Gamma(a)$ is the ascending Pochhammer symbol.

We can now use the above formulas to consider the OPEs of the su$(M)$ current $J^{(1)+}[t_a]$ dual to $1\otimes t_a$. Remembering the special definition $V^{(1)-}_{0}=\tfrac12(1-2\lambda+k)$, we define $J^{(1)-}[t_a]$ to be dual to $\tfrac12(1-2\lambda+k)\otimes t_a$. Here $k$ is, of course, the operator in the higher spin algebra, not the level. We then find that $J^{(1)+}[t_a]$ has a first order OPE with all operators except itself
\begin{align}\label{eq:umrotation}
    J^{(1)+}[t_a](z)J^{(s)\pm}[t_b](0)\sim\frac{{f_{ab}}^cJ^{(s)\pm}[t_c]}{z} \ ,
\end{align}
and
\begin{align}\label{}
    J^{(1)+}[t_a](z)J^{(1)+}[t_b](0)\sim\frac{2 k_{CS}\tr{t_at_b}}{\lambda(1-\lambda)M z^2}+\frac{{f_{ab}}^cJ^{(1)+}[t_c]}{z}\ ,
\end{align}
which follows from $\str(1)=-2/(\lambda(1-\lambda))$ and $\str(1-2\lambda +k)=0$. This fits perfectly with our currents. These were indeed the currents, which had only first order OPEs with the su$(M)$ current and transforming like \eqref{eq:umrotation}, see eqs. \eqref{eq:spinonechangeunderBRST} and \eqref{eq:changeunderNRST}. Also the fact that the found su$(M)$ current has level $k+N$ fits perfectly since
\begin{align}\label{}
    c\sim3\lambda(1-\lambda)M(N+k) \ .
\end{align}
We have here used that
\begin{align}\label{eq:BHrela}
    c=6k_{\text{CS}}\ ,
\end{align}
as we will see below.

We can see in general that a central term is only happening between two operators of the same spin and has the form
\begin{align}\label{}
    J^{(s)\sigma}[t_a](z)J^{(s)\sigma'}[t_b](0)\sim-\frac{(2s-1)c\str(V^{(s)\sigma'}_{s-1}V^{(s)\sigma}_{-s+1})\tr{t_at_b}}{6M z^{2s}}+\ldots\ .
\end{align}
For the second spin one current $J^{(1)-}[t_a]$, defined as above, we then easily get
\begin{align}\label{}
      J^{(1)-}[t_a](z)J^{(1)-}[t_b](0)\sim\frac{c\tr{t_at_b}}{3M z^2}+\frac{\lambda(1-\lambda){f_{ab}}^cJ^{(1)+}[t_c]+(1-2\lambda){f_{ab}}^cJ^{(1)-}[t_c]}{z}\ .
\end{align}
Here we have used $\tfrac14\str((1-2\lambda+k)(1-2\lambda+k))=-2$ and the commutator \eqref{eq:spinoneminuscommu}. This again fits perfectly with our currents \eqref{eq:spinonevminus}.  The OPE of $J^{(1)-}[t_a]$ with the higher spin operators take the form
\begin{align}\label{}
    J^{(1)-}[t_a](z)J^{(s)\pm}[t_b](0)\sim &\frac{{f_{ab}}^c((\tfrac12-\lambda)J^{(s)\pm}[t_c]+\tfrac12 J^{(s)\mp}[t_c])}{z}\qquad\textrm{for }s\in\mathbb{Z} \ ,\nonumber\\
\sim &\frac{(\tfrac12-\lambda){f_{ab}}^cJ^{(s)\pm}[t_c]+\tfrac12 {s_{ab}}^cJ^{(s)\mp}[t_c]}{z}\qquad\textrm{for }s\in\mathbb{Z}+\tfrac12\ ,\label{eq:opejminusone}
\end{align}
where ${s_{ab}}^c t_c=\{t_a,t_b\}$. This is indeed fulfilled for the fermionic currents \eqref{eq:spin32currents}.

We can now check the OPEs of the extended supercurrents $G^{\pm}[t_a]=(J^{(3/2)+}[t_a]\pm J^{(3/2)-}[t_a])/\sqrt{2}$. Following the linear approach we get
\begin{multline}\label{}
    G^{+}[t_a](z)G^{-}[t_b](0)\sim\frac{4 k_{CS}g_{ab}}{M z^3}
    +\frac{2\lambda(1-\lambda){f_{ab}}^cJ^{(1)+}[t_c]+((1-2\lambda){f_{ab}}^c+{s_{ab}}^c)J^{(1)-}[t_c]}{z^2}\\
    +\frac{{s_{ab}}^cJ^{(2)+}[t_c]+{f_{ab}}^c J^{(2)-}[t_c]}{z}\\
    +\frac{\del\Big(\lambda(1-\lambda){f_{ab}}^cJ^{(1)+}[t_c]+((\tfrac12-\lambda){f_{ab}}^c+\tfrac12{s_{ab}}^c)J^{(1)-}[t_c]\Big)+\mathcal{O}}{z} \ .
\end{multline}
Here the novelty compared to $M=1$ is that also $J^{(2)-}[t_c]$ appears, see appendix \ref{glM}. Starting at this spin non-linear terms can occur, and these are denoted by $\mathcal{O}$, whereas the above OPEs containing the spin-one currents cannot have non-linear terms. The non-linear terms come with a normalization of $1/k_{CS}$ and such non-perturbative terms are not captured by the above method, only the linear terms. Up to these non-linear terms, we now see that we have perfect agreement with \eqref{eq:opesupercurrent1} and \eqref{eq:opesupercurrent2}. From the central term, we see that the central charge need to be identified via the Brown-Henneaux relation \eqref{eq:BHrela}. Further, we can actually fix the non-linear terms for the stress-energy tensor, assuming that the proposed bulk-boundary correspondence holds:
\begin{align}\label{eq:nonlintermfromcft}
    J^{(2)+}[1_M]=&T_{\text{CFT}}-\frac{3}{2c}J^{(1)-}[1]J^{(1)-}[1]\nonumber\\
    &-\sum_{\text{su}(M)}\left(\frac{g^{de}I^{(M)}_d I^{(M)}_e}{2(N+M)}+\frac{g^{de}J^{(M)}_d J^{(M)}_e}{2(k+M)}-\frac{g^{de}J^{(1)+}[t_d]J^{(1)+}[t_e]}{2(N+k+M)}\right)\nonumber\\
    \approx&T_{\text{CFT}}-\frac{3M}{2c}\sum_{\text{u}(M)}g^{de}J^{(1)-}[t_d]J^{(1)-}[t_e]\ .
\end{align}
Here the last line is to leading $N$-order. To fix the terms we have used that we know from the bulk that $T_{\text{CFT}}$ has zero OPE with all $J^{(1)+}[t_a]$, and the same is true for $J^{(2)+}[1_M]$ via \eqref{eq:umrotation} and hence also for the non-linear terms relating $J^{(2)+}[1_M]$ and $T_{\text{CFT}}$. Further, we know from the CFT that the currents
$J^{(1)-}[t_a]$ are not quite primary and satisfy the OPE \eqref{jminusnonprim}, however, $J^{(1)-}[t_a]$ and $J^{(2)+}[1_M]$ should have zero OPEs by \eqref{eq:opejminusone}. The suggested non-linear terms solve this.

To find the non-linear terms, we can use the bulk equations of motion via the method developed in \cite{Gutperle:2011kf} and used very nicely in the $M=1$ case in \cite{Moradi:2012xd}. We first introduce the gauge field $a$ with the $\rho$-dependence adjointly removed
\begin{align}\label{}
  A=b^{-1}ab+b^{-1}db\ ,\qquad b=e^{\rho V^{(2)+}_0}\ .
\end{align}
We now consider the background with the current $-\int d^2 z\mu(z,\bar z) J^{(t)\delta}[t_b]$ turned on, where $\mu=\epsilon 2\pi\delta^{(2)}(z-w)$ when comparing to \eqref{eq:omegasol}. In this background the gauge field in the lowest weight gauge takes the form (including the AdS part)
\begin{align}\label{}
  a_z=& V^{(2)+}_1\otimes 1-\frac{1}{k_{CS}}\left(\langle J^{(s)\sigma}[t_a]\rangle_0+\langle J^{(s)\sigma}[t_a]\rangle_\mu\right) g^{V^{(s')\sigma'}_{-s'+1}\otimes t_{a'}\,V^{(s)\sigma}_{s-1}\otimes t_a}V^{(s')\sigma'}_{-s'+1}\otimes t_{a'}    \ , & \nonumber \\
  a_{\bar z}=& \sum_{m=-t+1}^{t-1} \mu^c_m V^{(t)\delta}_m\otimes t_c
\end{align}
where $\langle J^{(s)\sigma}[t_a]\rangle_0$ are zeroth order in $\mu$ and hence holomorphic, and $\langle J^{(s)\sigma}[t_a]\rangle_\mu$ is first order. $ g^{V^{(s')\sigma'}_{-s'+1}\otimes t_{a'}\,V^{(s)\sigma}_{s-1}\otimes t_a} $ is the inverse metric of the supertrace defined previously, and $\mu\equiv\mu^b_{t-1}$. The remaining $\mu^c_{t-1}$ for $c\neq b$ are zero, however the trailing terms $\mu^c_{m}$ with $m<t-1$ can and will be non-zero. The equations of motion
\begin{align}\label{}
\del a_{\bar z}-\bar\del a_z+[a_z,a_{\bar z}]=0
\end{align}
can be solved at linear order in $\mu$ for $\bar\del \langle J^{(s)\sigma}[t_a]\rangle_\mu$ which is equivalent to determine
\begin{align}\label{}
-\frac{1}{2\pi}\del_{\bar z}\int\, d^2 w\mu(w,\bar w)\langle J^{(s)\sigma}[t_a](z)J^{(t)\delta}[t_b](w)\rangle
\end{align}
on the CFT side. We can use this to confirm the above OPEs and to get the OPEs of the u$(M)$-extended supercharges $G^{\pm}[t_a]$. Firstly
\begin{align}\label{}
    G^{\pm}[t_a](z)G^{\pm}[t_b](0)\sim 0\ ,
\end{align}
which is certainly fulfilled by \eqref{eq:cftsupercurrents}. And the non-trivial OPEs are
\begin{align}\label{eq:supercurrentopebulk}
    G^{-}[t_a](z)G^{+}[t_b](0)\sim&\frac{4 k_{CS}g_{ab}}{M z^3}
    +\frac{2A_{ab}}{z^2}\\
    &+\frac{{s_{ab}}^cJ^{(2)+}[t_c]-{f_{ab}}^c J^{(2)-}[t_c]+\del A_{ab}+\frac{M}{2k_{CS}}{A^c}_bA_{ac}+\mathcal{O}(t_a,t_b)}{z}\nonumber
\end{align}
where
\begin{align}\label{}
    A_{ab}=\lambda(1-\lambda){f_{ab}}^cJ^{(1)+}[t_c]+((\tfrac12-\lambda){f_{ab}}^c-\tfrac12{s_{ab}}^c)J^{(1)-}[t_c]  \ .
\end{align}
The term $\mathcal{O}(t_a,t_b)$ is added due to the fact that this bulk can not know about the quantum normal ordering issues of the non-linear term. Due to the OPEs of the spin-one operators which each other, we see that $\mathcal{O}(t_a,t_b)$ can be a the derivative of some combination of the spin-one operators, and has a normalization of the order $1/k_{CS}$.

The result fits perfectly with the linear analysis. Also, looking at $t_a=t_b=1_M$ we see that we again get the $\mathcal{N}=2$ superalgebra with
\begin{align}\label{}
    T_{\textrm{bulk}}=J^{(2)+}[1_M]+\frac{3M}{4k_{CS}}\sum_{\text{u}(M)}g^{de}J^{(1)-}[t_d]J^{(1)-}[t_e]+\tfrac12\mathcal{O}(t_a,t_b)\ .
\end{align}
From \eqref{eq:nonlintermfromcft} we conclude that the stress-energy tensor from the bulk is the same as the CFT Virasoro tensor, at least up to the derivative of a spin-one current which amounts to a twist.

Finally, we can also compare for currents in general u$(M)$ directions. First, the relation between $J^{(2)\pm}[t_a]$ and the currents $J^{(2)}_{1,2}[t_a]$ found in last section in eqs. \eqref{eq:J21uM} and \eqref{eq:J22uM} needs to be determined. It turns out that the relation is fixed by \eqref{eq:umrotation} and \eqref{eq:opejminusone} with the normalizations fixed from comparison to \eqref{eq:suggestedfreecurrents} and demanding that the relation fits with the u(1) case for $t_a=1_M$. We then get
\begin{align}\label{}
  J^{(2)+}[t_a]&\approx\frac{1}{k+N}J^{(2)}_{1}[t_a]+\frac{1-\lambda}{2}J^{(2)}_{2}[t_a]-\frac{3M}{4c}\sum_{\text{u}(M)}{s^{bc}}_aJ^{(1)-}[t_b]J^{(1)-}[t_c]\ , \nonumber\\
    J^{(2)-}[t_a]&\approx-\frac{1}{k+N}J^{(2)}_{1}[t_a]+\frac{1-\lambda}{2}J^{(2)}_{2}[t_a]-\frac{3M}{4c}\sum_{\text{u}(M)}{s^{bc}}_aJ^{(1)-}[t_b]J^{(1)-}[t_c]\ ,
\end{align}
to leading order in the large $N$ limit. We can now compare the result from the boundary side, \eqref{eq:opesupercurrent1}, to the result from the bulk, \eqref{eq:supercurrentopebulk}. This indeed gives a match including the non-linear terms up to derivatives of spin-one currents scaling like $1/k$ which were undetermined from the bulk. This gives a strong confirmation of the proposed duality.

\section{Conclusion and discussions}
\label{conclusion}

We have proposed that a bosonic higher spin gravity on AdS$_3$ with $\text{U}(M)$
Chan-Paton factor in \cite{Prokushkin:1998bq} is dual to the two dimensional Grassmannian model \eqref{GC} (or \eqref{GC2} after su$(M)$ factor added)
in the 't Hooft limit with large $N,k$, but with $M$ and the  't Hooft parameter $\lambda_M$
\eqref{thooft} finite.
It was argued in \cite{Chang:2012kt} that a higher spin gravity on AdS$_4$ with U$(M)$
Chan-Paton factor is dual to three dimensional $\text{U}(N)_k \times \text{U}(M)_{-k}$ Chern-Simons theory
coupled with bi-fundamental matter $A_{i\alpha},B_{\beta j}$ for large $N,k$, but finite $M$.
Here we choose $i,j$ for the $\text{U}(N)$ index and $\alpha,\beta$ for the $\text{U}(M)$ index.
Then higher spin gauge fields with U$(M)$ Chan-Paton factor are  dual to operators of the form
$\sum_i A_{\alpha i}B_{i\beta}$. Thus the U$(N)$ invariant condition is assigned, but the
$\text{U}(M)$ invariance is somehow broken by a ``deconfinement.''
We have prepared a similar system with $NM$ complex free bosons
$\phi_{iA}$, $\phi^\dagger_{iA}$  in two dimensions
to construct operators \eqref{bw0m} dual to higher spin gauge fields. As in the higher dimensional
case, we assign only the U$(N)$ invariant condition and the U$(M)$ symmetry is treated as a global symmetry. We have shown that the free system has the same higher spin symmetry as the asymptotic
symmetry of bulk theory.
The free system arises as the $\lambda_M = 1$ limit of the Grassmannian model \eqref{GC}, and the
gravity partition function can be reproduced from the limit of the CFT.

We have also considered ${\cal N}=2$ holography between the higher spin supergravity on
 AdS$_3$ with $\text{U}(M)$ Chan-Paton factor in \cite{Prokushkin:1998bq} and the
 ${\cal N}=(2,2)$ Grassmannian  Kazama-Suzuki model \eqref{scoset} \cite{Kazama:1988qp,Kazama:1988uz} (or \eqref{scoset2} after su$(M)$ factor added).
 We again need to take the 't Hooft limit with large $N,k$, but with $M$ and the
't Hooft parameter $\lambda$ \eqref{thoofts} finite.
We have shown that a free system with $NM$ complex free bosons and
$NM$ complex free fermions has the desired higher spin symmetry, and
the $\lambda =0$ limit of the Grassmannian model reduces to the free system.
We have constructed low spin currents for the Grassmannian model at generic $\lambda$,
and discuss how  the U$(M)$ deconfinement occurs at the 't Hooft limit.
These currents are compared with those obtained from the bulk theory. The perfect match of the OPEs of the extended
supercharges between the bulk and boundary theories is a very strong hint that the symmetries of the two sides match at
all $\lambda$-values. The reason is that we expect the currents of spin-two, together with standard supersymmetry relations, to
generate all the higher spin symmetries, and from equation \eqref{eq:supercurrentopebulk} we see that the extended supercharges
in turn generate all the spin-two currents except $J^{(2)-}[1]$.
The one-loop partition function of the supergravity theory can be reproduced
by 't Hooft limit of the coset \eqref{scoset2}.

A motivation to study the supersymmetric case is to see a relation between higher spin
gauge theory and superstring theory. Supersymmetry can be introduced to the higher spin
gravity on AdS$_4$ and it was claimed in \cite{Chang:2012kt} that a supersymmetric
higher spin theory is dual to the  ${\cal N}=6$
$\text{U}(N)_k \times \text{U}(M)_{-k}$ Chern-Simons-matter theory (ABJ theory)
\cite{Aharony:2008ug,Aharony:2008gk} with large $N,k$, but finite $M$.
Since the ABJ theory with large $N,k,M$ is dual to a superstring theory, we can see
a relation between higher spin gauge theory  and superstring theory.
The states dual to superstrings are of the form $\text{tr}( AB \cdots AB)$, thus
we need a kind of U$(M)$ deconfinement phase transition  in terms of
$\lambda^\text{Bulk} = M/N$ to go to states dual to higher spin fields.
We have studied our holography in the expectation that a similar relation between higher spin
gauge theory and superstring theory can be seen even in the lower dimensional case.%
\footnote{See \cite{Gaberdiel:2013vva} for an example in this direction given quite recently.}
For this purpose, we first need to find a superstring dual of the Grassmannian model.
Moreover, we should investigate more on the  U$(M)$ confinement/deconfinement phase transition
of the Grassmannian model.

There are other open problems on the higher spin holography.
For instance,
it is natural to ask for a CFT dual to $\mathcal N=1$ higher spin supergravity
with O$(M)$ Chan-Paton factor. A candidate might be an appropriate orbifold of
\begin{align}
  \frac{\text{so}(2N+M)_k \oplus \text{so}(2NM)_1}{\text{so}(2N)_{k+M}
  \oplus \text{so}(M)_{k+2N}} \, .
 \end{align}
{}For $M=1$ the coset reduces to the one in \cite{Creutzig:2012ar}.
We are also interested in what would happen away from the 't Hooft limit
where we have to deal with quantum effects on the gravity side.
If a consistent quantum gravity can be described only by superstring theory,
then the Grassmannian coset \eqref{scoset} without extra su$(M)$ currents
should be used as the dual CFT. However, in three dimensions, there is a possibility
that the higher spin theory can be quantized consistently,
and in that case the duality between the higher spin theory and the coset
\eqref{scoset2} should hold even at finite $N$.

\subsection*{Acknowledgements}

We are grateful to T.~Eguchi, M.~R.~Gaberdiel, R.~Gopakumar and V.~Schomerus for useful discussions.
The work of YH was supported in part by JSPS KAKENHI Grant Number 24740170.
 The work of PBR is funded by AFR grant 3971664 from Fonds National de la Recherche, Luxembourg, and partial  support by the
Internal Research Project  GEOMQ11 (Martin Schlichenmaier),  University of Luxembourg,
is also acknowledged.

\appendix

\section{gl$(M)$ extended higher spin algebra}
\label{glM}

For the supersymmetric higher spin algebra $\text{shs}[\lambda]\oplus \mathbb{C}$ we will use the notation of appendix B in \cite{Creutzig:2012xb}. The generators are
$V_m^{(s)+}\equiv V_m^s$ and $ V_m^{(s)-}=k V_m^s$, where $k^2=1$, where $k$ is not to be confused with the level appearing in the dual CFT.
The star products among the generators $V_m^s$ can be expressed as
\begin{align}
 V_m^s * V_n^t = \frac12 \sum_{u=1,2,\cdots}^{s+t-|s-t| -1}
 g^{st}_u (m,n;\lambda_k) V^{s+t-u}_{m+n}
\end{align}
with $\lambda_k=(1-\nu k)/2$ where $\nu=1-2\lambda$. The star product for operators with $k$ follows trivially,
one simply has to remember that $k$ anti-commutes with the half-integer spin operators. The particular values of the structure coefficients that we need in this paper, can be found in \cite{Creutzig:2012xb}.

We know the following relations \cite{Creutzig:2012xb}
\begin{align}
    g^{st}_u (m,n;\lambda,k)=(-1)^{1+u}g^{ts}_u (n,m;\lambda,(-1)^{2(t+s)}k)\ ,
\end{align}
which means
\begin{align}\label{}
  [V_m^s , V_n^t]_{*}=\sum_{u=2,4,\cdots}^{s+t-|s-t| -1} g^{st}_u (m,n;\lambda_k) V^{s+t-u}_{m+n}\qquad \textrm{for }s,t\in \mathbb{Z}\ ,
\end{align}
and
\begin{align}\label{}
  \{V_m^s , V_n^t\}_{*}=\sum_{u=1,3,\cdots}^{s+t-|s-t| -1} g^{st}_u (m,n;\lambda_k) V^{s+t-u}_{m+n}\qquad \textrm{for }s,t\in \mathbb{Z}+1/2\ .
\end{align}
Further we expect that $g^{st}_u (m,n;\lambda,k)$ is independent of $k$ for $u$ odd. This would then imply
\begin{align}\label{}
  [V_m^s , V_n^t]_{*}=\sum_{u=2,4,\cdots}^{s+t-|s-t| -1} g^{st}_u (m,n;\lambda_k) V^{s+t-u}_{m+n}\qquad \textrm{for }s\in \mathbb{Z},t\in\mathbb{Z}+1/2\ .
\end{align}

We now want to consider the gl$(M)$ extension of this, $(\text{shs}[\lambda]\oplus \mathbb{C})\otimes \text{gl}(M)$. In this case we will, however, not find such nice relations. Denoting the generators of gl$(M)$ by $t_a$,
we define the product as
\begin{align}\label{}
  (V_m^s\otimes t_a)*(V_n^t\otimes t_b)=V_m^s*V_n^t\otimes t_a t_b\ .
\end{align}
We see that it is essential that gl$(M)$ is closed under matrix multiplication, and we could not have used e.g. sl$(M)$.
By using the standard (anti-)commutator $[[,]]_*$ we now get a Lie superalgebra. Denoting
\begin{align}\label{}
  [t_a,t_b]={f_{ab}}^c t_c\ ,\qquad \{t_a,t_b\}={s_{ab}}^c t_c\ ,
\end{align}
we have as example for the bosonic subalgebra
\begin{align}\label{}
  [V_m^s\otimes t_a , V_n^t\otimes t_b]_{*}=&\frac12\sum_{u=2,4,\cdots}^{s+t-|s-t| -1} g^{st}_u (m,n;\lambda_k) {s_{ab}}^c(V^{s+t-u}_{m+n},t_c)\nonumber \\
  &+\frac12\sum_{u=1,3,\cdots}^{s+t-|s-t| -1} g^{st}_u (m,n;\lambda_k) {f_{ab}}^c V^{s+t-u}_{m+n}\otimes t_c\qquad \textrm{for }s,t\in \mathbb{Z}\ .
\end{align}

As a particular example, not that $(1,t_a)$ is not central, but only $(1,1)$, and that the extension of the former only spin-one operator $\tfrac12(\nu+k)$, where $\nu=1-2\lambda$, now generates these elements
\begin{align}\label{eq:spinoneminuscommu}
  [(\nu+k)\otimes t_a,(\nu+k)\otimes t_b]_*={f_{ab}}^c (\nu^2+1+2\nu k)\otimes t_c\ .
\end{align}

\section{Extended $w_\infty [\lambda]$ algebras from ghost systems}
\label{shs}

In this paper, we have proposed that the Grassmannian coset \eqref{GC} is dual to the
bosonic higher spin theory with $\text{U}(M)$ Chan-Paton factor at generic $\lambda$.
Here we consider a bit different system with $NM$ sets of free bosons $(\gamma_{iA},\beta_{iA})$
where $i=1,2,\ldots,N$ and $A=1,2,\ldots , M$.
The conformal weights are $(\frac{1 - \lambda}{2}, \frac{1+\lambda}{2})$ and OPEs are
\begin{align}
 \gamma_{iA} (z) \beta_{jB} (0) \sim  \delta_{ij}  \delta_{AB} \frac{1}{z} \, .
\end{align}
{}From them we can define currents
\begin{align}
 & [J^{(s)} (z) ]_{AB}  = \sum_{i=1}^N \sum_{l=0}^{s-1} a^l (s, \lambda) \partial^{s - 1 - l} \{ (\partial^l \beta_{iA} ) \gamma_{iB} \} \, ,
\label{bmcurrent}
\end{align}
where
\begin{align}
 a^l (s , \lambda) = \begin{pmatrix} s-1 \\ i \end{pmatrix} \frac{(-  \lambda - s + 1)_{s-1-l}}{(s+l)_{s-1-l} } \end{align}
with $(a)_n = \Gamma (a+n)/ \Gamma (a)$. {}For $M=1$ they are  given in \cite{Bergshoeff:1991dz}.
We can construct the operators as
\begin{align}
[{\cal O} (z , \bar z )]_{AB} = \sum_{i=1}^N \gamma_{i A} (z) \otimes \bar \gamma_{iB} (\bar z)\, , \qquad
[ \tilde {\cal O}(z , \bar z ) ] _{AB}  = \sum_{i=1}^N \beta_{i A} (z) \otimes \bar \beta_{iB} (\bar z)\, ,
\end{align}
whose conformal weights are $h = \frac{1 - \lambda}{2}$ and $h = \frac{1 + \lambda}{2}$, respectively.
See \cite{Moradi:2012xd} for $M=1$.
The currents \eqref{bmcurrent}  generate a linear algebra $w_{1 + \infty}[\lambda] \otimes {\cal M} $,
where the wedge subalgebra of $w_{1+\infty}[\lambda]$ is $\text{hs}[\lambda] \oplus \text{u}(1)$.
However, outside the wedge subalgebra we cannot decouple the $\text{u}(1) \otimes 1_M$ sector, so
we cannot relate the asymptotic symmetry algebra of the dual gravity theory. In other words, the free theory
is closely related to, but different from the CFT dual to the higher spin theory with $\text{U}(M)$
Chan-Paton factor.

{}Similar generators can be constructed as in \eqref{eq:suggestedfreecurrents} for the ${\cal N}=2$
supersymmetric case by introducing $(b,c)$ systems along with $(\beta,\gamma)$ systems above.
The free system should have some direct relation to the Grassmannian coset \eqref{scoset}, see section \ref{sec:producttheory}.
In terms of the free system, the operators dual to the massive matter are expressed as
\begin{align}
 [ {\cal O}_B^+ (z , \bar z )]_{AB} = \sum_{i=1}^N \gamma_{i A} (z) \otimes \bar \gamma_{iB} (\bar z)\, , \qquad
  [ {\cal O}_F^+ (z , \bar z )]_{AB} = \sum_{i=1}^N c_{i A} (z) \otimes \bar \gamma_{iB}(\bar z)\, , \\
   [ {\cal O}_F^- (z , \bar z )]_{AB} = \sum_{i=1}^N \gamma_{i A} (z) \otimes \bar c_{i B} (\bar z) \, , \qquad
  [ {\cal O}_B^- (z , \bar z )]_{AB}= \sum_{i=1}^N c_{i A} (z) \otimes \bar c_{iB}(\bar z) \, , \nonumber
 \end{align}
and their duals
\begin{align}
 [ \tilde {\cal O}_B^+ (z , \bar z )]_{AB}= \sum_{i=1}^N \beta_{i A} (z) \otimes \bar \beta_{iB} (\bar z)\, , \qquad
   [ \tilde{\cal O}_F^+ (z , \bar z )]_{AB} = \sum_{i=1}^N b_{i A} (z) \otimes \bar \beta_{iB} (\bar z)\, , \\
  [\tilde {\cal O}_F^- (z , \bar z )]_{AB}  = \sum_{i=1}^N \beta_{i A} (z) \otimes \bar b_{iB}(\bar z) \, , \qquad
 [ \tilde {\cal O}_B^- (z , \bar z )]_{AB} = \sum_{i=1}^N b_{i A} (z) \otimes \bar b_{iB} (\bar z) \, . \nonumber
 \end{align}
For $M=1$, see \cite{Bergshoeff:1991dz,Moradi:2012xd}.

\section{Comparison of partition function}
\label{PF}

In this appendix we continue to compare the gravity partition function with
the CFT one. In particular, we  include the case with $\mathbf{\Lambda}_{N+M} \neq 0$, which
should be considered outside the free limit.
We focus on the supersymmetric case as it is more interesting,
but the bosonic case can be analyzed in almost the same way.
In the main context, we closely examined how decoupling of $\text{su}(M)$ in the
denominator occurs in the 't Hooft limit. Here we assume the decoupling from the
beginning, and deal with the coset \eqref{scoset2}
\begin{align}
 \frac{\text{su}(N+M)_k \oplus \text{so}(2NM)_1}
 {\text{su}(N)_{k+M} \oplus \text{u}(1)_\kappa}
\label{sWcoset}
\end{align}
with $\kappa = NM(N+M)(N+M+k)$ in the 't Hooft limit, where
$k,N \to \infty$ but $M$ and $\lambda = N/(N+M+k)$ finite.
Since we removed the $\text{su}(M)$,
the states of the coset are then given by the decomposition (see \eqref{decompose} without
the decoupling)
\begin{align}
 \Lambda_{N+M} \otimes \text{NS} = \bigoplus_{\Lambda_N , m}
(\Lambda_{N+M};\Lambda_{N},m) \otimes \Lambda_N \otimes m \, .
\label{decomp}
\end{align}
In the 't Hooft limit, the label $m$ is fixed
by the other labels as in \eqref{largeNu1}, thus the states
are labeled by $(\Lambda_{N+M};\Lambda_N)$.
We consider the diagonal modular invariant as
\begin{align}
 Z^\text{CFT}(q) = \lim_{N,k \to \infty} \sum_{\Lambda_{N+M},\Lambda_N} |sb^{N,M,k}_{(\Lambda_{N+M};\Lambda_N)} (q)|^2 \, ,
\end{align}
where $sb^{N,M,k}_{(\Lambda_{N+M};\Lambda_N)} (q)$ is the branching function of $(\Lambda_{N+M};\Lambda_N)$.
In appendix \ref{PFgravity} we rewrite the gravity partition function in the above form,
and in appendix \ref{PFCFT} we show that the CFT partition function in the 't Hooft limit
reproduces the gravity one.

\subsection{Higher spin partition function}
\label{PFgravity}

With U$(M)$ Chan-Paton factor the gravity partition function is given as \eqref{zmbulks}
\begin{align}
 Z^\text{Bulk}_M = ({\cal Z}_0)^{M^2} (Z^{(1)}_B)^{M^2 -1}
 |{\cal Z}^{\lambda/2}_\text{matter} {\cal Z}^{(1-\lambda)/2}_\text{matter} |^{M^2} \, ,
\label{zmbulk2}
\end{align}
where each contribution is expressed as in \eqref{sZ0}, \eqref{zbs} and \eqref{sZh0}.
In order to compare with the CFT partition function, it is necessary to rewrite the
matter part \eqref{sZh0}
\begin{align}
 \left( {\cal Z}^{h}_\text{matter} \right)^{1/2} = \prod_{n,m=0}^{\infty}\frac{(1 + q^{h+1/2 + n} \bar q^{h+m})(1 + q^{h + n} \bar q^{h+ 1/2 + m})}
 {(1 - q^{h + n} \bar q^{h+m})(1 - q^{h + 1/2 + n} \bar q^{h + 1/2 + m})}
\end{align}
in a suitable form as suggested in \cite{Gaberdiel:2011zw,Candu:2012jq}.

We utilize the formula (A.8) in \cite{Candu:2012jq}
\begin{align}
 \prod_{i,j=1}^\infty \frac{(1 - x_i \eta_j) (1 - y_i \xi_j) }{(1 - x_i y_j) (1 - \xi_i \eta_j)}
 = \sum_{\Lambda} s_\Lambda (x|\xi) s_{\Lambda} (y | \eta) \, .
\end{align}
Here $s_\Lambda (x|\xi) $ is defined as
\begin{align}
 s_\Lambda (x| \xi) = \sum_{T \in \text{STab}_\Lambda} \prod_{j \in T} X_{jj} (-1)^j
\end{align}
with $x=(x_1,x_2,\ldots),\xi=(\xi_1,\xi_2,\ldots)$ and
\begin{align}
 X_{2i,2i} = x_{i+1} \, , \qquad
X_{2i+1 , 2i+1} = \xi_{i+1} \, .
\end{align}
Introducing $M$ sets as $x^{(A)},\xi^{(A)},y^{(A)},\eta^{(A)}$ $(A=1,\ldots,M)$, we have
\begin{align}
& \prod_{A,B=1}^M \prod_{i,j=1}^\infty
 \frac{(1 - x^{(A)}_i \eta^{(B)}_j) (1 - y^{(A)}_i \xi^{(B)}_j) }{(1 - x^{(A)}_i y^{(B)}_j) (1 - \xi^{(A)}_i \eta^{(B)}_j)} \\
 &= \sum_{\Lambda} s_\Lambda (x^{(1)} \cup \cdots \cup x^{(M)}  |\xi^{(1)} \cup \cdots \cup \xi^{(M)}) s_{\Lambda}  (y^{(1)} \cup \cdots \cup y^{(M)}  |\eta^{(1)} \cup \cdots \cup \eta^{(M)}) \, .
\nonumber
\end{align}

Furthermore, we use (A.10) in \cite{Candu:2012jq}
\begin{align}
 s_{\Lambda} (x \cup y | \xi \cup \eta) = \sum_{\Xi , \Pi} c_{\Xi \Pi}^\Lambda
 s_{\Xi} (x | \xi) x_{\Pi} (y | \eta)
\label{sss0}
\end{align}
with $c_{\Xi \Pi}^\Lambda$ as the Clebsh-Gordan coefficients.
Applying the formula successively, we obtain
\begin{align}
 s_\Lambda (x^{(1)} \cup \cdots \cup x^{(M)}  |\xi^{(1)} \cup \cdots \cup \xi^{(M)})
 = \sum_{\Lambda_1 , \ldots , \Lambda_M}
   c^{\Lambda}_{\Lambda_1  \ldots  \Lambda_M}
	 \prod_{A=1}^{M} s_{\Lambda_A} (x^{(A)} | \xi^{(A)}) \, ,
	\label{sss}
\end{align}
where we have defined
\begin{align}
 c^{\Lambda}_{\Lambda_1  \ldots  \Lambda_M}
 = \sum_{\Xi_1 , \ldots , \Xi_{M-2}}    c^{\Lambda}_{\Lambda_1 \Xi_1} \left( \prod_{A=1}^{M-3} c^{\Xi_A}_{\Lambda_{A+1} \Xi_{A+1}} \right)
	c^{\Xi_{M-2}}_{\Lambda_{M-1} \Lambda_M} \, .
\end{align}
Setting
\begin{align}
 x^{(A)}_{i+1} = q^{h+i} \, , \quad \xi^{(A)}_{i+1} = - q^{h+1/2 + i} \, , \quad
 y^{(A)}_{i+1} = \bar q^{h+i} \, , \quad \eta^{(A)}_{i+1} = - \bar q^{h+1/2 + i}
\end{align}
for all $A$, we arrive at
\begin{align}
 ( {\cal Z}^h_\text{matter} )^{M^2/2} = \sum_\Lambda | {\cal B}^h_{\Lambda} |^2 \, , \quad
 {\cal B}^h_\Lambda =
   \sum_{ \Lambda_1 , \ldots , \Lambda_M}
   c^{\Lambda}_{\Lambda_1 \ldots \Lambda_M}
	 \prod_{A=1}^{M} \text{sch}_{\Lambda_A} ({\cal U} (h))
 \nonumber
\end{align}
with the supercharacter defined in \eqref{sschur}.
Thus the gravity partition function \eqref{zmbulk2} can be written as
\begin{align}
 Z^\text{Bulk}_M = ({\cal Z}_0)^{M^2} (Z^{(1)}_B)^{M^2 -1}
\sum_{\Lambda^l_{N+M} , \Lambda^r_{N+M} , \Lambda^l_{N} , \Lambda^r_{N}}
 \left |{\cal B}^{\lambda/2}_{\Lambda^l_{N+M}} {\cal B}^{\lambda/2}_{\Lambda^r_{N+M}}{\cal B}^{(1-\lambda)/2}_{\Lambda^r_{N}}{\cal B}^{(1-\lambda)/2}_{\Lambda^r_{N}} \right |^2 \, .
\label{zmbulk3}
\end{align}

\subsection{CFT partition function}
\label{PFCFT}

Now that we consider the coset \eqref{sWcoset} with the decomposition \eqref{decomp}, the branching function
$sb^{N,M,k}_{\tilde \Xi} (q)$ with $\tilde \Xi = (\Lambda_{N+M};\Lambda_N,m)$ is given by
\begin{align}
 \text{ch}_{\Lambda_{N+M}}^{N+M,k} (q , \imath_1 ( v,w)) \theta (q , \imath_2 (v,w))
= \sum_{\Lambda_N , m} sb^{N,M,k}_{\tilde \Xi} (q)
 \text{ch}_{\Lambda_{N}}^{N,k+M} (q , v )   \Theta^\kappa_m (q,w) \, ,
 \end{align}
where the characters are defined in \eqref{chsu}, \eqref{chu1} and \eqref{chtheta}.
Here the embeddings $\imath_i (v,w)$ $(i=1,2)$ are defined with $u=1$.
At large $k$, we have
\begin{align}
 \text{ch}^{N+M}_{\Lambda_{N+M}} (\imath_1 (v,w)) s\vartheta (q , \imath_2 (v,w)) z^{(1)}_B (q)
  = \sum_{\Lambda_N , m} sa^{N,M}_{\tilde \Xi}  (q) \text{ch}^{N}_{\Lambda_N} (v)
   w^m \, ,
\end{align}
with \eqref{chstheta} and
\begin{align}
 sb^{N,M,k}_{\tilde \Xi} (q) \sim q^{h^{N,M,k}_{\tilde \Xi}} sa^{N,M}_{\tilde \Xi} (q) \, , \qquad
 h^{N,M,k}_{\tilde \Xi} = h^{N+M,k}_{\Lambda_{N+M}} - h^{N,k+M}_{\Lambda_{N}}
 - h ^\kappa_m \, .
\end{align}
We have introduced
\begin{align}
 z^{(1)}_B (q) = \text{ch} ' _M (q,1)
 = \left( \prod_{n=1}^\infty \frac{1}{1 - q^n} \right)^{M^2 - 1}
\end{align}
as in \eqref{nmsum}.
Setting $m = N|\mathbf{\Lambda}_{N+M}|_- - (N+M)|\mathbf{\Lambda}_N|_- $ as in \eqref{largeNu1}, we can remove the factor
$w^m$ as
\begin{align}
 \widetilde{\text{ch}}^{N+M}_{\mathbf{\Lambda}_{N+M}} (\imath_1 (v,w)) s \vartheta (q , \imath_2 (v,w))
  = \sum_{\mathbf{\Lambda}_N} \widetilde{sa}^{N,M}_{\mathbf{\Lambda}_{M+N};\mathbf{\Lambda}_N }  (q) \widetilde{\text{ch}}^{N}_{\mathbf{\Lambda}_N} (v \bar w^{N+M})
\end{align}
by using U$(N)$ characters in \eqref{chu}.
Here we have introduced
\begin{align}
  sa^{N,M}_{\mathbf{\Lambda}_{M+N};\mathbf{\Lambda}_N}  (q) =
 z^{(1)}_B (q)\widetilde{sa}^{N,M}_{\mathbf{\Lambda}_{M+N};\mathbf{\Lambda}_N }  (q) \, .
\end{align}

With $\mathbf{\Lambda}_{M+N} = 0$, the above expression reduces to
\begin{align}
  s \vartheta (q , \imath_2 (v,w))
  = \sum_{\mathbf{\Lambda}_N, \Lambda_M } \widetilde{sa}^{N,M}_{0;\mathbf{\Lambda}_N }  (q) \widetilde{\text{ch}}^{N}_{\mathbf{\Lambda}_N} (v \bar w^{N+M}) \, .
\end{align}
Thus the function $ \widetilde{sa}^{N,M}_{0;\mathbf{\Lambda}_N }  (q)$ can be obtained by decomposing
free bosons and free fermions in terms of U$(N)$ representations. As in section \ref{scfl}, we obtain
\begin{align}
\widetilde{sa}^{M}_{0;0} = \lim_{N \to \infty} \widetilde{sa}^{N,M}_{0;0} = \prod_{s=2}^{\infty} \prod_{n=s}^\infty \left( \frac{(1+q^{n-1/2})^2}{(1-q^{n})(1-q^{n-1})}\right)^{M^2} \, .
\end{align}
For $\mathbf{\Lambda}_N \neq 0 $, we consider the Fock space generated by $(A=1,2,\ldots,M)$
\begin{align}
  \prod_{k=1}^{n_f} \psi^{b_k A}_{-s_k-\frac12}
 \prod_{m=1}^{n_b} j^{d_m A} _{- u_m - 1} | 0 \rangle \, ,
\end{align}
which produces a U$(N)$ tenser of the shape $\Lambda^l_A$.
Summing over all possible $s_k$ and $u_m$ with keeping
$n_f + n_b = |\Lambda^l_A|$, the contribution to the brunching function is written as
(see (3.69) of \cite{Candu:2012jq})
\begin{align}
 \text{sch}_{(\Lambda^l_A)^t} ({\cal U}(\tfrac12)) \, .
\end{align}
Products of characters with $A=1,\ldots,M$ have to be taken, and
the product of the representation $\Lambda^l_A$ should behave as $\Lambda^l_N$.
Thus we find
 \begin{align}
 \widetilde{sa}^{M}_{0;\mathbf{\Lambda}_N} = \lim_{N \to \infty}
 \widetilde{sa}^{N,M}_{0;\mathbf{\Lambda}_N} =
 {\cal B}_{(\Lambda_N^l)^t}^{1/2} {\cal B}_{(\Lambda_N^r)^t}^{1/2}\widetilde{sa}^{M}_{0;0} \, .
\end{align}

In order to consider the case with $\mathbf{\Lambda}_{N+M} \neq 0$,
it is convenient to	introduce the restriction functions $r^{N,M}_{\mathbf{\Lambda} \mathbf{\Phi}}$ as
\begin{align}
 \widetilde{\text{ch}}^{N+M}_{\mathbf{\Lambda}} (\imath_1 (v,w))
 = \sum_{\mathbf{\Phi}} r^{N,M}_{\mathbf{\Lambda} \mathbf{\Phi}} \widetilde{\text{ch}}^{N}_{\mathbf{\Phi}} (v \bar w^{N+M}) \, .
\end{align}
Then from (3.52) of \cite{Candu:2012jq} we have
\begin{align}
 \widetilde{sa}^{N,M}_{\mathbf{\Lambda}_{N+M};\mathbf{\Lambda}_N} (q)
 = \sum_{\mathbf{\Phi} , \mathbf{\Psi}} r^{N,M}_{\mathbf{\Lambda}_{N+M} \mathbf{\Phi}}
c^{(N) \bar{\mathbf{\Psi}}}_{\mathbf{\Phi} \bar{\mathbf{\Lambda}}_N}
 \widetilde{sa}^{N,M}_{0;\mathbf{\Psi}} (q) \, ,
\end{align}
where $c^{(N) \bar{\mathbf{\Psi}}}_{\mathbf{\Phi} \bar{\mathbf{\Lambda}}_N}$
are U$(N)$ Clebsh-Gordan coefficients. We will use later that
\begin{align}
 \lim_{N \to \infty}  r^{N,M}_{\mathbf{\Lambda} \mathbf{\Phi}}
 \to r_{\Lambda^l \Phi^l} r_{\Lambda^r \Phi^r} \, , \qquad
 \lim_{N \to \infty}  c^{(N) \bar{\mathbf{\Psi}}}_{\mathbf{\Lambda} \bar{\mathbf{\Phi}}}
 \to c_{\Lambda^l \Phi^r}^{\Psi^r} c_{\Lambda^r \Phi^l}^{\Psi^l} \, .
\end{align}
and
\begin{align}
\sum_{\Lambda} c^\Lambda_{\Xi \Phi } {\cal B}^{1/2}_\Lambda = {\cal B}^{1/2}_\Xi {\cal B}^{1/2}_\Phi \, ,
\end{align}
which comes from
\begin{align}
 & \sum_{\Lambda} c^\Lambda_{\Xi \Phi} s_\Lambda (x^{(1)} \cup \cdots \cup x^{(M)}  |\xi^{(1)} \cup \cdots \cup \xi^{(M)}) \\
 & = s_\Xi (x^{(1)} \cup \cdots \cup x^{(M)}  |\xi^{(1)} \cup \cdots \cup \xi^{(M)})
 s_\Phi (x^{(1)} \cup \cdots \cup x^{(M)}  |\xi^{(1)} \cup \cdots \cup \xi^{(M)}) \, . \nonumber
\end{align}

Let us introduce
\begin{align}
 x^{(M+1)} = (w_1, \ldots, w_M , 0 , 0, \ldots) \, , \qquad
 \xi^{(M+1)} = (0,0,\ldots) \, ,
\end{align}
 then $s_\Pi (x^{(M+1)}|\xi^{(M+1)})$ becomes a  U$(M)$ character.
Setting $w_A = 1$ for all $A$, we have
\begin{align}
& s_\Lambda (x^{(1)} \cup \cdots \cup x^{(M+1)}  |\xi^{(1)} \cup \cdots \cup \xi^{(M+1)}) \\
 & \qquad = s_\Lambda (( x^{(1)} \cup \{1\} ) \cup \cdots \cup ( x^{(M)} \cup \{1\}  )
 | \xi^{(1)} \cup \cdots \cup \xi^{(M)})\nonumber \\
 & \qquad = \sum_{ \Pi} r_{\Lambda \Pi}
s_\Pi (x^{(1)} \cup \cdots \cup x^{(M)}  |\xi^{(1)} \cup \cdots \cup \xi^{(M)}) \, , \nonumber
\end{align}
where \eqref{sss0} has been used and
$r_{\Lambda \Pi} = \sum_{\Xi} c^{\Lambda}_{\Xi \Pi} \, \text{dim} \, \Xi$ in this case.%
\footnote{With $w_A =1$ for all $A$, we have dim $\Xi = s_\Xi (x^{(M+1)}|0)$.}
With the second equality and (A.13) of \cite{Candu:2012jq}, we arrive at
\begin{align}
 {\cal B}^0_\Lambda = \sum_{\Xi} r_{\Lambda \Xi} {\cal B}^{1/2}_{\Xi^t} \, .
\end{align}
Noticing that
\begin{align}
\lim_{N,k,\to \infty} h^{N,M,k}_{\tilde \Xi}
 = \frac{\lambda}{2} (|\mathbf{\Lambda}_{N+M}| - |\mathbf{\Lambda}_{N}|)
\end{align}
and $|\Lambda | = \sum_{A=1}^M |\Lambda_A| $ for non zero $c^\Lambda_{\Lambda_1 \ldots \Lambda_A}$,
the branching function is computed as
\begin{align}
 \lim_{N,k \to \infty} b^{N,M,k}_{\mathbf{\Lambda}_{N+M}; \mathbf{\Lambda}_N}
 = {\cal B}^{\lambda/2}_{\Lambda^l_{N+M}} {\cal B}^{\lambda/2}_{\Lambda^r_{N+M}}
{\cal B}^{(1-\lambda)/2}_{(\Lambda^l_{N})^t} {\cal B}^{(1-\lambda)/2}_{(\Lambda_N^r)^t}
\widetilde{sa}^M_{0;0} (q) z^{(1)}_B (q)\, .
\end{align}
Considering the diagonal modular invariant
\begin{align}
 Z^\text{CFT} (q) = \lim_{N,k \to \infty}
 \sum_{\mathbf{\Lambda}_{N+M}; \mathbf{\Lambda}_N}|b^{N,M,k}_{\mathbf{\Lambda}_{N+M}; \mathbf{\Lambda}_N}|^2 \, ,
\end{align}
we can show that the CFT partition partition function defined above
reproduces the gravity one given in \eqref{zmbulk3}.

\end{document}